%% file: main_v16_submit_PNAS_arxiv_ver2.tex
\documentclass[reprint,preprintnumbers,amsmath,amssymb,aps,nofootinbib,showkeys, superscriptaddress,floatfix]{revtex4-2}

\usepackage{graphicx}
\usepackage{dcolumn} %
\usepackage{bm}      %
\usepackage{mathrsfs}
\usepackage{amsthm}
\usepackage{amsmath}  %
\usepackage{amsfonts}  %
\usepackage{latexsym}  %
\usepackage{amssymb,bbm}  %
\usepackage[table]{xcolor}
\usepackage{lineno}

\definecolor{SITableHeader}{HTML}{F0D5CF}
\definecolor{SITableHighlight}{HTML}{FFF7DE}
\definecolor{SITableRule}{HTML}{5C5C5C}

\usepackage{textcomp}
\usepackage[utf8]{inputenc}
\usepackage{booktabs}
\usepackage{algorithm}
\usepackage{algpseudocode}
\usepackage{listings}
\usepackage{upgreek}
\usepackage{braket}
\usepackage[colorlinks=true,allcolors=blue]{hyperref}  %
\usepackage{lipsum}  %
\usepackage{newunicodechar}
\newunicodechar{′}{\ensuremath{'} }
\usepackage{setspace}
\usepackage{musixtex}
\usepackage{ragged2e}
\usepackage{placeins}
\usepackage{caption}
\usepackage{array}

\makeatletter
\newif\ifsuppTOC
\suppTOCfalse  %

\let\orig@addcontentsline\addcontentsline
\renewcommand{\addcontentsline}[3]{%
  \ifsuppTOC
    \edef\tempa{#1}\edef\tempb{toc}%
    \ifx\tempa\tempb
      \orig@addcontentsline{stoc}{#2}{#3}%
    \else
      \orig@addcontentsline{#1}{#2}{#3}%
    \fi
  \else
    \orig@addcontentsline{#1}{#2}{#3}%
  \fi
}

\newcommand{\supplementtableofcontents}{%
  \phantomsection
  \pdfbookmark[1]{Contents}{supp-contents}%
  \begingroup
  \parskip=0pt
  \@starttoc{stoc}%
  \endgroup
}
\makeatother

\newcommand{\beginsupplement}{%
  \suppTOCtrue %
  \setcounter{table}{0}%
  \renewcommand{\tablename}{Table}%
  \renewcommand{\thetable}{T\arabic{table}}%
  \setcounter{figure}{0}%
  \renewcommand{\thefigure}{S\arabic{figure}}%
  \setcounter{section}{0}%
  \renewcommand{\thesection}{S\arabic{section}}%
  \setcounter{page}{1}%
}

\input{Commands3.tex}

\theoremstyle{plain}

\theoremstyle{definition}

\theoremstyle{remark}

\newcommand{\affA}{Department of Chemistry, University of California, Berkeley, Berkeley, CA 94720, USA.}
\newcommand{\affB}{Department of Chemical and Biomolecular Engineering, University of California, Berkeley,
CA 94720, USA.}

\newcommand{\affE}{Chemical Sciences Division,  Lawrence Berkeley National Laboratory,  Berkeley, CA 94720, USA.}
\newcommand{\affH}{Molecular Foundry,  Lawrence Berkeley National Laboratory,  Berkeley, CA 94720, USA.}

\begin{document}
\title{Room-temperature optically addressable and coherently controlled triplet spins in rapidly grown molecular microcrystals}

\author{Madhur Parashar}\affiliation{\affA}
\author{Guangzhao Chen}\affiliation{\affH}\affiliation{\affE}
\author{Emanuel Druga}\affiliation{\affA}\affiliation{\affE}
\author{Liang Z. Tan}\affiliation{\affH}
\author{Jeffrey Reimer}\affiliation{\affB}
\author{Ashok Ajoy}\email{ashokaj@berkeley.edu}\affiliation{\affA}\affiliation{\affE}

\begin{abstract}
Quantum sensing increasingly demands materials combining room-temperature optical spin addressability and coherent control with chemical tunability and accessible fabrication. Yet semiconductor platforms require specialized fabrication, while molecular systems often depend on tailored synthesis or involved crystal growth. Here we apply an ambient, minute-timescale growth process in which crystallization is directly observable by optical microscopy, forming thin organic molecular crystals that host photoexcited triplet-spin ensembles. Monitored \emph{in situ}, growth yields optically clear, stoichiometric single crystals of excellent quality that provide geometries suited to near-surface sensing and wide-field optical spin interrogation. Donor--acceptor charge--transfer cocrystals pairing anthracene or biphenyl with tetracyanobenzene, together with localized--exciton pentacene:pentacenequinone (P:PQ), exhibit room-temperature optically detected magnetic resonance (ODMR). Charge-transfer ODMR resonances reach linewidths below $4~\mathrm{MHz}$, consistent with motional narrowing of mobile triplet excitons, while P:PQ supports coherent ensemble spin control with $T_2=0.56~\mu\mathrm{s}$. Functionality across chemically and electronically distinct pairs establishes our growth--spectroscopy route for traversing the molecular triplet-spin design space, connecting molecular crystal engineering with chemically tunable spin materials and sensor geometries for quantum sensing and microscopy.
\end{abstract}

\maketitle

\medskip
\medskip
Quantum sensors translate minute perturbations of a quantum state into measurements of magnetic and electric fields, temperature and other local properties, enabling applications across condensed-matter physics, materials characterization and biology \cite{Awschalom2025QuantumHardwareChallenges,Aslam2023QuantumSensorsBiomedical,Degen2017QuantumSensing,Casola2018NVCondensedMatter}. Optically addressable spins are particularly powerful because optical initialization and readout can be combined with coherent control under ambient conditions. When confined to a thin layer, these spins can be brought close to a target to provide wide-field images of spatially heterogeneous systems \cite{Levine2018QuantumDiamondMicroscope}. The expanding reach of these measurements has created growing demand for sensing materials whose optical, spin and interfacial properties can be tailored to different operating environments.

Solid-state quantum sensing has so far been dominated by optically active defects, especially nitrogen-vacancy centers in diamond and related defects in wide-bandgap semiconductors such as silicon carbide \cite{Taylor2008DiamondMagnetometerNanoscale,Christle2015SiCMillisecondCoherence,Tesiman2026SurveyOpticallyAddressableSpinQubits,Awschalom2018OpticallyInterfacedSpins,Lukin2020SiCIntegratedPhotonics}. In parallel, optically addressable spins are being explored in silicon, van der Waals materials, chemically tunable organometallic complexes and organic molecular systems at cryogenic and room temperature \cite{Higginbottom2022OpticalSingleSpinsSilicon,Gottscholl2020hBNSpinDefectsRoomTemp,Bayliss2020OpticallyAddressableMolecularSpins,Laorenza2021Cr4MolecularColorCenters,Singh2025OrganicCrystalQuantumSensing,Mena2024MolecularSpinCoherentControl,Kopp2024LuminescentTripletDiradicals,Mann2025ChemicallyTunedPulsedODMR,Roggors2025CarbeneMolecularQubits,Li2026PentaceneACVectorSensing,Zheng2026SurfaceScaffoldedMolecularQubit,Zhou2026OpticallyAddressableMolecularSpins,Qiu2023NarrowLinewidthChromophoreRadicalQubit}. Diamond and related defect platforms combine favorable spin properties with ambient optical readout, but often require substantial materials and fabrication infrastructure. High-purity diamond and SiC crystals are expensive, while controlled creation of shallow defect layers generally requires specialized crystal growth, ion implantation, plasma processing and annealing \cite{Healey2020NVLayerFormation}. Related fabrication challenges accompany emerging defect spins in other semiconductors and van der Waals materials \cite{Kianinia2020hBNSpinDefectGeneration,Whitefield2026hBNOpticallyAddressableSpins}. The need for specialized infrastructure, together with the limited range of accessible hosts, makes it difficult to produce and systematically compare new quantum-sensing materials.

Molecular materials offer a fundamentally different opportunity. Chemical composition and crystal packing provide a vast design space in which optical and spin properties can be created or modified through molecular choice and intermolecular interactions \cite{Krzystek1993TripletExcitons,Sun2018MolecularCocrystals,Liu2023DonorAcceptorCocrystalsReview,Imahori2021DynamicExcitonCTStates,Wasielewski2020ChemistryQuantumInformation,Palmer2025PackingSuppressesHopping,Williams2022UltrafastTripletCocrystals}. These variables can tune emission colour \cite{Barman2023ColorTunableCocrystals,Zhu2026HierarchicalCocrystalAssemblies,Chen2022RainbowWhiteEmissionCocrystals}, triplet-formation pathways \cite{Yu2024NIROptoelectronicCocrystals,Bao2025HiddenTripletStates} and zero-field splitting \cite{Krzystek1993TripletExcitons,Attwood2025SpinRelaxationTripletMedia}. Recent observations of room-temperature optically detected magnetic resonance (ODMR) and coherent spin control in pentacene-based and related organic crystals illustrate this potential \cite{Singh2025OrganicCrystalQuantumSensing,Mena2024MolecularSpinCoherentControl,Li2026PentaceneACVectorSensing,Kopp2024LuminescentTripletDiradicals,Roggors2025CarbeneMolecularQubits,Mann2025ChemicallyTunedPulsedODMR,Zheng2026SurfaceScaffoldedMolecularQubit,Zhou2026OpticallyAddressableMolecularSpins,Qiu2023NarrowLinewidthChromophoreRadicalQubit}. Yet the breadth of chemical space also creates a practical bottleneck. Candidate systems can require tailored, multistep synthesis followed by molecule-specific bulk-crystal growth or vacuum deposition \cite{Jiang2013OrganicSemiconductorCrystalGrowth,Cui2020PentaceneDopedPTerphenylMaser}. Even when suitable molecules are commercially available, preparing them as thin, high-quality crystals on a substrate remains a separate optimization problem. Consequently, the large molecular design space remains difficult to screen in practice.

\begin{figure}[t]
    \centering
   \includegraphics[width=0.45\textwidth]{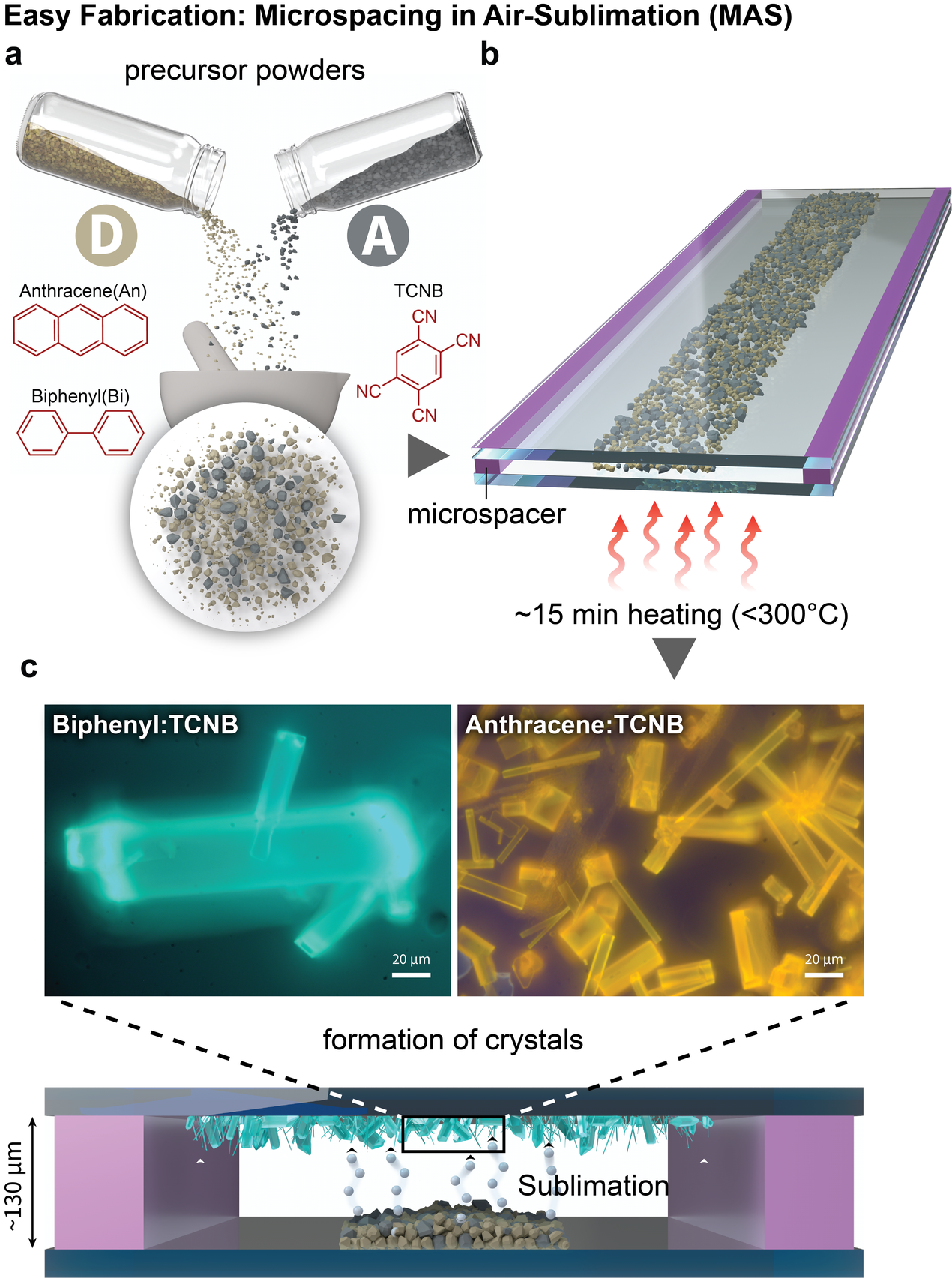}
    \caption{\justifying{\T{MAS growth of photoexcited triplet-hosting co-crystals.}
    (a) Anthracene (An) or biphenyl (Bi) donors are combined with 1,2,4,5-tetracyanobenzene (TCNB) acceptor and co-ground as a mixed precursor. D/A indicate precursors of donor(D):acceptor(A) or chromophore(D):lattice(A) co-crystals (b) Microgram-scale precursor is loaded between glass substrates separated by a micrometre-scale spacer. Heating for $\sim$15~min at $<300~^\circ$C drives co-sublimation and short-path transport across the confined air gap. (c) Crystallization on the upper substrate yields green-emitting Bi:TCNB and orange-emitting An:TCNB under 405~nm excitation. Cross-section shows substrate-bound crystals within a representative $\sim$130~$\mu$m gap. Scale bars, 20~$\mu$m.}}
    \label{fig:main_fig_1}
\end{figure}

\begin{figure*}[t]
    \centering
    \includegraphics[width=0.95\textwidth]{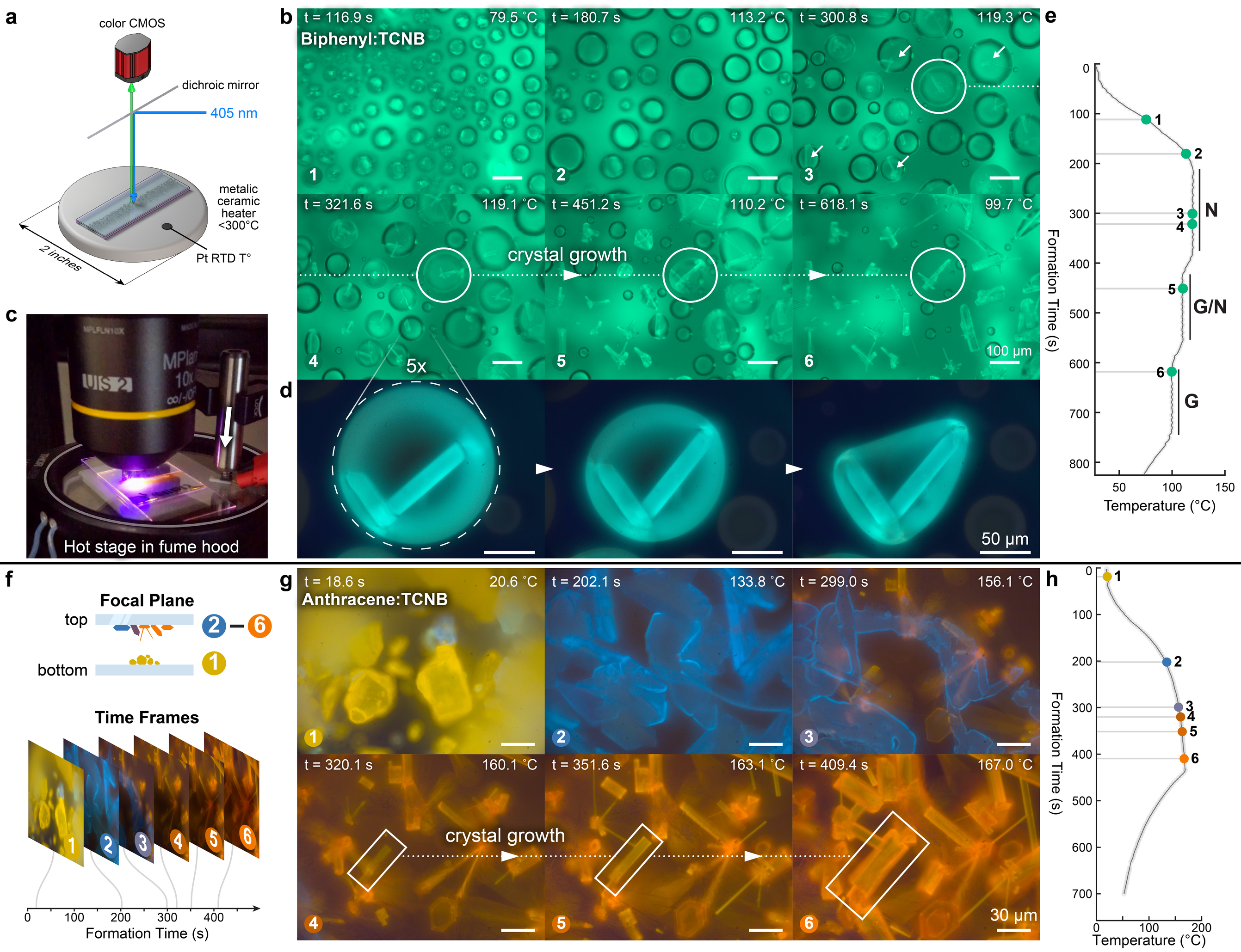}
    \caption{\justifying{\T{In situ tracking of temperature-programmed MAS co-crystal growth.}
    (a) Wide-field fluorescence microscope and hot-stage assembly with 405~nm excitation, colour CMOS detection and resistance temperature detector (RTD) readout. \T{\I{Droplet-mediated Bi:TCNB.}} (b) Frames 1--6 show melt condensation and ripening, droplet nucleation (white arrows), crystal expansion and subsequent cooling. (c) In-fume-hood implementation; arrow marks the RTD. (d) Fivefold-magnified Bi:TCNB crystal consuming its parent droplet. (e) Temperature trajectory for (b), with nucleation (N) and growth (G) regimes. \T{\I{Vapour-mediated An:TCNB.}} (f) Focal-plane and timing schematic: frame 1 images precursor on the lower substrate; frames 2--6 image deposition and growth on the upper substrate. (g) Frames show transient blue-emitting anthracene, emergence of orange An:TCNB and crystal expansion. (h) Temperature trajectory for (g). Scale bars as indicated.}}
    \label{fig:main_fig_2}
\end{figure*}

\begin{figure*}[t]
    \centering
    \includegraphics[width=0.95\textwidth]{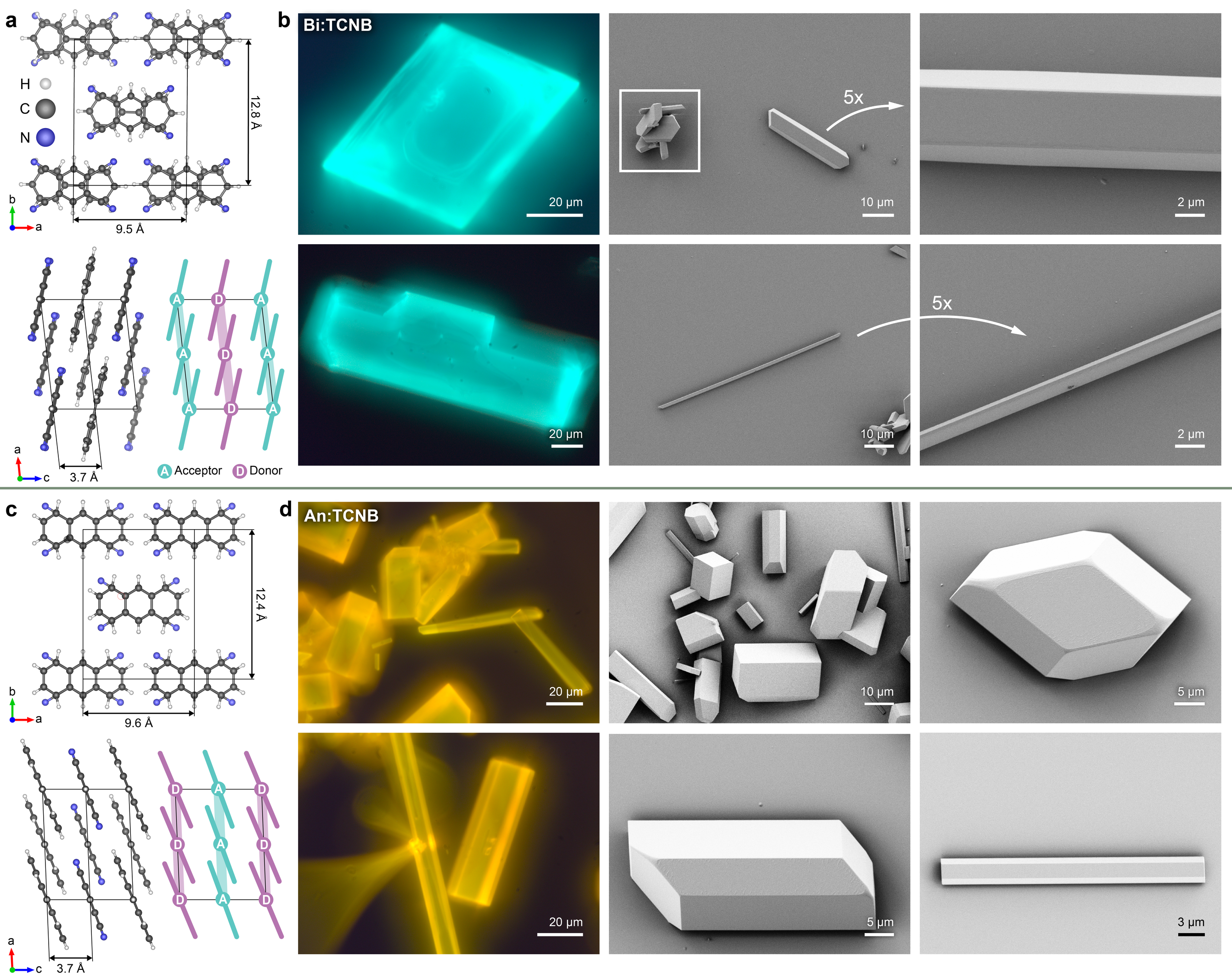}
    \caption{\justifying{\T{Crystal structure and morphology of MAS-grown co-crystals.}
    \textbf{(a,b) Biphenyl--tetracyanobenzene (Bi:TCNB).}
    (a) Triclinic $P\overline{1}$ 1:1 Bi:TCNB packing (CIF~1116344, \cite{Pasimeni1983BiphenylTCNBTripletExcitons}) viewed in the \textit{ab} and \textit{ac} planes; N atoms are blue, with unit-cell dimensions and donor--acceptor layer separation (\textit{c}/2) marked.
    (b) UV-excited fluorescence and SEM of MAS-grown plates and needles; rightmost panels show $\sim$5$\times$ magnified views and nanoscale-uniform surfaces.
    \textbf{(c,d) Anthracene--tetracyanobenzene (An:TCNB).}
    (c) Monoclinic $C2/m$ 1:1 An:TCNB packing (CIF~1103089, \cite{Lefebvre1989ATCNBNTCNBDisorder}) viewed in the \textit{ab} and \textit{ac} planes.
    (d) UV-excited fluorescence and SEM of faceted, mirror-symmetric crystals consistent with $C2/m$ symmetry. Scale bars as indicated.}}
    \label{fig:main_fig_3}
\end{figure*}

Here we show that microspacing in-air sublimation (MAS) provides a rapid and general route from molecular precursors to sensing-ready ODMR-active crystalline materials \cite{Ye2018MicrospacingSublimation,Ye2019CocrystalsMicrospacing}. In this ambient, minute-timescale process, molecular pairs co-sublime within a confined gap between glass substrates and crystallize on the top substrate, allowing nucleation and crystallization events to be observed directly by \emph{in situ} fluorescence microscopy and time-resolved emission spectroscopy  (Fig.~\ref{fig:main_fig_1} and Fig.~\ref{fig:main_fig_2}). This growth yields dense, stoichiometrically defined single crystals with smooth nanoscale-homogeneous surfaces and thicknesses spanning $\sim$100~nm--15~\textmu m, a form factor well matched to near-surface sensing and wide-field optical spin interrogation \cite{Tetienne2017GrapheneCurrentImaging,Levine2018QuantumDiamondMicroscope,Healey2023vdWHeterostructureQuantumMicroscopy,Allert2022NanoMicroscaleNMR,Xie2022BiocompatibleDiamondSensor,Briegel2025WidefieldNMRMicroscopy} (Fig.~\ref{fig:main_fig_3}).

\begin{figure*}[t]
    \centering
   \includegraphics[width=0.95\textwidth]{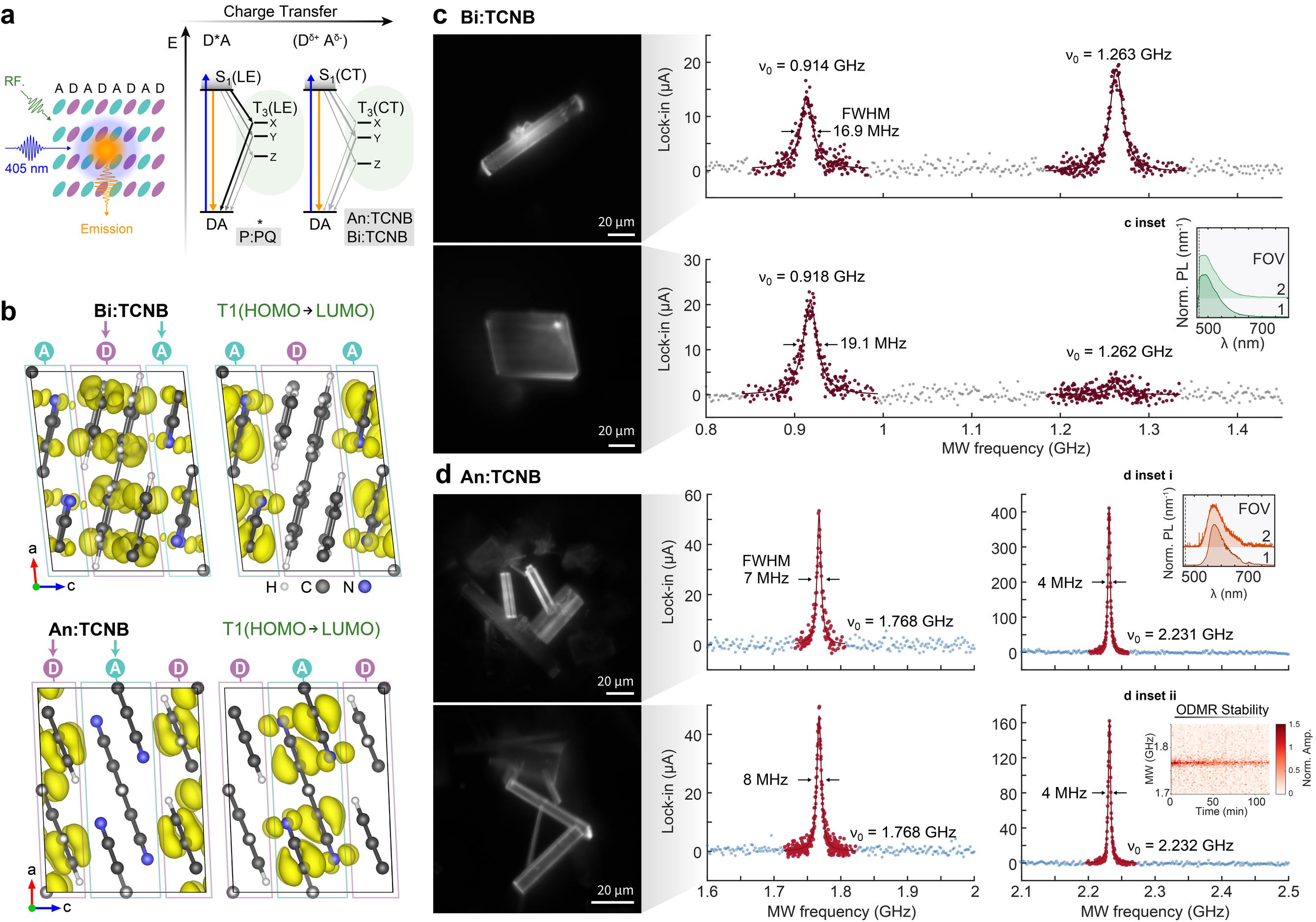}
    \caption{\justifying{\T{Room-temperature near-zero-field ODMR in MAS-grown co-crystals.}
    (a) \T{\I{Spin-optical readout.}} 405~nm excitation forms locally excited ($^1$LE) or charge-transfer ($^1$CT) singlets; intersystem crossing populates the corresponding zero-field-split triplets ($^3$LE or $^3$CT). Resonant microwaves redistribute triplet sublevels and modulate photoluminescence. P:PQ lies toward the localized limit, and An:TCNB and Bi:TCNB toward higher charge-transfer character.
    (b) \T{\I{Triplet electronic structure.}} Calculated HOMO and LUMO isosurfaces for the lowest triplet configuration; faded boxes mark donor (D) and acceptor (A) stacks. Bi:TCNB shows greater donor--acceptor delocalization than An:TCNB.
    (c) \T{\I{Bi:TCNB ODMR.}} Elongated crystal and plate fields of view (FOVs) show resonance pairs ($2|E|$/$D-|E|$) at 0.914/1.263 and 0.918/1.262~GHz, with lower-frequency linewidths of 16.9 and 19.1~MHz. Inset: green PL from both FOVs.
    (d) \T{\I{An:TCNB ODMR.}} Compact multicrystal and elongated-needle FOVs show pairs ($D-|E|$/$D+|E|$) at 1.768/2.231 and 1.768/2.232~GHz; lower-/higher-frequency linewidths are 7/4 and 8/4~MHz. Insets: (i) orange PL from both FOVs and (ii) $>60$~min ODMR stability. $D$ and $E$ are the triplet-spin axial and transverse zero-field splitting parameters;arrows mark linewidths; non-linear fits are Lorentzian profiles; scale bars, 20~\textmu m.}}
    \label{fig:main_fig_4}
\end{figure*}

\begin{figure*}[t]
    \centering
    \includegraphics[width=0.95\textwidth]{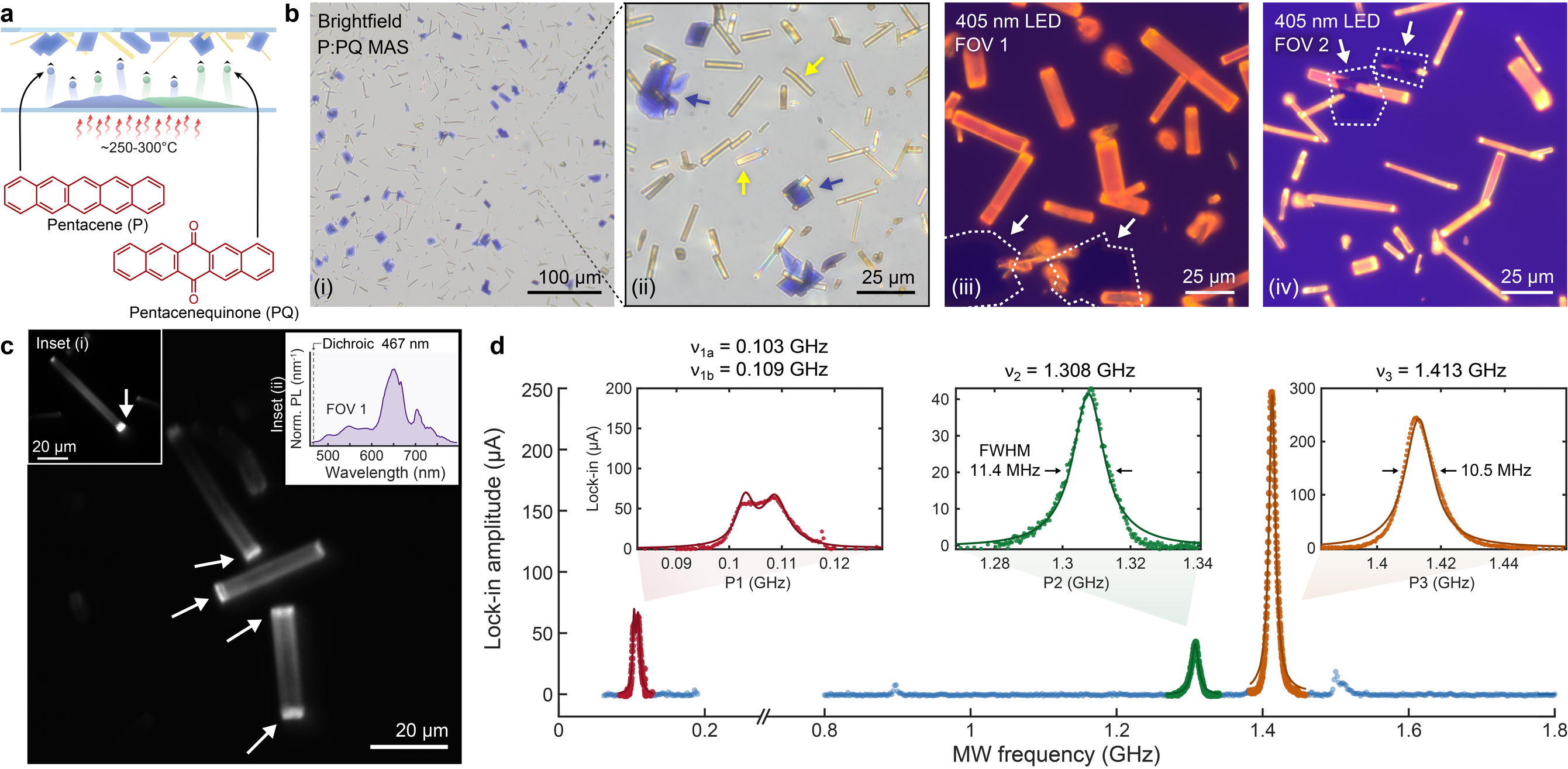}
    \caption{\justifying{\T{Room-temperature ODMR in MAS-grown pentacene:pentacenequinone microneedles.}
    \textbf{(a) MAS growth.}
    Co-sublimation of pentacene (P) and pentacenequinone (PQ) at $\sim$250--300~$^\circ$C yields known blue pentacene plates \cite{Ye2018MicrospacingSublimation} and distinct red-emissive microneedles.
    \textbf{(b) Phase-selective optical identification.}
    Bright-field images (i,ii) distinguish blue pentacene plates (blue arrows) from elongated yellow microneedles (yellow arrows); under 405~nm excitation (iii,iv), only microneedles emit bright red light, while dashed outlines mark non-emissive pentacene crystals.
    \textbf{(c) Localized-exciton optical response.}
    Fluorescence field of view (FOV) shows end-emission (white arrows); insets show (i) magnified waveguiding example and (ii) structured PL resembling pentacene doped in \textit{p}-terphenyl, consistent with an isolated-emitter-like pentacene environment.
    \textbf{(d) Room-temperature ODMR.}
    Same FOV shows three resonances: Lorentzian fits give (i) a bimodal feature ($2|E|$) at $\nu_{1a}=0.103$~GHz and $\nu_{1b}=0.109$~GHz, and (ii,iii) transitions ($D-|E|$/$D+|E|$) at $\nu_2=1.308$~GHz and $\nu_3=1.413$~GHz with linewidths of 11.4 and 10.5~MHz, respectively.}}
    \label{fig:main_fig_5}
\end{figure*}

\begin{figure*}[t]
    \centering
   \includegraphics[width=0.95\textwidth]{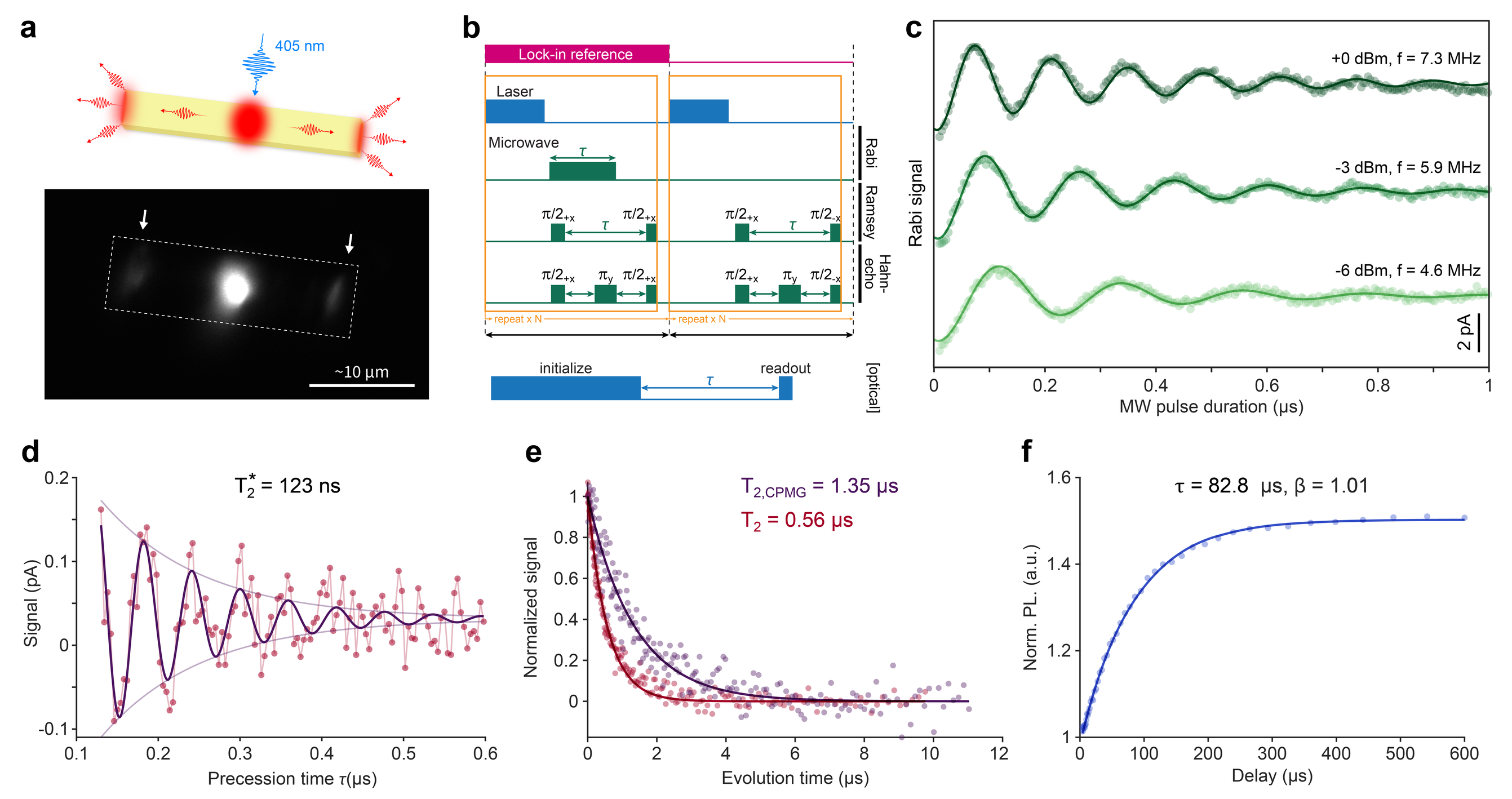}
    \caption{\justifying{\T{Room-temperature coherent control of photoexcited triplets in an MAS-grown P:PQ microneedle.}
    (a) \T{\I{Optical addressing and waveguiding.}} Focused 405~nm excitation addresses a $\sim$3~\textmu m region; red PL propagates to both needle ends (arrows). Scale bar, $\sim$10~\textmu m.
    (b) \T{\I{Pulse sequences.}} Upper: laser initialization, phase-controlled microwave manipulation and optical readout within a lock-in envelope for Rabi, Ramsey and Hahn echo. Lower: ground-state recovery after variable dark delay $\tau$.
    (c) \T{\I{Rabi oscillations.}} Rabi driving at three microwave input powers before amplification: $+0$~dBm (7.3~MHz), $-3$~dBm (5.9~MHz) and $-6$~dBm (4.6~MHz).
    (d) \T{\I{Ramsey fringes.}} A 17~MHz detuning gives $T_2^*=123$~ns.
    (e) \T{\I{Hahn echo and CPMG.}} Hahn echo gives $T_2=0.56~\mu$s; four-$\pi$-pulse CPMG extends coherence to $T_{2,\mathrm{CPMG}}=1.35~\mu$s.
    (f) \T{\I{Ground-state recovery.}} PL recovery gives $\tau=82.8~\mu$s and stretched-exponential $\beta=1.01$.}}
    \label{fig:main_fig_6}
\end{figure*}

We demonstrate room-temperature ODMR and coherent spin control across chemically and electronically distinct MAS-grown crystals. Anthracene:tetracyanobenzene (An:TCNB) and biphenyl:tetracyanobenzene (Bi:TCNB) donor--acceptor co-crystals exhibit ODMR from mobile triplet excitons with linewidths as narrow as 4~MHz (Fig.~\ref{fig:main_fig_4}); their distinct zero-field splittings reflect variable triplet-spin delocalization across molecular stacks. Pentacene:pentacenequinone (P:PQ) microneedles host localized triplet excitons, exhibit ODMR (Fig.~\ref{fig:main_fig_5}) and support coherent ensemble spin control (Fig.~\ref{fig:main_fig_6}), including Rabi oscillations, Ramsey fringes, Hahn-echo coherence of $T_2=0.56~\mu$s and $T_{2,\mathrm{CPMG}}=1.35~\mu$s under dynamical decoupling. Together, these results establish ODMR and coherent control across distinct molecular motifs, while identifying triplet localization and mobility as molecular-design variables for balancing narrow room-temperature ODMR with persistent coherent control. More broadly, this growth--spectroscopy process can be extended to other organic co-crystal coformers \cite{Ye2019CocrystalsMicrospacing,Sun2018MolecularCocrystals,Liu2023DonorAcceptorCocrystalsReview,Corpinot2019PracticalGuideMolecularCrystals,Jiang2021GNNCocrystalDiscovery}.

\vspace{0.5em}

\noi{\normalsize\bfseries In situ growth reveals distinct co-crystallization pathways\par}
\noi
Figs.~\ref{fig:main_fig_1} and~\ref{fig:main_fig_2} first introduce MAS using donor--acceptor charge-transfer co-crystals formed by anthracene (An) or biphenyl (Bi) with 1,2,4,5-tetracyanobenzene (TCNB), denoted An:TCNB and Bi:TCNB, respectively. These optically distinct systems allow their growth to be followed directly. Fig.~\ref{fig:main_fig_1}a--c summarizes the three-step workflow. The donor and acceptor are co-ground into a mixed precursor powder (Fig.~\ref{fig:main_fig_1}a), and a microgram-scale quantity is placed between glass coverslips separated by a micrometre-scale spacer (Fig.~\ref{fig:main_fig_1}b). Heating the unsealed assembly for $\sim$15~min drives co-sublimation across the confined air gap. Crystallization on the upper coverslip yields green-fluorescent Bi:TCNB and orange-fluorescent An:TCNB microcrystals (Fig.~\ref{fig:main_fig_1}c). Once the temperature window is known, the process requires only glass slides, microspacers and a hot plate inside a fume hood, without vacuum or inert-gas handling. Substrate preparation and assembly are detailed in SI Secs.~\ref{sec:SI_OTS} and~\ref{sec:SI_bottom}.

To resolve crystal formation and optimize pairs with unknown co-crystallization behaviour, we built a fume-hood-compatible wide-field fluorescence microscope around a programmable hot stage (Fig.~\ref{fig:main_fig_2}a). Co-registered fluorescence and temperature readouts assign individual nucleation and growth events to the applied trajectory, while in situ emission spectroscopy independently reports co-crystal formation. Instrument construction and time-resolved measurements are described in SI Secs.~\ref{sec:SI_scope} and~\ref{sec:SI_spectrum} and Figs.~\ref{fig:supple_fumehood_scope} and~\ref{fig:mas_cocrystallization_pl}. Notably, the same apparatus directly resolves two distinct crystallization pathways: nucleation from a condensed droplet melt and direct nucleation from the vapour phase \cite{Ye2018MicrospacingSublimation,Ye2019CocrystalsMicrospacing}.

Bi:TCNB, not previously prepared by MAS to our knowledge, predominantly follows the droplet-mediated pathway (Fig.~\ref{fig:main_fig_2}b--e). Co-sublimed material first condenses as isolated liquid droplets at $79.5~^\circ$C (frame 1, Fig.~\ref{fig:main_fig_2}b). The droplets grow larger by Ostwald ripening as the temperature reaches $113.2~^\circ$C (frame 2) \cite{Ye2018MicrospacingSublimation}. Near the $119~^\circ$C maximum, nuclei emerge within individual droplets (frames 3 and 4), which then grow laterally as the crystals consume their parent droplet until most have fully absorbed it (frames 5 and 6); subsequent cooling is then initiated. The magnified sequence captures single-crystal growth by consumption of the parent melt (Fig.~\ref{fig:main_fig_2}d), while the thermal trace maps its narrow nucleation and growth windows (Fig.~\ref{fig:main_fig_2}e). Bi:TCNB is therefore sensitive to thermal history: premature cooling traps needles within a solidified precursor matrix, whereas prolonged heating compromises single-crystalline order. An ambient-phase octadecyltrichlorosilane (OTS) nanolayer prepared entirely in a fume hood without vacuum deposition alters condensate wetting to form more circular, spatially separated droplets, providing a simple handle over nucleation density and placement. These surface effects and thermal failure modes are detailed in SI Secs.~\ref{sec:SI_OTS_effect} and~\ref{sec:SI_fragile} and Figs.~\ref{fig:supp_ots_growth_effect}, \ref{fig:supp_precool_btcnb} and~\ref{supp_fig_sem_broken}.

An:TCNB instead follows a vapour-mediated pathway (Fig.~\ref{fig:main_fig_2}f--h). Fig.~\ref{fig:main_fig_2}f distinguishes the precursor below (frame 1) from deposition and growth on the top surface (frames 2--6). Blue-emitting anthracene flakes appear at $133.8~^\circ$C (frame 2, Fig.~\ref{fig:main_fig_2}g), before orange An:TCNB emerges near $156~^\circ$C (frame 3). Directly capturing this transient anthracene intermediate reveals sequential co-crystallization otherwise obscured by post-growth characterization. Faceted crystals then nucleate near $160~^\circ$C and expand to $167~^\circ$C (frames 4--6), placing both nucleation and growth on the rising ramp (Fig.~\ref{fig:main_fig_2}h). At lower loading, An:TCNB forms without the resolved blue phase; spacer thickness and loading select isolated plates or multi-needle floral structures (SI Sec.~\ref{sec:SI_videos} and Supplementary Videos~5--8).

These pathway-specific rules provide a practical strategy for screening new molecular pairs. For vapour-mediated growth, heating continues until a new fluorescent phase or faceted crystal appears; for droplet-mediated growth, cooling begins at nucleation. An:TCNB tolerates substantial variation in this trajectory, whereas Bi:TCNB requires tighter control of its nucleation and cooling windows. By linking phase formation and crystal growth to temperature, in situ imaging makes growth observable and rapidly optimizable.

\vspace{0.5em}
\noi{\normalsize\bfseries Crystal structure and optical waveguiding in MAS-grown co-crystals\par}
\noi
We next examine the structure and morphology of the MAS-grown Bi:TCNB and An:TCNB microcrystals used for ODMR (Fig.~\ref{fig:main_fig_4}). Fig.~\ref{fig:main_fig_3} combines established crystallographic models with fluorescence microscopy and SEM. Molecular packing defines the photoexcited triplet environment, while low strain, optical clarity and smooth surfaces are critical for quantum sensing. We demonstrate that MAS realizes these bulk motifs directly as substrate-bound, single-crystalline microcrystals with strong emission and intrinsic waveguiding.

Fig.~\ref{fig:main_fig_3}a first shows the triclinic $P\overline{1}$ Bi:TCNB lattice (CIF~1116344) \cite{Pasimeni1983BiphenylTCNBTripletExcitons}. Alternating 1:1 donor--acceptor layers form ordered, triplet-hosting stacks separated by $\sim$3.7~\AA{}. Fig.~\ref{fig:main_fig_3}b then links this packing to the MAS-grown morphology. Green fluorescence reveals plates and elongated needles, with intense edge and corner emission demonstrating excellent intrinsic optical waveguiding. SEM resolves isolated plates, high-aspect-ratio needles and multi-nucleated ``floral'' assemblies. Magnified views show nanoscale-uniform basal planes and sharply defined sidewalls, consistent with high single-crystal quality. The coexisting plate and needle habits are consistent with packing-directed anisotropic growth, whereas floral assemblies reflect competing nucleation events.

Fig.~\ref{fig:main_fig_3}c next shows the monoclinic $C2/m$ An:TCNB lattice (CIF~1103089) \cite{Lefebvre1989ATCNBNTCNBDisorder}. Its alternating layers retain this $\sim$3.7~\AA{} spacing within a distinct unit cell. Fig.~\ref{fig:main_fig_3}d then shows bright-orange plates and needles with mirror-symmetric facets and preferential axial growth, highlighting anisotropic growth along preferred crystallographic directions consistent with the underlying $C2/m$ symmetry. High-magnification SEM again reveals smooth faces and sharply defined edges. Thus, the same ambient, minute-scale process accommodates distinct triclinic and monoclinic lattices while preserving their symmetry-dependent morphologies and surface quality. The confined MAS geometry naturally yields this sensing-compatible microscale form factor, which can be thinned further through reduced precursor loading \cite{Guo2020UltrathinOrganicSingleCrystals,LeCoutre2025GrowthFewLayer}; additional morphologies are shown in SI Sec.~\ref{sec:SI_addSEM} and Fig.~\ref{fig:supp_additional_sem}.

\vspace{0.5em}
\noi{\normalsize\bfseries Room-temperature ODMR of charge-transfer triplets\par}
\noi

We now investigate the spin-optical properties of the MAS-grown crystals. Fig.~\ref{fig:main_fig_4}a outlines the readout principle. Excitation at 405~nm creates locally excited (LE) or charge-transfer (CT) singlets. Intersystem crossing then prepares non-equilibrium populations within the corresponding zero-field-split triplet manifold. Resonant microwaves redistribute these populations, changing the total photoluminescence detected by continuous-wave, lock-in ODMR near zero field (see Methods). The molecular pair and its packing set the triplet position along the localized-to-CT continuum (Fig.~\ref{fig:main_fig_4}a). An:TCNB and Bi:TCNB support weak-CT triplet excitons, whereas P:PQ, discussed below (Fig.~\ref{fig:main_fig_5}), lies closer to the localized limit. The same growth and readout scheme therefore accesses both regimes (see triplet-spin Hamiltonian, SI Sec.~\ref{sec:SI_ODMR_background}, where $D,E$ denote the axial, transverse zero-field-splitting parameters). In molecular systems like Bi:TCNB and An:TCNB, where the triplet states are dense and sufficiently mobile, ODMR emerges through an additional triplet--triplet annihilation process, following generation of photoexcited triplet states (SI Sec.~\ref{sec:SI_DelayEmmission}, SI Sec.~\ref{sec:SI_ODMR_background}) \cite{Krzystek1993TripletExcitons,Keevers2016TripletTripletAnnihilationODMR}.

Fig.~\ref{fig:main_fig_4}b next connects this photophysical picture to the electronic structure. Calculated HOMO and LUMO isosurfaces for the lowest triplet configuration show appreciable donor--acceptor delocalization in Bi:TCNB, particularly for the HOMO, whereas the An:TCNB HOMO and LUMO orbitals are concentrated more strongly on anthracene. This agrees with reported CT fractions of $\sim$0.5 and $\sim$0.05, respectively \cite{Mohwald1974MobileTrappedTripletStates,PonteGoncalves1977TripletExcitons,Krzystek1993TripletExcitons}, and identifies triplet localization as a chemically tunable variable within analogous donor--acceptor stacks. Constrained-state energetics supporting optically accessible triplets are provided in SI Sec.~\ref{sec:SI_TDDFT} and Fig.~\ref{fig:dft_atcnb_btcnb}. The two systems provide complementary tests of MAS. Room-temperature triplet-exciton magnetic resonance was known in bulk Bi:TCNB \cite{Corvaja1988BiphenylTCNBODMR,Agostini1991BiphenylTCNBXTrapODMR} (Table~\ref{tab:SI_ODMR_comparison}), whereas prior bulk An:TCNB studies established triplet dynamics and magnetic-resonance signatures but not room-temperature ODMR from thin crystals \cite{VonSchutz1980ATCNBDelayedFluorescenceODMR,Steudle1978ATCNBMobileTripletExcitons,Frankevich1978ATCNBRYDMRTripletPairs} (Table~\ref{tab:SI_ODMR_comparison_ATCNB}). Notably, the majority of these prior studies were carried out at cryogenic temperature or at high magnetic fields, in contrast to near-zero-field room temperature measurements reported here. Results in Fig.~\ref{fig:main_fig_4}c-d confirm that MAS growth both preserves established triplet spin physics and extends it to a sensing-compatible geometry and ODMR readout regime. At the outset, this is slightly surprising because An:TCNB and Bi:TCNB are unencapsulated 1:1 donor--acceptor microcrystals, and triplet migration, annihilation and broad CT fluorescence make narrow spin-dependent contrast far from assured.

Fig.~\ref{fig:main_fig_4}c turns first to Bi:TCNB. An elongated crystal (Fig.~\ref{fig:main_fig_4}c, top panel) gives two well-resolved single-shot ODMR transitions ($2|E|$/$D-|E|$) at 0.914 and 1.263~GHz. A plate in a separate field of view (FOV) (Fig.~\ref{fig:main_fig_4}c, bottom panel) reproduces them at 0.918 and 1.262~GHz despite its distinct habit. The lower-frequency linewidths are 16.9 and 19.1~MHz, with a maximum contrast of 0.054\%. The inset shows broad, largely featureless green PL bands from both FOVs, consistent with CT emission. Bright emission concentrated at the crystal edges and ends further reveals intrinsic optical waveguiding. Thus, the same near-zero-field resonances are recovered directly from the entire collected PL band across distinct Bi:TCNB morphologies.

Fig.~\ref{fig:main_fig_4}d shows the corresponding result for An:TCNB. A compact multicrystal FOV (Fig.~\ref{fig:main_fig_4}d top panel) and elongated needles (Fig.~\ref{fig:main_fig_4}d bottom panel) show similarly broad orange CT emission and pronounced edge- and end-emission. Both FOVs yield the same transition pair ($D-|E|$/$D+|E|$) at 1.768 and $\sim$2.23~GHz, with lower-/higher-frequency linewidths of 7/4 and 8/4~MHz, respectively, and a maximum contrast of 0.060\%. The $\sim$4~MHz lines are remarkably narrow for an undiluted 1:1 donor--acceptor lattice, and are well described by Lorentzian lineshapes (Fig.~\ref{fig:main_fig_4}d). We tentatively attribute this counterintuitive narrowing to rapid triplet hopping, which motionally averages variations in the local spin environment \cite{PonteGoncalves1977TripletExcitons,Anderson1953ExchangeNarrowing}. Prior reports of mobile An:TCNB triplets and our room-temperature delayed-emission measurements, consistent with triplet--triplet annihilation (TTA), support this interpretation \cite{VonSchutz1980ATCNBDelayedFluorescenceODMR,Steudle1978ATCNBMobileTripletExcitons,Frankevich1978ATCNBRYDMRTripletPairs} (SI Sec.~\ref{sec:SI_DelayEmmission}; Fig.~\ref{fig:supp_delayed_emission}).

Additional multicrystal regions reproduce both material-specific ODMR transition pairs across FOVs and crystal habits (SI Sec.~\ref{sec:SI_additional_ODMR}; Fig.~\ref{fig:supp_additional_btcnb_atcnb_fov}), while solvent-grown bulk controls give the same ODMR frequencies (SI Fig.~\ref{fig:supp_bulk_atcnb_btcnb_odmr}). This supports an intrinsic triplet response rather than an exceptional crystal or growth artefact. Fig.~\ref{fig:main_fig_4}d(inset ii) shows that the An:TCNB resonance remains resolved for more than 60~min under repeated optical and microwave interrogation, although its amplitude gradually decreases. This stability is notable because the crystals are unencapsulated and measured directly on their glass growth substrate without post-growth processing (SI Sec.~\ref{sec:SI_odmr_stable}; Fig.~\ref{fig:odmr_stable}).

Together, these results constitute, to our knowledge, the first room-temperature ODMR from MAS-fabricated charge-transfer co-crystals. ODMR readout from the broad CT-emission band, combined with intrinsic waveguiding, narrow resonances and reproducibility across FOVs, establishes MAS-grown crystals as thin, optically readable triplet-spin layers. A detailed comparison with the bulk-crystal literature is provided in SI Sec.~\ref{sec:SI_ODMR_literature}.

\vspace{0.5em}
\noi{\normalsize\bfseries Room-temperature ODMR of localized triplets\par}
\noi
To extend our results beyond donor--acceptor CT co-crystals to localized-triplet hosts, we applied it to pentacene (P) and pentacenequinone (PQ), denoted P:PQ (Fig.~\ref{fig:main_fig_5}a). Heating at $\sim$250--300~$^\circ\mathrm{C}$ yields blue pentacene plates and a novel, distinct population of high-aspect-ratio microneedles (Fig.~\ref{fig:main_fig_5}b(i),(ii)) \cite{Ye2018MicrospacingSublimation}. Under 405~nm excitation, only the microneedles emit bright red PL (Fig.~\ref{fig:main_fig_5}b(iii),(iv)), directly selecting the distinct microneedle population within a heterogeneous growth.
Fig.~\ref{fig:main_fig_5}c next reveals intrinsic optical waveguiding: emission propagates along each needle and emerges intensely from its ends (Fig.~\ref{fig:main_fig_5}c(i)). Unlike the broad, largely featureless CT bands in Fig.~\ref{fig:main_fig_4}, the spectrum in Fig.~\ref{fig:main_fig_5}c(ii) is structured and resembles pentacene doped in \textit{p}-terphenyl. This supports a localized, isolated-emitter-like pentacene environment and is consistent with reduced intermolecular quenching relative to pure pentacene \cite{Jundt1995PentaceneExcitonDynamics,Singh2025OrganicCrystalQuantumSensing,Mena2024MolecularSpinCoherentControl}. Single-crystal X-ray diffraction of a representative yellow microneedle resolves a pentacenequinone host lattice (SI Sec.~\ref{sec:SI_PPQ_structure} and Fig.~\ref{fig:SI_PQ_structure}). The pentacene-like emission and triplet ODMR of the emissive microneedles are consistent with pentacene incorporation into this host.
Fig.~\ref{fig:main_fig_5}d then shows the room-temperature spin fingerprint. A bimodal feature ($2|E|$) at 0.103/0.109~GHz (Fig.~\ref{fig:main_fig_5}d(i)) accompanies transitions ($D-|E|$/$D+|E|$) at 1.308/1.413~GHz with linewidths of 11.4/10.5~MHz (Fig.~\ref{fig:main_fig_5}d(ii),(iii)). The low-frequency doublet lies near the high-frequency separation, yielding a self-consistent three-transition fingerprint. The shifted frequencies relative to pentacene-doped \textit{p}-terphenyl place a localized pentacene triplet in a distinct lattice environment (SI Table \ref{tab:SI_PPQ_ZFS_comparison_table}) \cite{Singh2025OrganicCrystalQuantumSensing,Mena2024MolecularSpinCoherentControl}. The 10--11~MHz lines are broader than the $\sim$4~MHz An:TCNB features in Fig.~\ref{fig:main_fig_4}. Differences in local spin dynamics and excitation-induced broadening, together with potentially higher chromophore doping enabled by the similar molecular size of pentacene and pentacenequinone, suggest an explanation for this comparison. Additional FOVs reproduce the resonances (SI Sec.~\ref{sec:SI_additional_ODMR}; Fig.~\ref{fig:supp_ppq_additional_fov}). We observe that unencapsulated microneedles retain ODMR for days and yield single-shot scans in $<$2~min with $\sim500~\mu W$ of optical power. These measurements demonstrate stable spin readout from localized P:PQ triplets. More broadly, across this heterogeneous MAS growth, individual yellow microneedles likely span a range of pentacene incorporation, from variable doping to potential co-crystallization of pentacene and pentacenequinone. The simplicity of MAS growth for room-temperature ODMR-active crystals, together with its wide span of accessible organic-molecule coformers, encourages the development of a same-crystal, correlated growth--ODMR--XRD experimental flow---enabled by sharing a common crystal-holding microloop across all three experimental setups.

\vspace{0.5em}
\noi{\normalsize\bfseries Coherent control of localized triplet spins\par}
\noi
We next investigated whether P:PQ microneedles support coherent spin control. Their resolved pentacene-like localized triplet manifold provides a direct comparison with pentacene-doped \textit{p}-terphenyl (PDP), an established dilute-host benchmark \cite{Singh2025OrganicCrystalQuantumSensing,Mena2024MolecularSpinCoherentControl}.
Fig.~\ref{fig:main_fig_6}a shows focused 405~nm excitation of a $\sim$3~\textmu m region. Red emission propagates from this spot to both needle ends, demonstrating that the microcrystal acts as spin host and optical waveguide. Fig.~\ref{fig:main_fig_6}b outlines the protocol performed at room temperature: optical preparation, phase-controlled MW manipulation at the $D-|E|$ transition (1.306~GHz) and PL readout are repeated within a lock-in envelope. ODMR spectra from this same microneedle, including the transition selected for coherent driving, are shown in SI Fig.~\ref{fig:supple_ppq_odmr_figure6}.
Fig.~\ref{fig:main_fig_6}c shows cycles of power-tunable Rabi oscillations, reaching 7.3~MHz. Independently grown microneedles reproduce this coherent response (SI Sec.~\ref{sec:SI_PPQ_additional_coherence}; Figs.~\ref{fig:supp_ppq_fov2_full_coherence} and~\ref{fig:supp_ppq_fov_rabi_only}). Fig.~\ref{fig:main_fig_6}d then resolves free phase evolution through Ramsey fringes, giving $T_2^*=123$~ns. Despite the nominally high chromophore density (linewidth $>15$~MHz), this coherence remains remarkably close to the PDP benchmark (SI Fig.~\ref{fig:supp_pdp_reference}, Table~\ref{tab:SI_PQ_PDP_comparison_table}).
Fig.~\ref{fig:main_fig_6}e gives $T_2=0.56~\mu$s under Hahn echo; a four-$\pi$-pulse CPMG train extends this 2.4-fold to 1.35~$\mu$s. PDP measured using the same microwave apparatus gives $T_2^*=246$~ns and $T_2=1.4~\mu$s, larger by factors of only 2.0 and 2.5. The comparatively modest CPMG-driven $T_2$ enhancement, relative to the dilute PDP reference, suggests a distinct decoherence channel dominated by high-frequency fluctuations \cite{Singh2025OrganicCrystalQuantumSensing}.
Fig.~\ref{fig:main_fig_6}f shows an enhanced, near-single-exponential ground-state recovery of 82.8~$\mu$s (average $|T\rangle \rightarrow |S_0\rangle$), compared with 58.6~$\mu$s for PDP (SI Fig.~\ref{fig:supp_pdp_reference}; Table~\ref{tab:SI_PQ_PDP_comparison_table}).
Under analogous conditions, we did not resolve near-zero-field room-temperature coherence in mobile triplets of An:TCNB or Bi:TCNB, consistent with dynamical dephasing from stochastic incoherent triplet hopping, fluctuating exchange/dipolar couplings and triplet--triplet encounters. We recognize similarities between the room-temperature ODMR and persistent spin coherence observed in P:PQ microneedles and a recent study of bulk, millimeter-scale dihydropentacene:pentacene (DHP:PT) cocrystal grown by physical vapor transport, which retain room-temperature ODMR at a 33.3~mol\% pentacene concentration \cite{DSouza2026DensePentaceneCocrystalCoherentControl}.

\vspace{0.5em}
\noi{\normalsize\bfseries Discussion and outlook\par}
\noi

Taken together, these results establish a materials-discovery route to sensing-ready molecular spin layers. The ambient, \emph{in situ} workflow accesses charge-transfer triplets in An:TCNB and Bi:TCNB and localized pentacene triplets in P:PQ microneedles. Because the active medium is a co-crystal, donor--acceptor or chromophore--host composition is set by co-crystallization rather than dilute doping. This provides high densities of photoexcited molecular triplets while retaining optical clarity, single-crystalline nature and near-surface access. Reproducible ODMR fingerprints, consistent with bulk controls, enable comparisons across pairs beyond one established host or exceptional crystal \cite{Ye2018MicrospacingSublimation,Ye2019CocrystalsMicrospacing}. This growth--spectroscopy route's central, novel contribution is access to discrete stoichiometric (1:1 or 1:2) co-crystals with high, well-defined triplet density (unlike other fabrication approaches), while further extending the possibility to a dopant regime for similarly sized stack-compatible molecules, as for example pentacene in pentacenequinone (P:PQ), or phenanthrene in An:TCNB \cite{Sun2014MixedCocrystalMicrotubes}.

The contrasting ODMR and coherence behavior of mobile and localized triplets reveals an important design tension. Room-temperature ODMR collected across the full PL band is not assured in undiluted 1:1 crystals: high active-molecule density, triplet migration and annihilation, broad CT emission and exposed surfaces could all obscure spin-dependent contrast. Yet An:TCNB yields narrow ODMR resonances but no resolved pulsed coherence, whereas P:PQ supports ensemble Rabi oscillations and microsecond coherence. A consistent, although not yet definitive, picture is that rapid motion averages static inhomogeneity in continuous-wave ODMR, while stochastic hopping, fluctuating couplings and triplet encounters dephase pulsed coherence. This tension identifies localization and mobility as molecular-design variables for quantum sensing that are not readily accessed in conventional dilute defect-sensor architectures. 

Looking forward, in situ fluorescence can identify new phases and growth windows, while electronic-structure calculations and ODMR test emission, intersystem crossing, triplet lifetime, zero-field splitting (ZFS), mobility and coherence (SI Secs.~\ref{sec:SI_scope}, \ref{sec:SI_spectrum} and~\ref{sec:SI_TDDFT}). Because the ZFS tensor is tied to molecular orientation and crystal packing, donor--acceptor and chromophore--host libraries could seek stacks whose principal axes span different molecular and crystallographic directions \cite{Krzystek1993TripletExcitons,Attwood2025SpinRelaxationTripletMedia,Li2026PentaceneACVectorSensing}, allowing novel vector-sensing geometries.

Preliminary single-shot static field sensitivities reach $0.11~\mu\mathrm{T}/\sqrt{\mathrm{Hz}}$ without sensitivity optimization (SI Sec.~\ref{sec:SI_field_sensitivity}; Fig.~\ref{fig:fig_field_sensitivity}). Pulsed-ODMR could further improve ODMR contrast by enabling efficient initialization and readout of triplet population differences \cite{Singh2025OrganicCrystalQuantumSensing}. In mobile charge-transfer co-crystals, time-gated collection of TTA-dependent delayed emission could additionally reject prompt fluorescence, further enhancing ODMR contrast. Thinner growth, deterministic placement, integrated microwave and photonic structures, and encapsulation could improve stand-off, optical collection and stability \cite{Guo2020UltrathinOrganicSingleCrystals}.
Moreover, crystal habit offers distinct integration routes: templated or epitaxially guided growth could organize waveguiding needles into aligned arrays, whereas controlled lateral growth of plates could approach large-area single-crystal or tiled thin films using inexpensive substrates and microgram feedstocks \cite{Lin2023OrientedOrganicSingleCrystallinePatterns}. These advances could support wide-field magnetic imaging of low-dimensional materials and biological interfaces. Molecular control of spin density and charge-transfer character could also be used to pursue electric-field and local-environment sensing. More broadly, our results complement defect sensors by making molecular composition, crystal packing and exciton motion active elements of quantum-sensor design.

\section*{Methods}\label{sec11}
\vspace{0.5em}
\noi{\normalsize\bfseries Materials and precursor preparation\par}
\noi
Anthracene, biphenyl, 1,2,4,5-tetracyanobenzene (TCNB), pentacene and 6,13-pentacenequinone were used as received from commercial suppliers (TCI, Sigma-Aldrich). For each MAS growth, MAS candidate pair molecules D and A were combined in equal proportions by mass and mixed in a single glass vial to prepare a stock precursor, which was then co-ground using mortar and pestle to form a fine ground mixture, at times accompanied by a noticeable colour change. For a single MAS growth per slide, small amounts of the co-ground mixture ($\leq$1~mg) were sprinkled onto the bottom slide.

\vspace{0.5em}
\noi{\normalsize\bfseries Substrate preparation and microspacer assembly\par}
\noi
The MAS growth chamber was assembled from Corning \#1 glass coverslips, forming a simple sandwich of a bottom glass slide (amorphous precursor loading), a spacer glass slide (\#1 thickness), and a top glass slide (crystal growth). Bottom and top substrates were used directly from new boxes, owing to their known thickness. Corning \#1 glass slides (0.12--0.16~mm, quoted in the main text as $\sim$130~\textmu m) were used directly as the spacer layer, owing to their known thickness; all data presented in this work were obtained using this Corning Inc. \#1-thickness spacer. High-precision-thickness coverglass alternatives (Thorlabs \#0: CG00E1; \#1.5: CG15E1; 3~mm diameter) were also used for some variational experiments. The spacer was affixed to the bottom substrate using a UV-curing optical adhesive (e.g., Norland), then cured under an off-the-shelf UV lamp for $>$15~min. Optionally, the bottom glass surface was rendered additionally hydrophilic using a low-cost handheld plasma tool for a few minutes. In practice, a small layer ($\sim 10\mu$m) of UV cured glue and, at times, large ground precursor powder can modify the spacing for some MAS experiments, though more precise protocols can be easily adapted (SI Sec.~\ref{sec:SI_videos}).

\vspace{0.5em}
\noi{\normalsize\bfseries Ambient-phase OTS coating of the growth substrate\par}
\noi
In prior MAS-growth work, the top glass slide was coated with hydrophobic octadecyltrichlorosilane (OTS) using vacuum deposition, as OTS and moisture are highly reactive. To simplify this process, we adopted an ambient-phase OTS nanolayer protocol \cite{Zhang2021SuperhydrophobicSilanization} (SI Sec.~\ref{sec:SI_OTS}) to optionally render the top coverslip hydrophobic, controlling condensate wetting and nucleation density. Briefly, 1~mL OTS (Sigma Aldrich) was transferred into a glass vial inside a fume hood and stoichiometric 20~\textmu L deionized water was added. Immediately after water addition, the mixture was vortexed (10~s), ultrasonicated (10~s) and vortexed again (10~s) to form a stabilized, faint white-coloured emulsion. The vial was then left undisturbed for $\sim$40~min incubation (inside a fume hood), after which the emulsion was diluted with 19~mL dry \textit{n}-hexane to achieve $\sim$5\% v/v OTS. Coverslips were left in the prepared diluted OTS coating solution overnight to coat them with an OTS nanolayer. When a soft whitish aggregate formed during OTS coating of glass coverslips, it was removed with additional hexane rinsing to yield a clear glass slide coated with an OTS layer. Hydrophobicity was checked qualitatively by placing a 5~\textmu L DI-water droplet on the surface and imaging from the side; on half-coated slides (Fig.~\ref{fig:supp_ots_formation}e), the water droplet spreads on uncoated region and remains more spherical on the OTS-coated region.

\vspace{0.5em}
\noi{\normalsize\bfseries Microspacing in-air sublimation growth\par}
\noi
After plasma treatment of the bottom substrate, the co-ground precursor mixture was placed on the bottom slide between the pre-attached spacers. The (optionally OTS-coated) top coverslip was cleaned with hexane, placed on the spacers to form an unsealed glass--spacer--glass microgap (typically 120--200~\textmu m), and gently pressed to ensure conformal contact with the spacers. The assembled microgap chamber was placed on a hot plate or PID-controlled ceramic heater and heated in ambient atmosphere (typically 100--300~$^\circ$C, depending on the materials pair). Co-crystal growth proceeds via a droplet-mediated pathway or a vapour-phase-mediated pathway, depending on the materials pair. Temperature trajectories were tuned empirically by in situ monitoring; for the hot-stage geometry described below, Bi:TCNB nucleation typically occurred near $\sim$115--120~$^\circ$C, An:TCNB co-crystal formation was observed near $\sim$159--160~$^\circ$C, and pentacene:pentacenequinone (P:PQ) microneedle formation occurred near $\sim$250--300~$^\circ$C, with more pure pentacene crystals forming in the $\sim$300~$^\circ$C range. Across these systems, temperature and microspacing (set here by the \#1 glass coverslip thickness) are the two critical parameters governing cocrystal growth outcome; further microspacing optimization may improve the crystal growth pattern to a more isolated single-crystal distribution over the top substrate. We also observe growth of the above co-crystals using sapphire and silicon wafers as the top substrate. Once co-crystals formed on the cooler top surface above the source; the top substrate with MAS single crystals was removed after cooling for post-growth characterization, and moved to the ODMR and spin-spectroscopy setup without any additional encapsulation. See SI Sec.~\ref{sec:SI_videos} for growth videos and detailed comments.

\vspace{0.5em}
\noi{\normalsize\bfseries In situ fluorescence microscopy of MAS growth\par}
\noi
To monitor nucleation and growth in real time and guide growth optimization, MAS chambers were imaged on a compact wide-field epi-fluorescence microscope built from off-the-shelf components (Thorlabs) and installed inside a chemical fume hood; full instrument and parts details are provided in SI Sec.~\ref{sec:SI_scope}. Samples were excited with a 405~nm LED and imaged onto a color CMOS camera; image scale bars were calibrated using a USAF 1951 resolution test target (Thorlabs). The chamber was heated on a ceramic hot plate under closed-loop PID control (Thorlabs, custom control software), with local surface temperature monitored via a platinum resistance temperature detector (RTD) positioned near the sample. Custom MATLAB acquisition software time-stamped and temperature-stamped each recorded frame, allowing growth movies to be directly correlated with the applied temperature trajectory, as shown for Bi:TCNB and An:TCNB in Fig.~\ref{fig:main_fig_2} (see also annotated growth recordings in SI Sec.~\ref{sec:SI_videos}). The same collection path was optionally coupled to a UV--Vis spectrometer for time-resolved in situ emission spectroscopy during growth, correlating growth temperature and time profiles with cocrystal nucleation-induced emission changes, as exemplified for An:TCNB (SI Sec.~\ref{sec:SI_spectrum}). After growth, the same microscope was used to inspect MAS-grown microcrystals for morphology (plate/needle habits), optical clarity and waveguided facet emission.

\vspace{0.5em}
\noi{\normalsize\bfseries Scanning electron microscopy\par}
\noi
Scanning electron microscopy (SEM) was used to assess crystal habit (needle or plate morphology) and cross-verify crystal system (e.g., monoclinic or triclinic) against \textit{a priori} diffraction-based crystallographic information (CIF), as well as to assess the approximate surface homogeneity and single-crystal quality of MAS-grown structures. As-grown crystals on glass substrates were mounted on SEM stubs for imaging after applying a 2~nm Au/Pd coating to reduce charging-induced saturation of SEM images. Representative SEM images in the main text (Fig.~\ref{fig:main_fig_3}) and SI (Sec.~\ref{sec:SI_addSEM}, Fig.~\ref{fig:supp_additional_sem}) illustrate the range of morphologies (needles, plates and multi-nucleated ``floral'' assemblies) accessible under variable temperature and microspacer conditions, and comparisons of optimized versus overheated growth (SI Sec.~\ref{sec:SI_fragile}, Fig.~\ref{supp_fig_sem_broken}) highlight the importance of controlled cool-down for preserving smooth, single-crystalline surfaces.

\vspace{0.5em}
\noi{\normalsize\bfseries Density functional theory calculations\par}
\noi

DFT calculations were performed using VASP\cite{Kresse1996EfficientIterativeSchemes} with the PBE functional\cite{Perdew1996GeneralizedGradientApproximation}.
Long-range van der Waals interactions were included using Grimme's D3 dispersion correction with Becke--Johnson damping\cite{Grimme2011DFTD3BJ}. 
This correction is important for describing molecular packing in organic co-crystals.
Spin-polarized calculations used $\Gamma$-point Brillouin zone sampling, a plane-wave cutoff energy of 900 eV, Gaussian smearing of 0.01 eV, and an electronic convergence criterion of $10^{-6}$ eV. Unit cells of An:TCNB and Bi:TCNB each contained $n_D=2$ donor molecules and $n_A=2$ acceptor molecules.
Excited-state geometries were obtained by first constraining the occupations of selected frontier orbitals. 
Geometry optimization was then performed under the fixed electronic configuration, with ionic total energy converged to a threshold value of {\(10^{-5}\)} eV.
The optimized singlet and triplet structures were used to analyze the highest occupied molecular orbital (HOMO), the lowest unoccupied molecular orbital (LUMO), and spin-density localization.

\vspace{0.5em}
\noi{\normalsize\bfseries Optically detected magnetic resonance measurements\par}
\label{sec:Methods_ODMR}
\noi
\textbf{\textit{ODMR apparatus.}}
All MAS-crystal ODMR measurements were performed under 405~nm CW excitation, with a flip mount used to switch between the 405~nm (Coherent OBIS) and 532~nm (Coherent Verdi) excitation lines on a custom microscope (Mad City Labs RM21 body, with a 20$\times$ or 50$\times$ objective for photoluminescence collection and a CMOS camera (Teledyne Kinetix) for wide-field imaging); the 532~nm line was used for NV-diamond calibration (DNVB1 sample, Thorlabs) and for measurements on custom lab-grown 0.1\% pentacene-doped para-terphenyl (PDP) crystals. Excitation light was separated from the emission using dichroics (Thorlabs): 463~nm for the 405~nm line and 567~nm for the 532~nm line. The emission line of the microscope had switchable laser-line notch filters for 405~nm and 532~nm to minimize laser noise into the detection photodiode. Continuous-wave ODMR spectra were recorded by sweeping the microwave frequency while monitoring the lock-in amplitude (SRS SR830) of the microwave-amplitude-modulated photoluminescence signal. Emission light was detected on a photodiode (Thorlabs DET100A2) for all ODMR and spin-coherence measurements, coupled into a multimode fiber and directed to a UV--Vis spectrometer (Ocean Optics HR-6UVV330-25) for emission-spectrum measurements, or fiber-coupled to an APD (Excelitas) for fast delayed-emission photon-counting measurements (SI Sec.~\ref{sec:SI_DelayEmmission}). Microwaves were delivered in a near-field microscope-compatible geometry using a custom 50~$\Omega$ microstrip PCB resonator (0.51~mm-thick Rogers RO4350B laminate) with an optical imaging aperture, driven by an SRS SG384 source, Mini-Circuits microwave switch, a 30~W high-power amplifier, and switchable circulator-protected delivery lines (0.8--2~GHz and 2--3~GHz); a separate, lower-gain 25~dB wideband amplifier without a circulator was used for P:PQ. ODMR signal was reported as the lock-in amplitude corresponding to the microwave-amplitude-modulated photon signal; resonance frequencies and linewidths were extracted by fitting Lorentzian line shapes.

\medskip
\noindent\textbf{\textit{ODMR acquisition conditions.}}
Laser lateral spread is visible via the scale of the excitation footprint relative to the MAS crystal fields of view shown in Fig.~\ref{fig:main_fig_4} and Fig.~\ref{fig:main_fig_5}, both acquired using back-focal-plane illumination. Representative single-shot acquisition conditions for the room-temperature fields of view shown in the main text were as follows: for Bi:TCNB, 500~\textmu W excitation and $\sim$10~W microwave power, with a 40~ms acquisition time per microwave point (approximately four times the lock-in low-pass-filter settling time), each example spectrum completed in $<$3~min; for An:TCNB, 3~mW excitation and $\sim$10~W microwave power, with each spectrum representing $n=10$ repeated single-shot scans separated by a 40~s laser-off/microwave-off waiting window to dissipate absorbed heat and minimize crystal damage; and for P:PQ, $\sim$500~\textmu W excitation and $\sim$1~W microwave power, with a 40~ms lock-in integration time per microwave point, full wideband single-shot scans completed in $\sim$2~min. For all MAS-crystal ODMR responses, the resonant lock-in response was additionally verified as a genuine spin resonance rather than an instrumental artefact by reversibly perturbing the signal with a nearby bar magnet while keeping the laser, microwave drive and detection conditions unchanged, ruling out coupled electronic artifacts as the source of the observed resonances. MAS samples were positioned on the microstrip resonator using waveguided emission as a coarse visual guide: when the excitation beam struck a MAS-grown crystal, the emission waveguided along the plate/needle and to nearby crystals, eventually reaching the edges of the full glass coverslip (the top substrate of the MAS growth process), with visual colour changes aiding placement (e.g., blue anthracene flakes shifting to orange at An:TCNB crystals) before fine field-of-view adjustment. Additional setup details are provided in Supplementary Section~\ref{sec:SI_ODMR_setup}.

\vspace{0.5em}
\noi{\normalsize\bfseries Optically detected spin-coherence measurements\par}
\noi
Spin-coherence measurements were detected via synchronized laser and phase-controlled microwave switching in the same ODMR setup, as shown in Fig.~\ref{fig:main_fig_6}b; referential PDP measurements verified the phase-controlled microwave pulse timings. The 532~nm line was switched via an acousto-optic modulator, and the 405~nm line was switched via intrinsic fast-diode modulation (TTL lines) built into the Coherent OBIS laser. Optically read Rabi, Ramsey and Hahn-echo sequences were wrapped in a lock-in envelope and detected via the lock-in amplifier for high noise rejection. Ground-state recovery experiments were instead recorded via photodiode current, amplified by a preamplifier and digitized on a data-acquisition device. Lock-in-envelope-wrapped spin-coherence signals are shown as the in-phase component, in which the sign of the pentacene-triplet ODMR contrast is visible (Fig.~\ref{fig:main_fig_6}; SI Sec.~\ref{sec:SI_PPQ_additional_coherence}).

\vspace{0.5em}
\noi{\normalsize\bfseries Acknowledgments\par}
\noi
We acknowledge technical contributions from N. D'Souza and W. Ng and funding from U.S. DOE (DE-FOA-0001968, DE-SC0025524), NNSA (LB24-NV center 13C quantum sensor-PD3Ta, LB26-nuclear-spin quantum sensing-PD3Td),ONR (N00014-20-1-2806), AFOSR YIP (FA9550-23-1-0106), Dreyfus and Sloan Foundations, and instrumentation support from AFOSR DURIP (FA9550-22-1-0156) and NSF MRI (2320520) . Work at the Molecular Foundry was supported by the Office of Science, Office of Basic Energy Sciences, of the U.S. Department of Energy under Contract No. DE-AC02-05CH11231. This work used resources of the National Energy Research Scientific Computing Center (NERSC), a Department of Energy User Facility using NERSC allocation m4269, under the award BES-ERCAP0032327. We thank Misung Kang and Danielle Jorgens for their advice and assistance at the Electron Microscopy Laboratory, University of California, Berkeley, which is supported by NIH S10 Grant No. S10OD030258-01. 

\vspace{0.5em}
\noi{\normalsize\bfseries Author Contributions\par}
\noi
M.P. built the crystal-growth and optical spin-spectroscopy instrumentation, performed measurements, and analyzed the data. G.C. performed the density-functional-theory calculations, supervised by L.Z.T. E.D. assisted with the growth instrumentation. M.P. and A.A. wrote the initial draft of the manuscript. All authors reviewed and contributed to the final manuscript. A.A. supervised the project.

\vspace{0.5em}
\noi{\normalsize\bfseries Data Availability\par}
\noi
The authors declare that the data supporting the findings of this study are available within the paper and its supplementary information files. Source data and other raw data files are available from the corresponding author upon reasonable request.

\bibliographystyle{naturemag-MAS}
\bibliography{references}

\clearpage

\beginsupplement

\setcounter{topnumber}{4}
\setcounter{dbltopnumber}{3}
\setcounter{totalnumber}{6}
\renewcommand{\topfraction}{0.95}
\renewcommand{\dbltopfraction}{0.95}
\renewcommand{\textfraction}{0.05}
\renewcommand{\floatpagefraction}{0.80}
\renewcommand{\dblfloatpagefraction}{0.80}
\makeatletter
\setlength{\@fptop}{0pt}
\setlength{\@fpbot}{0pt plus 1fil}
\setlength{\@dblfptop}{0pt}
\setlength{\@dblfpbot}{0pt plus 1fil}
\makeatother

\begin{widetext}
\begin{center}
\textbf{\large\textit{Supplementary Information}}\\[0.5em]
\textbf{Room-temperature optically addressable and coherently controlled triplet spins in rapidly grown molecular microcrystals}\\[1em]
\textbf{CONTENTS}
\end{center}
\end{widetext}

\setcounter{tocdepth}{2}
\supplementtableofcontents

\section*{Guide to the supplementary evidence and workflow}
\label{sec:SI_organization}

This Supplementary Information (SI) is organized in three parts. We first provide the fabrication and measurement methods: vacuum-free ambient OTS functionalization (Sec.~\ref{sec:SI_OTS}), minimal-processing glass microchamber assembly (Sec.~\ref{sec:SI_bottom}), the fume-hood microscope for real-time growth monitoring (Sec.~\ref{sec:SI_scope}) and the integrated optical--microwave platform for ODMR and coherent control (Sec.~\ref{sec:SI_ODMR_setup}). Descriptions of Supplementary Videos~1--8 follow these methods and document pathway-specific growth, optimization and failure modes (Sec.~\ref{sec:SI_videos}).

We next present supplementary evidence for growth and crystal quality. This includes OTS-controlled condensate wetting and nucleation (Sec.~\ref{sec:SI_OTS_effect}), the narrow thermal window required to preserve Bi:TCNB single-crystal quality (Sec.~\ref{sec:SI_fragile}), in situ fluorescence spectroscopy of sequential An:TCNB formation (Sec.~\ref{sec:SI_spectrum}) and additional SEM morphologies (Sec.~\ref{sec:SI_addSEM}). The remaining sections establish the spin physics and sensing performance: charge-transfer triplet context and prior benchmarks (Sec.~\ref{sec:SI_ODMR_background}; Tables~\ref{tab:SI_ODMR_comparison} and~\ref{tab:SI_ODMR_comparison_ATCNB}), comparison with bulk-crystal ODMR (Sec.~\ref{sec:SI_ODMR_literature}), constrained DFT (Sec.~\ref{sec:SI_TDDFT}), reproducibility across additional MAS crystals and bulk controls (Sec.~\ref{sec:SI_additional_ODMR}), delayed emission and long-term ODMR stability (Secs.~\ref{sec:SI_DelayEmmission} and~\ref{sec:SI_odmr_stable}), single-crystal X-ray structural resolution of the P:PQ microneedle phase (Sec.~\ref{sec:SI_PPQ_structure}), coherent-control reproducibility and comparison with pentacene-doped \textit{p}-terphenyl (Sec.~\ref{sec:SI_PPQ_additional_coherence}; Table~\ref{tab:SI_PQ_PDP_comparison_table}), and preliminary magnetic-field sensitivity (Sec.~\ref{sec:SI_field_sensitivity}).

\section{Ambient OTS nanolayers enable vacuum-free surface functionalization}\label{sec:SI_OTS}

\subsection*{Vacuum-free hydrophobic surface preparation}

Ambient OTS coating cannot be applied directly because OTS reacts violently with moisture in-air, producing explosive HCl formation. For this reason, conventional OTS coating relies on silane vapor deposition under vacuum inside microfabrication facilities, where the absence of 
moisture prevents hydrolysis reaction pathways. These vacuum-based workflows are 
infrastructure-intensive and not compatible with routine benchtop crystal growth studies \cite{Ye2018MicrospacingSublimation,Dong2006OTSCVDMonolayers}.

In the ambient-phase approach reproduced here (from Zhang \textit{et al.} \cite{Zhang2021SuperhydrophobicSilanization}), a specific 
stoichiometric addition of water into OTS followed immediately by rapid sonication 
and vortexing produces microscopic emulsions of water inside OTS. This stabilizes a 
reaction that would otherwise be highly explosive in significant volumes and enables formation of a uniform 
water-in-OTS emulsion at room temperature. The method allows us to prepare hydrophobic 
glass substrates for biphenyl--TCNB and anthracene--TCNB thin-crystal growth without 
vacuum deposition. Our adaptation of this process allows further simplification of the 
MAS quantum sensor production workflow.

The description below follows exactly what we observed experimentally, including rapid 
mixing requirements, visible behavior of the emulsion, and the typical 
$\sim 40$~min incubation used in our trials, during which the mixture thickens and swells 
consistent with formation of OTS nanofibers \cite{Zhang2021SuperhydrophobicSilanization}. Despite the shorter 
incubation time, we consistently observed considerable hydrophobicity on the coated 
glass substrates.

\begin{figure*}[!t]
    \centering
   \includegraphics[width=0.65\textwidth]{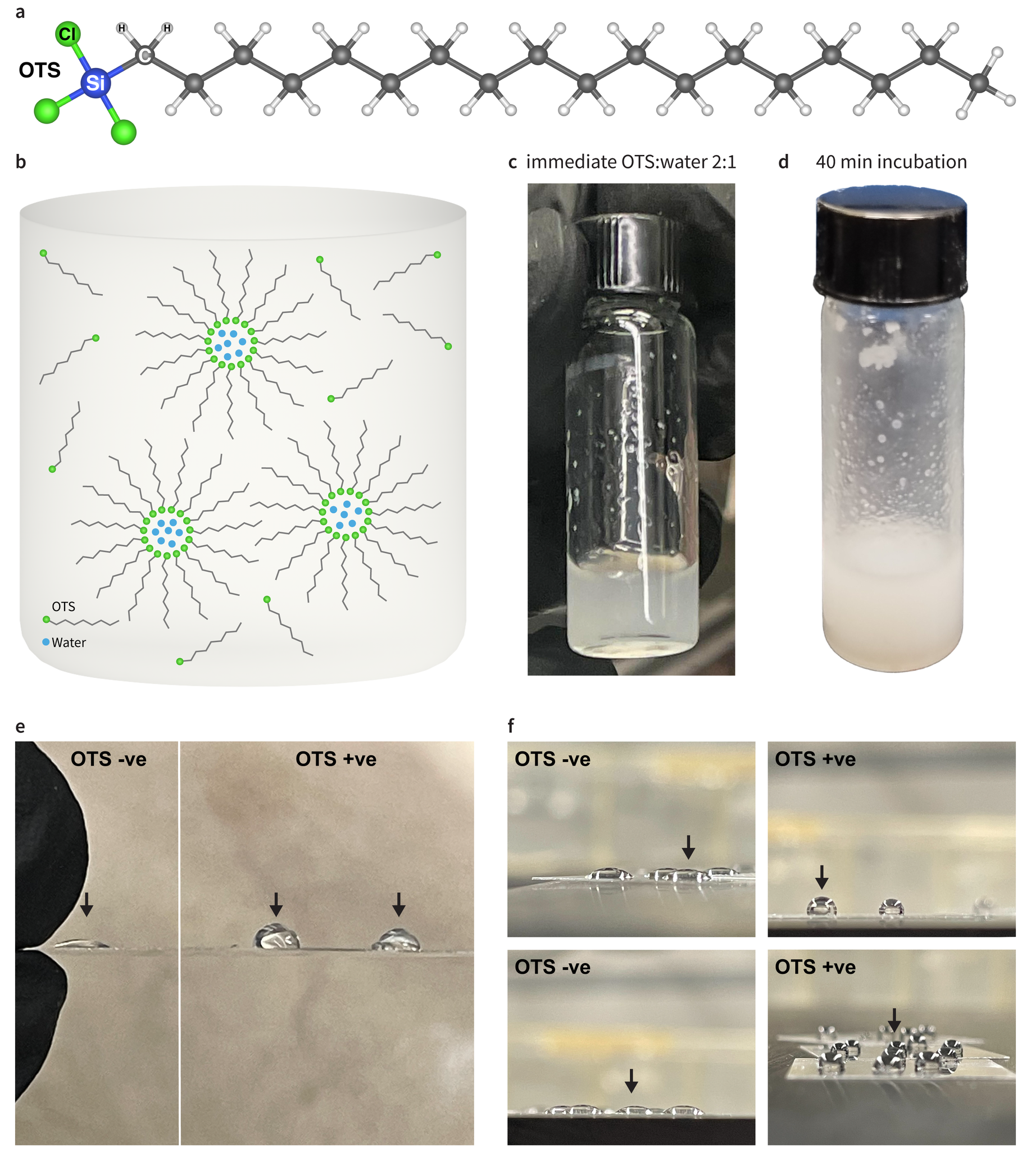}
    \caption{\justifying
    \textbf{Formation of octadecyltrichlorosilane (OTS) nanoaggregates on glass substrates.}
    This figure illustrates the formation of octadecyltrichlorosilane (OTS) nanoaggregates on glass substrates, following the process adapted from Zhang \textit{et al.} \cite{Zhang2021SuperhydrophobicSilanization}. This procedure enables hydrophobic OTS-derived coating of glass coverslips inside a fume hood without the need for vacuum-deposition facilities.
    \textbf{(a)} Molecular structure of octadecyltrichlorosilane, comprising an 18-carbon alkyl chain terminated by a silicon atom bonded to three chlorine atoms, forming a hydrophilic headgroup attached to a long hydrophobic chain.
    \textbf{(b)} Schematic depiction of OTS nanomicelle formation when OTS and water are mixed at a defined stoichiometric ratio (2:1, OTS:water), immediately vortexed, and sonicated. Water droplets (blue) are surrounded by the hydrophilic Si--Cl termini of OTS, while the hydrophobic alkyl chains extend outward. Rapid vortexing and sonication stabilize the otherwise highly reactive hydrolysis of OTS with water.
    \textbf{(c)} Image of the OTS:water mixture immediately after addition of 20~$\mu$L of water to 1~mL of OTS (2:1 molar ratio), followed by vortexing and sonication, showing a relatively stabilized dispersion.
    \textbf{(d)} After 40~min of incubation, the mixture becomes increasingly swollen, accompanied by visible HCl accumulation within the vial, produced as a by-product of the OTS--water hydrolysis reaction.
    \textbf{(e)} Glass coverslips incubated in this mixture---even for a few minutes---become noticeably hydrophobic. A coverslip half-coated with OTS and half uncoated shows a clear wetting contrast: a 5~$\mu$L water droplet spreads on the uncoated half while remaining more spherical on the OTS-coated half
    \textbf{(f)} Additional examples demonstrate the wetting behaviour of water droplets on uncoated versus OTS-coated glass. Droplets on uncoated glass exhibit low contact angles, whereas droplets on glass coated overnight in the OTS mixture show markedly increased contact angles. Although the short incubation time used here ($\approx$40~min) is below optimal durations reported in the literature, the differential wetting contrast remains immediately evident.}
    \label{fig:supp_ots_formation}
\end{figure*}

Fig.~\ref{fig:supp_ots_formation}(a) shows the OTS molecule, while Fig.~\ref{fig:supp_ots_formation}(b) schematically illustrates the stabilized water-in-OTS nanomicelle/emulsion that enables safe ambient processing. Photographs of the mixture immediately after rapid mixing and after $\sim$40~min incubation are shown in Fig.~\ref{fig:supp_ots_formation}(c,d).

\subsection*{Ambient OTS-emulsion preparation and glass coating}
\begin{enumerate}
    \item \textit{OTS preparation.}
    Fresh $1\,\mathrm{mL}$ OTS (Sigma Aldrich) was transferred into a glass vial inside a functioning 
    fume hood. OTS is a severe eye-damage agent; sealed eye protection is required.

    \medskip

    \item \textit{Water addition.}
    Exactly $20\,\mu\mathrm{L}$ deionized water was added to the OTS. We used Drummond calibrated glass pipets to sample this volume. 
    This water loading corresponds to the controlled OTS:water stoichiometry used to form stabilized
    water-in-OTS emulsions (schematized in Fig.~\ref{fig:supp_ots_formation}(b)).

    \medskip

    \item \textit{Rapid emulsion formation (must be executed quickly).}
    Immediately after adding water, the sample was:
    \begin{itemize}
        \item vortexed 10~s,
        \item ultrasonicated 10~s,
        \item vortexed 10~s.
    \end{itemize}
    A proper emulsion appears uniform and clear, without visible precipitate. 
    This sequence must be performed in quick succession.
    A representative ``cleared'' dispersion immediately after rapid mixing is shown in
    Fig.~\ref{fig:supp_ots_formation}(c).

    \medskip

    \item \textit{Incubation.}
    The vial was left undisturbed with the glass cap slightly open. 
    Our usual trials used $\sim 40$~min incubation (instead of optimal 2~h in Zhang \textit{et al.} \cite{Zhang2021SuperhydrophobicSilanization}), during which the mixture 
    began to thicken and swell, consistent with polymerization and formation of 
    OTS nanofibers. 
    During incubation, vial cap opening leads to sudden release of vapors of HCl
    and a swollen/gel-like appearance of the emulsion (Fig.~\ref{fig:supp_ots_formation}(d)); this step
    should therefore be performed entirely inside a fume hood.

    \medskip

    \item \textit{Dilution.}
    The emulsion was diluted with $19\,\mathrm{mL}$ dry n-hexane to achieve 
    $\sim 5\%$ v/v OTS.

    \medskip

    \item \textit{Coating.}
    Clean Corning Inc.\ \#1 cover glass slides were incubated in the stabilized OTS:water emulsion overnight. Although, we observe immediate 10 seconds dips of glass coverslips in the stable OTS:water emulsion can considerably change hydrophobicity of the coverslips.   
    A soft whitish aggregate sometimes appeared on the surface and was removed using 
    hexane, producing a transparent glass coverslip coated with OTS nanoscopic layer.
    For quick qualitative wetting checks, we also prepared ``half-coated'' slides by dipping only
    one half of a coverslip in the OTS solution and leaving the other half uncoated, enabling
    a direct side-by-side comparison (Fig.~\ref{fig:supp_ots_formation}(e)).

\end{enumerate}

\subsection*{Wetting contrast confirms OTS functionalization}

A $5\,\mu\mathrm{L}$ DI-water droplet was added onto each surface (or onto each half of a half-coated slide)
and imaged from the side to qualitatively assess the apparent contact angle.  (Fig.~\ref{fig:supp_ots_formation}(e)).

\medskip

\begin{itemize}
    \item \textbf{Uncoated slides:} the water droplet spreads.
    \item \textbf{OTS-coated slides:} the water droplet shows higher curvature and an enhanced 
    contact angle, demonstrating enhanced hydrophobicity even with the shorter 
    $\sim 40$~min incubation period.
\end{itemize}

Additional representative wetting comparisons (including strongly hydrophobic slides prepared
with longer incubation/coating times) are shown in Fig.~\ref{fig:supp_ots_formation}(f).

\subsection*{Surface control for MAS growth}

These ambient-OTS-treated slides enable improved wetting control and thin-layer formation for biphenyl--TCNB and anthracene--TCNB crystals, supporting simplified MAS-based thin-crystal fabrication workflows. The resulting changes in condensate wetting and droplet patterning are illustrated in Sec.~\ref{sec:SI_OTS_effect} (Fig.~\ref{fig:supp_ots_growth_effect}).

\section{Minimal-processing glass microchambers for MAS growth}\label{sec:SI_bottom}

This section describes the preparation of the bottom glass substrate, which forms the hydrophilic 
base of the microscopic growth chamber. The top hydrophobic OTS-coated glass slides are later 
placed above this hydrophilic substrate to create the complete chamber assembly. Typically, new 
boxes of Corning Inc.\ \#1 glass slides are opened and used as the bottom substrates, which are 
then batch-fabricated. These slides may also be cleaned by immersion in ethanol followed by 
approximately 5 minutes of sonication. Importantly, we do not perform any potassium-hydroxide-based glass cleaning procedures commonly used in single-molecule biophysics, as our aim is to maintain a simple and accessible MAS quantum sensor fabrication process and demonstrate that the method functions effectively under minimal preparation requirements \cite{Chandradoss2014SurfacePassivationSingleMolecule}. Notably, MAS co-crystals can be reproducibly formed on top substrate not coated with OTS, although OTS coating might offer additional control over patterning and lateral size of co-crystals like Bi:TCNB that form via droplet-mediated pathway.  

\medskip
The same Corning Inc.\ \#1 glass coverslips (approximately $130\,\mu$m thickness) are used as 
spacers. These spacers are positioned on the bottom glass slide and secured using a UV-curing optical 
adhesive (e.g., Norland), then cured under an off-the-shelf UV lamp for $>15$~min. 
After curing, the exposed glass surfaces are rendered hydrophilic using a low-cost handheld plasma tool (optional step) for several minutes. The resulting glass-sandwich assembly 
defines the microspacing used throughout this work (see also Sec.~\ref{sec:SI_videos} for practical 
examples of how spacer thickness and excess precursor powder can affect crystal growth features).

\medskip

After UV curing of the spacer and optional hand-held plasma treatment, a small quantity ($<1$\,mg) of crushed and ground D + A organic amorphous co-crystal powder is sparsely distributed across the bottom glass slide using metal forceps. The OTS-coated and hexane-cleaned top glass slide is then placed on top of this prepared substrate to enable growth of the organic co-crystals. The complete assembly is transferred to a hot plate to initiate the MAS quantum sensor fabrication process.

\medskip
We note that the low-cost handheld plasma tool can be replaced with a commercial high-quality oxygen plasma cleaning chamber, which would further improve surface uniformity. We also note that the glass substrate may be replaced with silicon, quartz, or sapphire wafers, all of which provide more homogeneous heating of the plates and improved thermal conduction compared to glass. In our work, glass substrates are used to enable direct optical imaging of the MAS growth process. Prior reported development of MAS crystal growth described by Ye \textit{et al.} \cite{Ye2019CocrystalsMicrospacing,Ye2018MicrospacingSublimation}, have demonstrated that MAS quantum sensor fabrication can be performed successfully on both glass cover-slips and silicon--silicon dioxide wafer substrates.

\section{A fume-hood microscope for real-time MAS growth and spectroscopy}\label{sec:SI_scope}

\begin{figure}[!t]
    \centering
    \includegraphics[width=0.49\textwidth]{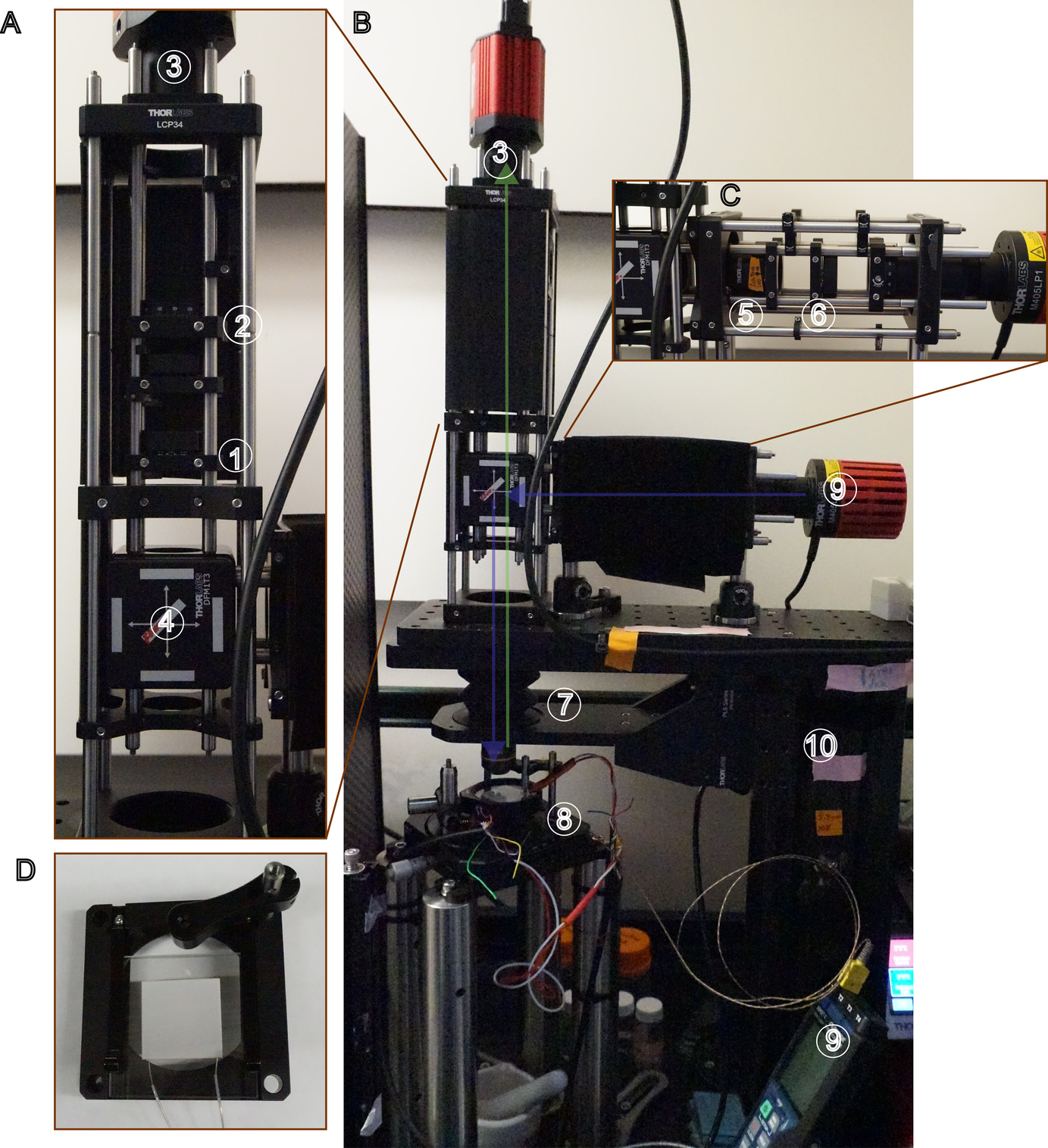}
    \caption{\justifying{\textbf{Off-the-shelf microscope for MAS-triplet crystal growth and in situ optical monitoring.}
\textbf{(a)} Emission and imaging path. \textbf{(b)} Photograph of the microscope installed inside a chemical fume hood. \textbf{(c)} Excitation path (components uncovered). \textbf{(d)} Hot-plate arrangement used for MAS growth.}}
\label{fig:supple_fumehood_scope}
\end{figure}

The microscope was built to image, optimize, and provide direct evidence of droplet-based and vapor-phase-based growth during the MAS-triplet crystal formation process. In principle, MAS crystal growth can be performed using only a hot plate, glass coverslips, and an amorphous organic D/A precursor mix placed in a glass vial---and can even be implemented using a standard soldering hot plate with post-growth optical inspection. The microscope enables real-time monitoring, controlled optimization of the growth dynamics and rapid determination of MAS-crystallization temperature points for unknown organic molecule pairs. The system was assembled using Thorlabs (TL) Cerna components and customized for installation inside a standard chemical fume hood. Organic precursor molecules are used in extremely small microgram amounts in MAS growth, however, many organic (polyaromatic hydrocarbons) vapors involved in MAS growth are carcinogenic \cite{Jameson2019PAHOccupationalExposures} and therefore, require operation inside standard fume-hood. The microscope is constructed on a compact aluminum breadboard with a footprint of 12~in $\times$ 24~in (TL: \textbf{MBH1224}), supported on Sorbothane vibration-isolation pads. The excitation and emission paths follow a wide-field microscope geometry. Both the camera and LED light source were selected to operate with passive cooling inside fume-hood during crystal imaging. Owing to the limited space inside the fume hood, the microscope was built in a vertically tall configuration, which is more susceptible to mechanical vibrations, motivating the use of passively cooled components with structural thermal fins.
In addition to documenting the droplet-mediated and vapor-phase growth pathways, this microscope enables \emph{time-resolved, in situ} growth \emph{and} characterization of MAS-triplet crystals as they form. In particular, wide-field fluorescence imaging can be combined with optional in situ emission spectroscopy through the same collection path (Sec.~\ref{sec:SI_spectrum}), enabling direct correlation between nucleation events observed on the camera and evolving spectral signatures of the condensed phases. The compact footprint and vertical layout were intentionally chosen so that the entire instrument (including the hot stage) can be installed and operated inside a standard chemical fume hood for safe handling of volatile organics. Although home-built, the design is based on low-cost off-the-shelf components and a straightforward mechanical layout; we therefore provide assembly details and a parts list so that the microscope can be replicated in many laboratories.
The microscope body is based on a 400~mm-tall Cerna modular frame with epi arm (TL: \textbf{CEA1400}). Focusing along the vertical ($z$) axis is achieved using a motorized dovetail translation module (TL: \textbf{PLSZ}), mounted via an angle bracket (TL: \textbf{PLSZ2}) and driven by a single-axis stepper controller (TL: \textbf{MCM301}). Lateral positioning of the sample is provided by a manual $X$--$Y$ translation stage with integrated rotation (TL: \textbf{XYR1}), which supports the custom hot-plate assembly described below.

\medskip
\noindent\textbf{(a) Emission and imaging path (Fig.~\ref{fig:supple_fumehood_scope}(a)).}
Fluorescence or bright-field light is collected after either a dichroic mirror (463~nm cut-on; TL: \textbf{DMLP463R}, marked 4) or a beam splitter for bright-field imaging. The emission path includes a tube lens (achromatic lens, $f = 200$~mm; TL: \textbf{AC254-200-A}, marked 1), mounted in SM2 lens tubing (TL: \textbf{SM2L15}, \textbf{SM2L20}), two optional rotation joints (TL: \textbf{CRM1T}) to enable cross-polarization microscopy, and a color CMOS camera (TL: \textbf{CS126CU}, marked 3) connected via a USB~3.0 right-angle cable (TL: \textbf{CABU32}). The panel also shows the motorized objective head consisting of a single-objective Cerna arm (TL: \textbf{CSN100}) holding a 10$\times$ objective (Edmund Optics: \textbf{88-333}, marked 7) via an RMS adapter (TL: \textbf{M32RMSS}). From the emission line, a 90-degree optical adapter can be used to split the collected light and couple it into a UV--Vis spectrometer, enabling direct observation of emission spectral changes correlated with nucleation events of fluorescent co-crystals.

\medskip
\noindent\textbf{(b) Full microscope inside fume hood (Fig.~\ref{fig:supple_fumehood_scope}(b)).}
Photograph of the complete microscope installed inside a standard chemical fume hood. Sorbothane pads were placed beneath the base aluminum breadboard to minimize vibration coupling from the fume hood environment. Structural support is provided using standard 1.5-inch-diameter pedestal posts and clamps (TL: \textbf{P10}, \textbf{PB4}, \textbf{PF175B}, \textbf{CL8}).

\medskip
\noindent\textbf{(c) Excitation path (Fig.~\ref{fig:supple_fumehood_scope}(c)).}
Excitation path with optical components uncovered. The excitation line consists of a UV LED light source (TL: \textbf{M405LP1}; alternately, TL: \textbf{SOLIS-1D}, marked 9), driven by a passive-cooled LED driver (TL: \textbf{DC20}), followed by a ground-glass diffuser (TL: \textbf{DG20-1500}), a back-focal-plane lens (achromatic lens, $f = 200$~mm; TL: \textbf{AC254-200-A}), and an optical slot (marked 6) for insertion of additional filters. An adjacent rotation cage mount can hold polarizers, enabling cross-polarization microscopy of the crystals.

\medskip
\noindent\textbf{(d) Hot-plate arrangement (Fig.~\ref{fig:supple_fumehood_scope}(d)).}
Example hot-plate arrangement used for MAS growth. A rectangular ceramic heater (TL: \textbf{HT24S}) is sandwiched between two high-temperature glass plates (Edmund Optics BOROFLOAT\textsuperscript{\textregistered} windows; \textbf{39-274} for circular and \textbf{29-392} for square plates), rather than standard glass slides, which tend to fracture during repeated heating and cooling cycles. The hot-plate assembly (marked 8) is mounted on the manual $X$--$Y$ translation stage for imaging. Temperature is monitored using a platinum RTD (TL: \textbf{TH100PT}), with an additional slow-response Kapton-adhesive K-type temperature probe attached to the underside of the hot plate to provide an independent temperature cross-check. Heating is driven using a dual-channel PID temperature controller (TL: \textbf{TC300B}) to supply controlled current profiles to the ceramic heater.

\section{An integrated optical and microwave platform for ODMR and coherent control}\label{sec:SI_ODMR_setup}

\noindent\textbf{Microscope and collection geometry.}
ODMR and spin-coherence measurements were performed on a custom-built microscope based on a Mad City Labs RM21 custom TIRF microscope body. The microscope was equipped with a micron-resolution microdrive and a nanopiezo for focusing the sample along the Z axis. Excitation light was coupled from the back port of the microscope and directed onto the sample through a Thorlabs dichroic beamsplitter, using either a 463~nm or 567~nm dichroic beamsplitter. Photoluminescence emitted from the sample crystals was collected through either a 20$\times$ long-working-distance objective [ Olympus LWD MPLANFL 20X OBJ] or a 50$\times$ objective [Olympus MXPlan Semi-Apo 50X OBJ, NA 0.8, WD 3~mm], transmitted through the dichroic beamsplitter, and directed into the RM21 emission path.

\medskip
\noindent\textbf{Dual-wavelength excitation and switching.}
Two laser excitation lines were implemented: a 532~nm Coherent Verdi green laser and a 405~nm Coherent OBIS laser. For the 532~nm excitation path, the beam diameter was reduced using a lens to pass through an Isomet acousto-optic modulator [model: M1133-aQ80L-1.5; A/R-coated for operation at 532~nm] driven by an Isomet digital driver [model: 522F-4], enabling optical switching with an approximate switching speed of 150~ns. The beam was then expanded and directed into the excitation cage assembly of the RM21 microscope. A 150~mm achromat lens in the excitation cage assembly focused the excitation beam into the objective back aperture, providing back-focal-plane illumination at the sample. For the 405~nm excitation path, the beam diameter was adjusted using a telescope lens arrangement to obtain a usable field of view and was coupled to the same resonator geometry for ODMR measurements. The 405~nm laser was switched directly using the TTL modulation available from the OBIS laser line, specified for digital switching up to 50~MHz. A motorized flip mount in the excitation path was used to select between the laser excitation lines. NV ODMR and spin-coherence measurements on a DNVB1 sample were used to verify ODMR and spin-coherence operation of the microscope; the green laser line was used for NV ODMR verification, while the 405~nm laser line was switched in for measurements on MAS-grown crystals.

\medskip
\noindent\textbf{Emission and detection paths.}
In the emission path, the collected photoluminescence was focused using a tube lens and then split between camera imaging and detection channels. A 10\% fraction of the emission light was directed to a Teledyne Kinetix CMOS camera for imaging, while the remaining 90\% was directed to the detection and spectroscopy path. A motorized flip mount controlled whether this 90\% emission path was directed into an Ocean Optics spectrometer for emission-spectrum measurements or into the photodetector path for photodiode-based ODMR measurements. In the detector path, the emission was either focused directly onto a photodiode using a 25~mm plano-convex lens, coupled into a multimode fiber connected to a Thorlabs APD, or coupled into a single-photon detector from Excelitas. Laser-line notch filters were placed in the emission path to minimize excitation-laser leakage into the detector. A 405~nm laser-line notch filter was used for 405~nm excitation, and a 532~nm laser-line notch filter was used for 532~nm excitation. For photodiode-based measurements, the photodiode signal was either sampled using an NI USB-6343 DAQ or connected to an SRS SR830 lock-in amplifier for lock-in-based ODMR detection.

\medskip
\noindent\textbf{Through-aperture microstrip resonator.}
For microwave delivery, a custom PCB resonator was fabricated on a Rogers RO4350B laminate substrate [RO4350B; $D_k = 3.48$, $D_f = 0.0037$]. A thin 0.51~mm board was used to allow through-board optical collection from the sample field of view. The PCB contained an imaging aperture, around which an impedance-matched 50~$\Omega$ microstrip line was routed in a curved geometry. The microstrip was designed to wrap around an approximately 1.6~mm diameter hole, allowing optical imaging through the aperture while delivering microwave fields to the sample region.

\medskip
\noindent\textbf{Waveguiding-assisted crystal positioning.}
Glass slides containing MAS-triplet crystals grown using the fume-hood microscope setup were placed onto the microstrip resonator. The laser beam diameter was adjusted so that the excitation beam passed collimated through the microstrip resonator imaging aperture. The beam was aligned to pass through a narrow region of the aperture, minimizing microwave-drive gradients across the illuminated field of view during coherence experiments. MAS crystals often populated or grew over millimeter-scale regions of the glass slide, so when the laser beam hit a single crystal, the emission could shine strongly and waveguide through the slide, providing a useful coarse visual guide for placement. This visual guidance was especially useful while moving the glass slide on the microstrip platform with the laser aligned through the imaging aperture. For systems such as anthracene:TCNB, leftover blue anthracene flakes could make the coverslip appear blue. When the laser reached an An:TCNB crystal, the emission visibly changed to orange at the laser spot and at the coverslip edges. These visible color and waveguiding changes were essential guides for positioning the glass slide and locating MAS-triplet crystals before fine field-of-view adjustment by the microdrive and camera for ODMR measurements.

\medskip
\noindent\textbf{Microwave drive and resonator protection.}
The microwave path consisted of an SRS SG384 signal generator [part number: SG384], followed by a Mini-Circuits absorptive SPDT microwave switch [part number: ZASWA-2-50DRA+] for microwave gating, microwave amplification using a Mini-Circuits high-power amplifier [part number: ZHL-30W-252-S+], and delivery to the PCB microstrip resonator. The gated and amplified microwave signal was coupled into the microstrip to drive spin resonance transitions during ODMR and spin-coherence measurements. Microwave circulators were included in the amplified microwave line to protect the amplifier and manage reflected microwave power from the resonator. Circulators covering the 0.8--2~GHz and 2--3~GHz frequency ranges were used switchably depending on the resonance frequency being addressed. For the 0.8--2~GHz line, a Pasternack high-power circulator [part number: PE83CR1000; 100~W, SMA female, 18~dB isolation] was used. For the 2--3~GHz line, a second circulator was used.

\medskip
\noindent\textbf{Phase control and pulse synchronization.}
For pulsed measurements, the in-phase and quadrature control voltages were routed to the IQ inputs of the SRS SG384 signal generator to control the RF phase. Two high-speed microwave switches [part number: ZASWA-2-50DRA+] were used in the IQ-control circuit to toggle the I and Q control voltages. The resulting IQ-controlled microwave output from the SG384 was then routed through the microwave delivery path. The experiment timing was synchronized using a Swabian Pulse Streamer, which gated the laser, the microwave switch used for absolute microwave delivery, and the microwave DC switches used for IQ control. The Pulse Streamer also provided the measurement scan clock and start trigger for the NI USB-6343 DAQ during pulsed measurements. Custom-written MATLAB code was used to control the instruments and coordinate the pulse timing for the ODMR and spin-coherence experiments. For continuous-wave ODMR measurements, amplitude-modulated microwave excitation and lock-in-amplifier detection were used.

\FloatBarrier
\section{Growth videos reveal pathway-specific optimization and failure modes}\label{sec:SI_videos}
The broad goal of this section is to provide comments along with example growth video recordings to enable communicating all aspects (non-ideal experiments as well) of the MAS-triplet crystal growth processes. 

\T{Supplementary Video 1.}
\I{Fluorescence microscopy recording of Bi:TCNB growth. (Left) Fluorescence microscopy frames, with time and temperature for each frame printed in the top corner. (Right) Real-time temperature marker: a vertical bar, with faded variants indicating the temperature of prior frames.}

This growth recording image the top glass slide; an amorphous, ground mixture of biphenyl and tetracyanobenzene (TCNB) is located on the bottom glass slide ($\sim$120~$\upmu$m below) and exhibits green fluorescence upon UV excitation, due to the potentially polycrystalline formation of Bi:TCNB during grinding.

Heating induces sublimation from the bottom slide and condensation on the top slide due to the temperature gradient between them. Droplets of biphenyl/TCNB appear and grow via Ostwald ripening, transforming smaller droplets into larger ones. The periodic droplet pattern arises from the OTS coating on the top slide. Non-uniform loading on the bottom substrate is reflected in spatial variations in droplet size on the top slide, with larger droplets forming above regions with higher source material due to increased local sublimation particle flux.

Around 230~s, nucleation occurs within the droplets, appearing as small optically clear single-crystal nuclei growing out of the droplet melt. Their optical contrast is consistent with single-crystal formation, supported by SEM images showing a distinctive triclinic morphology. The measured temperature ($\sim$115--120$^\circ$C for this hot-stage geometry) is critical for enabling nucleation. For unknown co-crystals growing from the droplet melt, observing the process live is recommended, as the temperature of heating indicates a deviation from the optimal growth profile. After nucleation, the temperature is lowered and held, allowing the nuclei to grow while drawing material from their parent droplets.

\T{Supplementary Video 2.}
\I{Fluorescence microscopy recording of Bi:TCNB growth. (Left) Fluorescence microscopy frames, with time and temperature for each frame printed in the top corner. (Right) Real-time temperature marker: a vertical bar, with faded variants indicating the temperature of prior frames.}

The nucleation and growth process is largely similar to that in Supplementary Video~1, with two differences. First, the stage was cooled prematurely, causing the single crystals to become frozen within a polycrystalline biphenyl/TCNB melt rather than being sustained long enough in the hot droplet for continued growth. Second, contact of the amorphous ground source material with the top glass slide leads to polycrystalline formation upon heating, indicating a deviation from the optimal growth profile. In this acquisition, the fume-hood directional inward air-currents are visible around 190s timepoint, creating an angle to the droplet patterns.   

\T{Supplementary Video 3.}
\I{Fluorescence microscopy recording of Bi:TCNB growth. (Left) Fluorescence microscopy frames, with time and temperature for each frame printed in the top corner. (Right) Real-time temperature marker: a vertical bar, with faded variants indicating the temperature of prior frames.}

This recording shows growth behavior similar to that observed in Supplementary Videos~1 and~2 for Bi:TCNB; however, it highlights increased heterogeneity in the sizes of the condensed droplets. Some droplets become relatively large and reach an equilibrated size after droplet Ostwald ripening ceases as the temperature rises. Within these larger droplets, multiple nucleation events can occur, leading to the formation of several needle-shaped single crystals originating from a single parent droplet.

\T{Supplementary Video 4.}
\I{Fluorescence microscopy recording of Bi:TCNB growth. (Left) Fluorescence microscopy frames, with time and temperature for each frame printed in the top corner. (Right) Real-time temperature marker: a vertical bar, with faded variants indicating the temperature of prior frames.}

This recording captures a case in which the hot stage was pre-warmed to approximately 50$^\circ$C prior to inserting the glass-slide sandwich; this acquisition was performed by error, while in most usual cases we ensure the stage returns to room temperature before sample insertion. As a result, small floral-like structures are already present in the first recorded frame, arising from premature heating of biphenyl and TCNB molecules that condensed on the top glass slide. As the temperature increases further, the usual and repeatable Bi:TCNB growth process is initiated, with the formation of approximately periodic Bi:TCNB melt droplets condensing on the top slide. These droplets subsequently enlarge and eventually exhibit nucleation events in the temperature range of approximately 115--120$^\circ$C.

\T{Supplementary Video 5.}
\I{Fluorescence microscopy recording of An:TCNB growth. (Left) Fluorescence microscopy frames, with time and temperature for each frame printed in the top corner. (Right) Real-time temperature marker: a vertical bar, with faded variants indicating the temperature of prior frames.}

Recordings of the formation of An:TCNB crystals imaged through a 50$\times$ long--working-distance objective, focused near the MAS glass-slide assembly. An amorphous mixture of anthracene and TCNB was ground together, and less than 1~mg of the powder was lightly sprinkled onto the bottom glass substrate. The sample was assembled in a sandwich geometry using a microspacing (\#1 cover glass spacer), with the top slide consisting of an OTS-coated glass substrate.

Recordings were initiated with the microscope focus on the bottom substrate, where the amorphous ground mixture is visible. Simultaneously, the PID-controlled current driving the hot stage was monitored to provide an approximate estimate of the temperature at which An:TCNB co-crystallization occurs. The focus was then moved to the top glass slide. Around 130$^\circ$C, blue fluorescence of the top slide under UV excitation is observed, originating from anthracene crystallization. At slightly elevated temperatures, UV-excited orange An:TCNB co-crystals begin to appear as nucleation sites and grow in size, primarily forming needle-like crystals and occasionally plate-like morphologies. The stage was allowed to cool after reaching approximately 160$^\circ$C.

During growth, prominent side facets consistent with a monoclinic single-crystal structure of An:TCNB are visible. Due to excess material loaded on the bottom substrate, in addition to the An:TCNB co-crystals, a poly-amorphous anthracene film covering the top slide remains at the end of Supplementary Video~5. This excess deposition film can be easily minimized by reducing the bottom-substrate loading of source anthracene and TCNB.

\T{Supplementary Video 6.}
\I{Fluorescence microscopy recording of An:TCNB growth. (Left) Fluorescence microscopy frames, with time and temperature for each frame printed in the top corner. (Right) Real-time temperature marker: a vertical bar, with faded variants indicating the temperature of prior frames.}

In this acquisition, the amount of amorphous source material loaded onto the bottom slide was reduced. As a result, the recording does not exhibit a residual blue--purple film of excess anthracene at the end of the growth process. Instead, a clearer representation of the nucleation and growth of single-crystalline An:TCNB co-crystals is observed.

\T{Supplementary Video 7.}
\I{Fluorescence microscopy recording of An:TCNB growth. (Left) Fluorescence microscopy frames, with time and temperature for each frame printed in the top corner. (Right) Real-time temperature marker: a vertical bar, with faded variants indicating the temperature of prior frames.}

These recordings also show well-defined needle-shaped anthracene crystals forming in parallel with the An:TCNB co-crystals. Prior studies have reported MAS formation of pure anthracene crystals, consistent with these observations.

\T{Supplementary Video 8.}
\I{Fluorescence microscopy recording of An:TCNB growth. (Left) Fluorescence microscopy frames, with time and temperature for each frame printed in the top corner. (Right) Real-time temperature marker: a vertical bar, with faded variants indicating the temperature of prior frames.}

In this case, the relative vertical spacing between the glass slides was higher than usual, possibly due to a ground mixture thickness greater than the spacer thickness, thereby increasing the effective vertical spacing. As a result, many floral-like crystal structures are observed. In general, excessively large vertical spacing leads to the formation of small, floral single-crystal morphologies, while excessively small vertical spacing can result in no crystallization.

In the present work, optimization of the vertical spacing was guided by the requirement to demonstrate ODMR from MAS-grown crystals. Further optimization may be pursued from a nanoscale device deployment perspective. We also note that for extremely sparse loading of source material on the bottom substrate, no anthracene blue film is observed, and instead only sparse, ultra-thin ($<1~\upmu$m) An:TCNB co-crystals grow directly, typically at a similar temperature of around 160$^\circ$C.

\section{OTS nanolayers control condensate wetting and nucleation geometry}\label{sec:SI_OTS_effect}
\noindent
Using top substrates prepared as described in Sec.~\ref{sec:SI_OTS}, during MAS growth of Bi:TCNB, co-sublimed biphenyl and TCNB condense on the (cooler) top substrate as a droplet melt driven by the small vertical temperature gradient between the bottom and top slides \cite{Ye2018MicrospacingSublimation}. The surface chemistry of the top slide strongly influences the droplet wetting profile and therefore the spatial patterning of subsequent nucleation events \cite{Ye2019CocrystalsMicrospacing,Ye2018MicrospacingSublimation}.

\begin{figure*}[!t]
\centering
\includegraphics[width=0.6\textwidth]{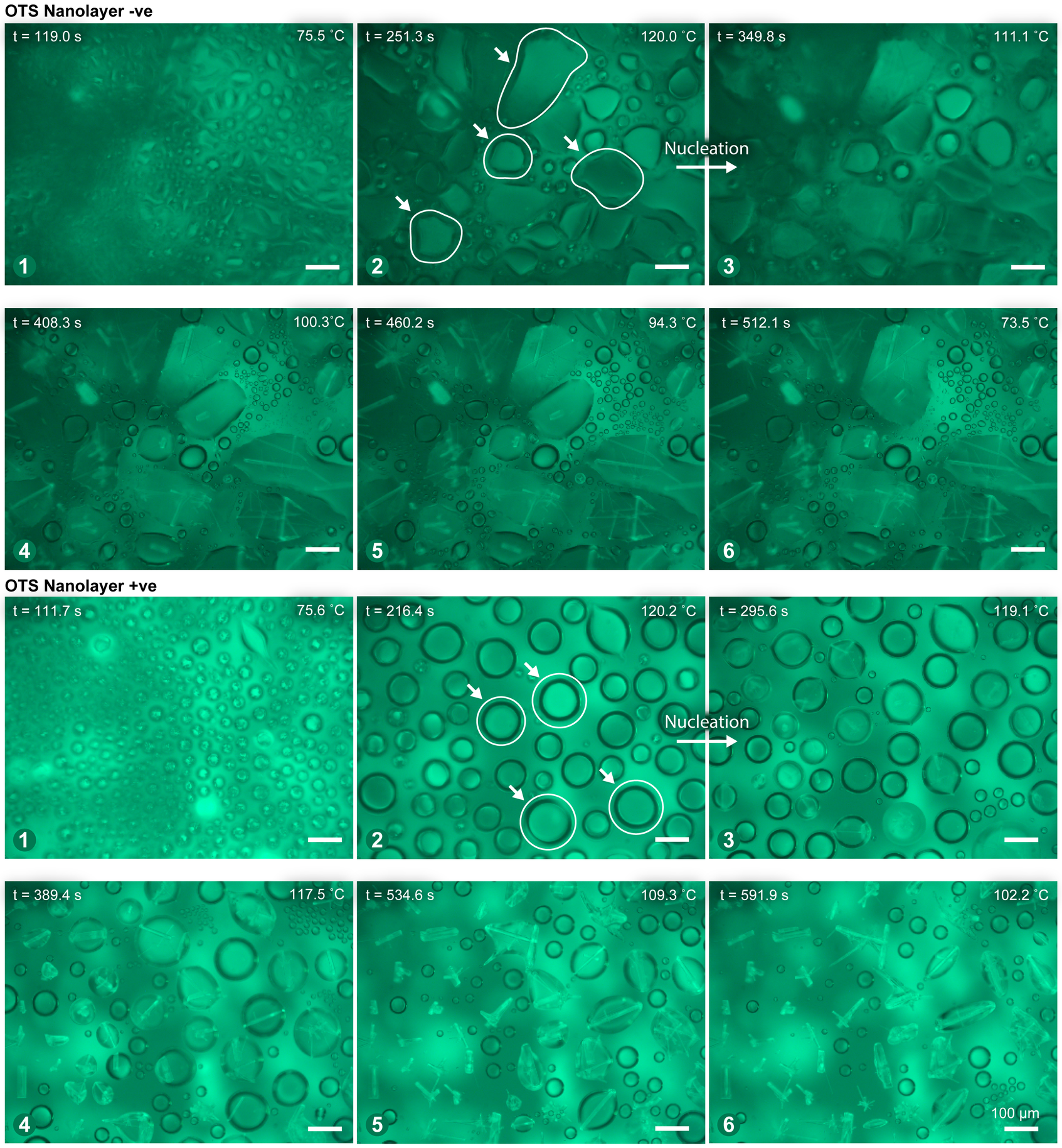}
\caption{\justifying{\textbf{OTS nanolayer alters condensate wetting and nucleation of MAS-grown biphenyl--TCNB crystals.}
Representative growth sequences on \textbf{(a)} an uncoated top slide and \textbf{(b)} an OTS-coated top slide. In \textbf{(a)}, droplet boundaries in the second time frame (examples are marked by arbitrarily shaped white outline to droplets) show a non-periodic, random wetting pattern; since nuclei grow to the size of their parent droplets, this yields large heterogeneity in lateral crystal size. In \textbf{(b)}, the second time frame instead shows circular and uniformly placed droplet condensates on the top substrate, leading to more homogeneous sized Bi:TCNB needles. Scale bars are $100\,\mu\mathrm{m}$.}}
\label{fig:supp_ots_growth_effect}
\end{figure*}

On an \textit{uncoated} top slide, the condensate wets the glass and adopts irregular lateral geometries (Fig.~\ref{fig:supp_ots_growth_effect}(a)). Upon heating to the nucleation temperature (here $\approx 111\,^\circ\mathrm{C}$), Bi:TCNB nucleation occurs within these droplets and needle-like crystals extend laterally in proportion to the dimensions of their parent droplet melt (Fig.~\ref{fig:supp_ots_growth_effect}(a)). In contrast, with an OTS-coated top slide the condensate forms more circular, periodically spaced droplets (Fig.~\ref{fig:supp_ots_growth_effect}(b)), consistent with the increased hydrophobicity/contact angle. Nucleation occurs at a similar temperature, but the more regular droplet array provides a straightforward handle to modulate nucleation density and spatial placement of MAS-grown crystals. In both cases, droplet growth proceeds via Ostwald ripening, while crystals expand by drawing material from their parent droplets during the growth and cool-down stages. For temperature-sensitive systems such as Bi:TCNB, representative failure modes and optimized cool-down profiles are discussed in Sec.~\ref{sec:SI_fragile}, and annotated growth recordings are provided in Sec.~\ref{sec:SI_videos}.

\section{Thermal-window control preserves Bi:TCNB single-crystal quality}\label{sec:SI_fragile}
\noindent
Biphenyl--TCNB (Bi:TCNB) is comparatively sensitive to temperature history during MAS growth, especially during the period immediately following nucleation in the droplet-mediated pathway described in Sec.~\ref{sec:SI_OTS_effect} (Fig.~\ref{fig:supp_ots_growth_effect}). We find that \textit{small deviations in the cool-down profile} can lead to either (i) premature solidification of the parent droplet melt or (ii) degradation of single-crystalline order from prolonged exposure at elevated temperature.

\begin{figure*}[!t]
    \centering
    \includegraphics[width=0.8\textwidth]{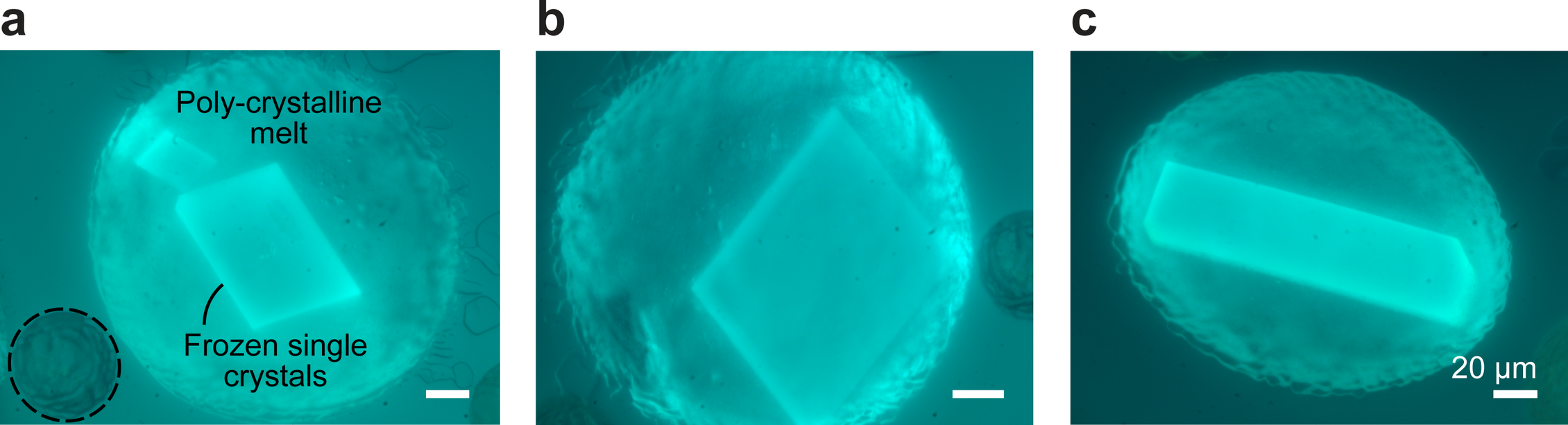}
    \caption{\justifying{\textbf{Premature cool-down can freeze Bi:TCNB needles inside the droplet melt.}
Representative micrographs showing \textbf{(a)} a solidified polycrystalline melt with embedded needles and \textbf{(b,c)} additional examples. Scale bar: $20\,\mu\mathrm{m}$.}}
    \label{fig:supp_precool_btcnb}
\end{figure*}

\textbf{Premature cooling.} If the stage is cooled too early, Bi:TCNB needles nucleate but become frozen within a solidified, polycrystalline biphenyl/TCNB droplet melt rather than continuing to grow while drawing material from the liquid phase. Fig.~\ref{fig:supp_precool_btcnb}(a) shows a representative case where multiple needles are embedded in a polycrystalline matrix; additional examples are shown in Fig.~\ref{fig:supp_precool_btcnb}(b,c).

\begin{figure*}[!t]
\centering
\includegraphics[width=0.9\textwidth]{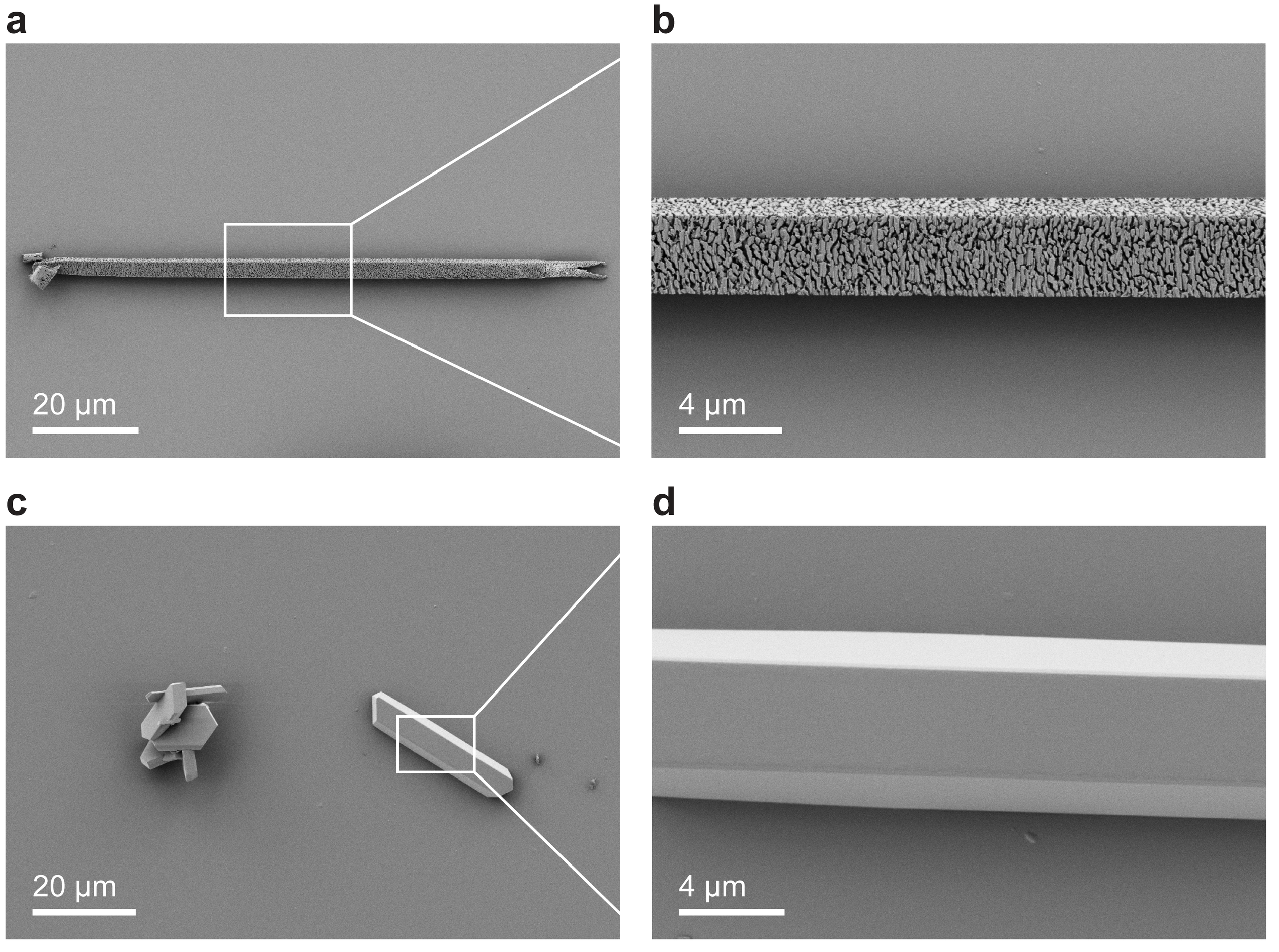}
\caption{\justifying{\textbf{SEM comparison of overheated versus optimized Bi:TCNB growth.}
\textbf{(a,b)} Example of a Bi:TCNB needle whose single-crystalline integrity was compromised by delayed cooling/overheating. \textbf{(c,d)} Optimized MAS growth and controlled cool-down preserve smooth single-crystalline surfaces.}}
\label{supp_fig_sem_broken}
\end{figure*}

\textbf{Delayed cooling / overheating.} Conversely, if crystals are held at elevated temperature longer than required after nucleation and growth, their single-crystalline integrity can be compromised. SEM images in Fig.~\ref{supp_fig_sem_broken}(a,b) show an overheated needle that retains its overall macroscopic morphology but develops nanometer-scale heterogeneity and apparent polycrystalline domains. Under an optimized MAS temperature profile with prompt, controlled cooling, Bi:TCNB needles maintain smooth, homogeneous surfaces (Fig.~\ref{supp_fig_sem_broken}(c,d)). In our experience, An:TCNB is substantially more thermally robust than Bi:TCNB and retains its single-crystalline morphology even when the cooling dynamics are modestly perturbed.

\section{In situ fluorescence resolves sequential An:TCNB co-crystallization}\label{sec:SI_spectrum}

To correlate the visually observed nucleation and growth dynamics with the evolving molecular composition on the top substrate, we coupled the microscope emission path to a UV--Vis spectrometer (via a 90$^{\circ}$ optical pick-off described in Sec.~\ref{sec:SI_scope}). During MAS growth, the sample was excited under the same wide-field 405~nm LED illumination used for imaging, and the emitted light was filtered by the 463~nm dichroic prior to spectral acquisition. This configuration rejects scattered excitation while preserving the fluorescence bands relevant to anthracene and the An:TCNB co-crystal.

\begin{figure}[!t]
    \centering
    \includegraphics[width=0.49\textwidth]{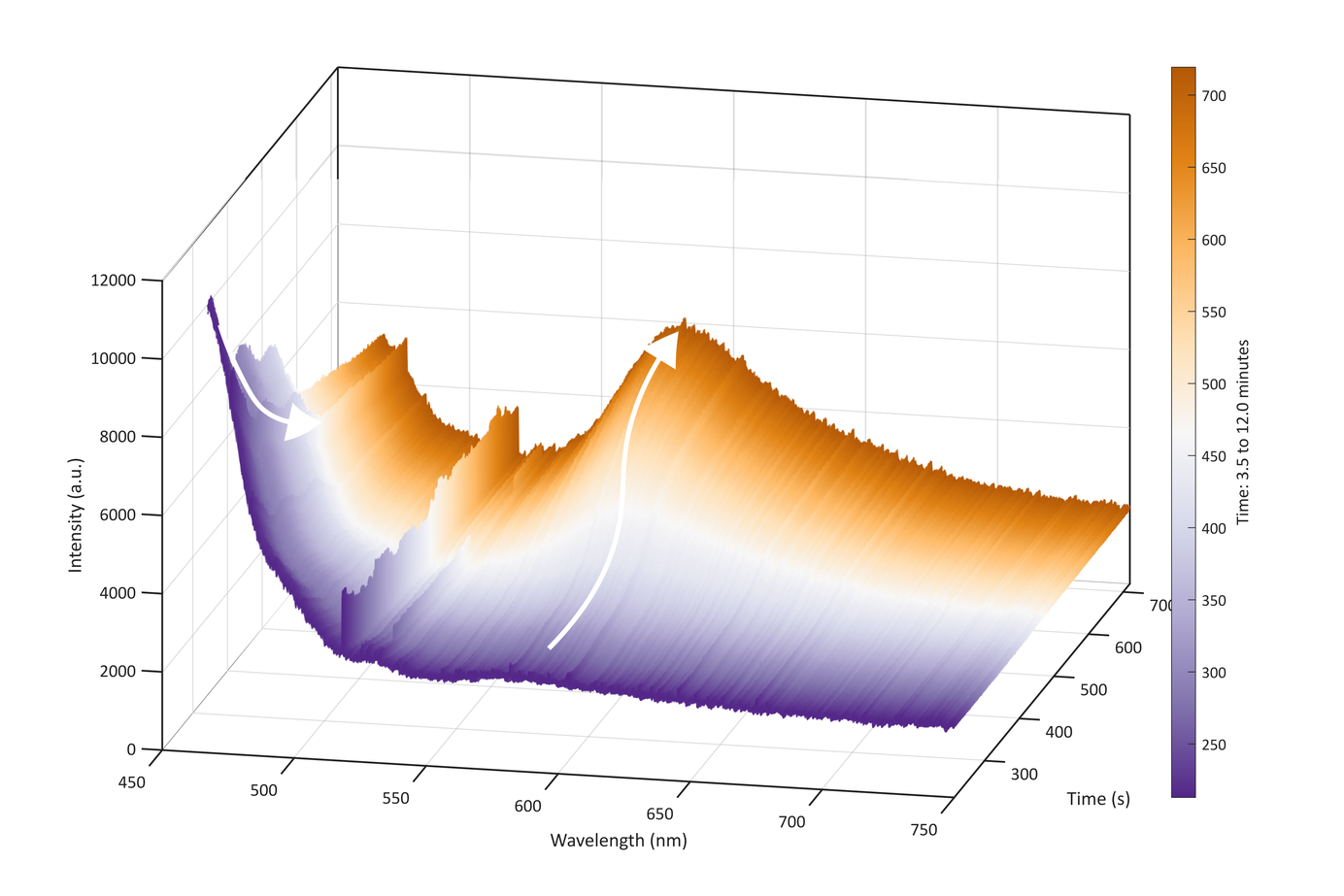}
    \caption{\justifying
    \textbf{In situ PL spectroscopy during MAS co-crystallization of An:TCNB (405~nm excitation).}
    Time-resolved emission spectra recorded through the microscope spectrometer port (filtered above 463~nm) show the decay of an anthracene-related band (463--500~nm) and the concurrent emergence of the orange An:TCNB co-crystal emission during nucleation and growth.}
    \label{fig:mas_cocrystallization_pl}
\end{figure}

Fig.~\ref{fig:mas_cocrystallization_pl} shows a representative time-resolved photoluminescence (PL) spectrum acquired during An:TCNB formation under the temperature profile used for the growth recordings in the main text. At early times ($\sim$250~s), the emission is relatively featureless, consistent with minimal condensed material on the top slide. As pure anthracene crystallites/films transiently populate the top substrate, a broad band appears in the 463--500~nm region under 405~nm excitation; this band subsequently decreases between $\sim$250~s and 450~s, coincident with the melting/restructuring of the anthracene flake-like features observed in the growth videos.

Beyond $\sim$350~s, as donor--acceptor co-crystallization proceeds, a new orange-emitting feature emerges and grows, corresponding to the An:TCNB co-crystal emission. The onset and growth of this orange peak provide a direct spectroscopic signature of co-crystal nucleation and growth that complements the optical micrographs. More broadly, this time-resolved emission monitoring provides a simple, non-contact diagnostic for MAS screening: for any candidate MAS-triplet crystal with a known (or expected) emission spectrum, the emergence of new spectral features can confirm successful co-crystallization \emph{in situ} and guide optimization of temperature ramps, hold times, and loading densities \cite{Ye2019CocrystalsMicrospacing}.

\section{MAS growth accesses diverse co-crystal morphologies}\label{sec:SI_addSEM}

\begin{figure}[!t]
    \centering
   \includegraphics[width=0.49\textwidth]{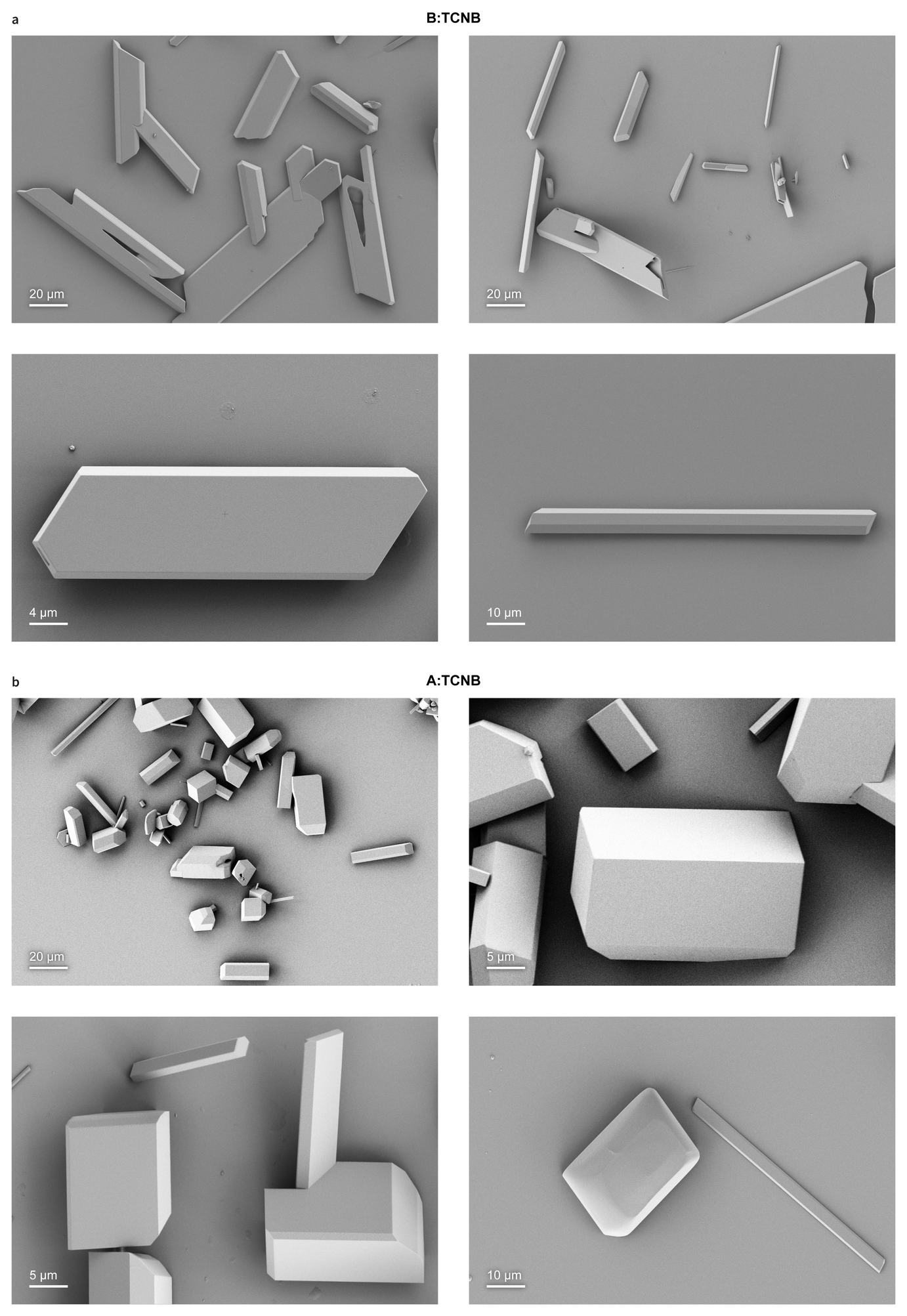}
    \caption{\justifying{\textbf{Additional SEM images of MAS-grown co-crystals.}
\textbf{(a)} Additional Bi:TCNB crystals (triclinic) with occasional residual TCNB flakes visible in the field of view. \textbf{(b)} Additional An:TCNB crystals (monoclinic), showing representative needle-like and plate-like morphologies.}}
    \label{fig:supp_additional_sem}
\end{figure}

\noindent
To provide additional examples of the morphological diversity accessible with MAS growth, we include representative SEM micrographs for both co-crystal systems discussed in the main text. For Bi:TCNB, Fig.~\ref{fig:supp_additional_sem}(a) shows additional triclinic crystals across wider fields of view, where residual flakes of TCNB that did not co-crystallize with biphenyl can sometimes be observed adjacent to the co-crystals. For An:TCNB, Fig.~\ref{fig:supp_additional_sem}(b) shows additional monoclinic crystals, where needle-like and plate-like morphologies can coexist within a single MAS growth event. In practice, bottom-substrate loading density, grinding fineness, vertical spacing, and temperature control can be tuned to bias growth toward isolated needles or larger plates.

\FloatBarrier
\section{Charge-transfer character and triplet-exciton dynamics in An:TCNB and Bi:TCNB}
\label{sec:SI_ODMR_background}
Weak organic charge-transfer (CT) co-crystals such as anthracene--tetracyanobenzene (An:TCNB) and biphenyl--tetracyanobenzene (Bi:TCNB) provide established reference systems for connecting donor--acceptor packing, CT admixture, intersystem crossing and photoexcited triplet magnetic resonance \cite{Krzystek1993TripletExcitons}. Photoexcitation produces singlet excited states with mixed local and CT character through partial electron transfer from the donor HOMO, anthracene or biphenyl, to the TCNB acceptor LUMO \cite{Krzystek1993TripletExcitons,VonSchutz1980ATCNBDelayedFluorescenceODMR,Corvaja1988BiphenylTCNBODMR}. The triplet CT admixture differs strongly between the two systems: An:TCNB is reported near the localized anthracene-triplet limit, whereas Bi:TCNB has been estimated to have approximately 0.5 CT character \cite{Pasimeni1983BiphenylTCNBTripletExcitons,Krzystek1993TripletExcitons}.
Following optical excitation, intersystem crossing (ISC) into the triplet manifold is influenced by the acceptor component and produces non-Boltzmann triplet-sublevel populations \cite{Krzystek1993TripletExcitons,Lower1966TheTS}. In An:TCNB, this mechanism generates mobile triplet excitons at room temperature, with motion primarily along molecular stacks and zero-field-splitting (ZFS) parameters that evolve with temperature due to anthracene librations \cite{VonSchutz1980ATCNBDelayedFluorescenceODMR,Steudle1978ATCNBMobileTripletExcitons,Frankevich1978ATCNBRYDMRTripletPairs}. ODMR and ESR studies of bulk An:TCNB crystals have resolved triplet magnetic-resonance signatures under high-field or low-temperature delayed-fluorescence detection conditions, with triplet dynamics governed by trapping and triplet--triplet annihilation across temperature \cite{VonSchutz1980ATCNBDelayedFluorescenceODMR,Steudle1978ATCNBMobileTripletExcitons,Frankevich1978ATCNBRYDMRTripletPairs}. In bulk Bi:TCNB crystals, related studies indicate room-temperature triplet excitons with intermediate CT character, mixed biphenyl/TCNB triplet character, acceptor-influenced ISC and observable magnetic-resonance contrast \cite{Agostini1991BiphenylTCNBXTrapODMR,Corvaja1988BiphenylTCNBODMR}. Thus, weak CT interaction stabilizes triplet excitons without fully ionizing the donor--acceptor pair, enabling ODMR through spin-dependent emission while retaining mobility and lifetime for room-temperature detection \cite{Krzystek1993TripletExcitons}.
For MAS-triplet materials, the earliest materials requirement is co-crystallization. Once a stable molecular lattice forms, optical triplet readout can emerge either from a donor--acceptor CT pair \cite{Sun2018MolecularCocrystals} or from a triplet-hosting photoexcited chromophore co-crystallized with a lattice host in a dense stoichiometric ratio \cite{Xue2024FullColorTADFCocrystals,Ye2018MolecularBarrierCocrystals}. This co-crystallization requirement is governed by noncovalent interactions, including $\pi$--$\pi$ stacking, hydrogen bonding, halogen bonding, charge-transfer interactions and van der Waals forces, which determine co-crystal formation and packing \cite{Zhu2026HierarchicalCocrystalAssemblies}. In such ordered co-crystals, fixed intermolecular geometry and orbital overlap can support triplet mobility \cite{Yost2012TripletSingletEnergyTransfer}. MAS is compatible with a wide range of co-crystals, including donor--acceptor systems, pharmaceutical co-crystals and twisted PAH--TCNB complexes \cite{Ye2019CocrystalsMicrospacing}, motivating its use here as a screening route for optically addressable triplet-spin layers.
The triplet magnetic-resonance spectra are described, to first order, by the spin-1 ZFS Hamiltonian, expressed in frequency units,
\[
H/h=(g\mu_B/h)\,\mathbf{B}\cdot\mathbf{S}+D[S_z^2-S(S+1)/3]+E(S_x^2-S_y^2),
\]
where $\mathbf{S}$ is the spin-1 operator, $g$ is the effective $g$-factor, $\mu_B$ is the Bohr magneton, $h$ is Planck's constant, $\mathbf{B}$ is the applied magnetic field, and $D$ and $E$ are the axial and transverse ZFS parameters, in frequency units, \cite{Krzystek1993TripletExcitons}. Electron--electron dipolar interactions lift the triplet-sublevel degeneracy, generating zero-field transitions in the MHz--GHz range accessible by ODMR.
Optical excitation populates the triplet manifold through ISC, producing non-equilibrium populations of the triplet eigenlevels, $P_i$ (labeled $m_S$ in the high-field limit). The detected photoluminescence follows spin-selective decay,
\[
I_{\mathrm{PL}}\propto \sum_{i} k_i P_i,
\]
so microwave redistribution of triplet-eigenlevel populations appears as ODMR contrast \cite{Krzystek1993TripletExcitons}. In mobile regimes, triplets diffuse across donor--acceptor sites and delayed emission becomes coupled to exciton hopping and triplet--triplet encounters, which can be described phenomenologically as
\[
\frac{dT}{dt}=G-k_TT-\gamma T^2,
\]
where $\gamma$ is an effective bimolecular triplet--triplet annihilation coefficient \cite{Bardeen2014MolecularExcitons}.

\begin{table*}[!t]
\centering
\caption{Comparison of ODMR studies in biphenyl--tetracyanobenzene (Bi:TCNB)}
\label{tab:SI_ODMR_comparison}
\small
\renewcommand{\arraystretch}{3}

\begingroup
\arrayrulecolor{SITableRule}
\setlength{\arrayrulewidth}{0.45pt}
\begin{tabular}{|l|>{\columncolor{SITableHighlight}}l|l|l|}
\hline
\rowcolor{SITableHeader}
\multicolumn{1}{|c|}{\textbf{Property}} &
\multicolumn{1}{c|}{\textbf{This work}} &
\multicolumn{1}{c|}{\shortstack{\textbf{Corvaja et al.\ (1988)}\\\cite{Corvaja1988BiphenylTCNBODMR}}} &
\multicolumn{1}{c|}{\shortstack{\textbf{Agostini et al.\ (1991)}\\\cite{Agostini1991BiphenylTCNBXTrapODMR}}} \\
\hline

Crystal (form) &
\shortstack[l]{Thin MAS-grown\\microcrystals} &
\shortstack[l]{Bulk single\\crystal ($>$mm)} &
\shortstack[l]{Bulk single\\crystal ($>$mm)} \\
\hline

Temperature &
\shortstack[l]{Room\\temperature} &
\shortstack[l]{4.2--300 K} &
\shortstack[l]{1.25--4.2 K\\RT ODMR reported} \\
\hline

Optical excitation &
\shortstack[l]{405 nm\\CW-laser} &
\shortstack[l]{360 nm CW,\\337 nm pulsed} &
\shortstack[l]{400 nm\\filtered lamp} \\
\hline

Emission collected &
\shortstack[l]{Green emission; \\ Total fluorescence ($>$463nm)} &
\shortstack[l]{Delayed fluorescence,\\Phosphorescence} &
\shortstack[l]{Phosphorescence (low T)\\Fluorescence (RT)} \\
\hline

ODMR method &
\shortstack[l]{CW-ODMR\\(zero-field)} &
\shortstack[l]{X-band high\\field-swept ODMR} &
\shortstack[l]{Zero-field\\ODMR} \\
\hline

Additional notes &
\shortstack[l]{MAS thin crystals, \\ RT ODMR observable} &
\shortstack[l]{TTA dominant at RT,\\X-trap phosphorescence\\at 4.2 K} &
\shortstack[l]{RT ODMR at\\919 and 1265 MHz,\\linewidth $\sim$3 MHz} \\
\hline

\end{tabular}
\endgroup

\end{table*}

\begin{table*}[!t]
\centering
\caption{Comparison of ODMR and optical studies in anthracene--tetracyanobenzene (An:TCNB)}
\label{tab:SI_ODMR_comparison_ATCNB}
\small
\renewcommand{\arraystretch}{2.7}

\begingroup
\arrayrulecolor{SITableRule}
\setlength{\arrayrulewidth}{0.45pt}
\begin{tabular}{|l|>{\columncolor{SITableHighlight}}l|l|l|l|}
\hline
\rowcolor{SITableHeader}
\multicolumn{1}{|c|}{\textbf{Property}} &
\multicolumn{1}{c|}{\textbf{This work}} &
\multicolumn{1}{c|}{\shortstack{\textbf{Frankevich et al.\ (1978)}\\\cite{Frankevich1978ATCNBRYDMRTripletPairs}}} &
\multicolumn{1}{c|}{\shortstack{\textbf{Steudle et al.\ (1978)}\\\cite{Steudle1978ATCNBMobileTripletExcitons}}} &
\multicolumn{1}{c|}{\shortstack{\textbf{Sch{\"u}tz et al.\ (1980)}\\\cite{VonSchutz1980ATCNBDelayedFluorescenceODMR}}} \\
\hline

Crystal (form) &
\shortstack[l]{Thin MAS-grown\\microcrystals} &
\shortstack[l]{Bulk single\\crystal ($>$mm)} &
\shortstack[l]{Bulk single\\crystal ($>$mm)} &
\shortstack[l]{Bulk single\\crystal ($>$mm)} \\
\hline

Temperature &
\shortstack[l]{Room\\temperature} &
\shortstack[l]{Room\\temperature} &
\shortstack[l]{1.2--300 K\\(low-T emphasized)} &
\shortstack[l]{1.2 K} \\
\hline

Optical excitation &
\shortstack[l]{405 nm\\CW laser} &
\shortstack[l]{Xe lamp\\350--530 nm (CW)} &
\shortstack[l]{457.9 nm Ar-ion\\laser; dye laser} &
\shortstack[l]{514 nm laser;\\triplet-band excitation} \\
\hline

Emission collected &
\shortstack[l]{Orange emission;\\total fluorescence$>$463nm} &
\shortstack[l]{Total fluorescence\\($>$550 nm);\\delayed fluorescence} &
\shortstack[l]{Delayed fluorescence;\\phosphorescence\\below 40 K} &
\shortstack[l]{Delayed fluorescence\\(from TTA)} \\
\hline

ODMR method &
\shortstack[l]{CW-ODMR\\(zero-field)} &
\shortstack[l]{High-field RYDMR\\(X-band $\sim$9.5 GHz)} &
\shortstack[l]{Optical\\spectroscopy only} &
\shortstack[l]{Zero-field\\DF-ODMR} \\
\hline

Additional notes &
\shortstack[l]{MAS thin crystals;\\optically clear;\\RT ODMR observable} &
\shortstack[l]{TTA via short-lived\\bitriplet state;\\polarized triplets} &
\shortstack[l]{Triplet on anthracene;\\TTA-dominated DF;\\mobile triplets at 1.2 K} &
\shortstack[l]{Triplet on anthracene;\\TTA-dominated;\\Davydov splitting} \\
\hline
\end{tabular}
\endgroup
\end{table*}
\section{MAS extends charge-transfer ODMR to substrate-bound microcrystals at room temperature}\label{sec:SI_ODMR_literature}

Tables~\ref{tab:SI_ODMR_comparison} and~\ref{tab:SI_ODMR_comparison_ATCNB} place our ODMR results in the context of prior magnetic-resonance and optical studies on biphenyl--tetracyanobenzene (Bi:TCNB) and anthracene--tetracyanobenzene (An:TCNB). Earlier work on these donor--acceptor co-crystals established the existence of photoexcited triplet manifolds and identified spin-dependent emissive pathways (delayed fluorescence and/or phosphorescence) that enable ODMR readout. However, these studies predominantly focused on bulk, millimeter-scale crystals and (especially for zero-field ODMR) often relied on cryogenic conditions to access long-lived triplet populations and narrow linewidths.

\textbf{Bi:TCNB.} Corvaja \textit{et al.} performed field-swept X-band ODMR on bulk single crystals over 4.2--300~K, using near-UV excitation and detecting triplet signatures primarily through delayed fluorescence/phosphorescence channels (Table~\ref{tab:SI_ODMR_comparison}) \cite{Corvaja1988BiphenylTCNBODMR}. Agostini \textit{et al.} reported zero-field ODMR on bulk crystals at 1.25--4.2~K and noted room-temperature ODMR frequencies (919 and 1265~MHz) with linewidths of order a few MHz \cite{Agostini1991BiphenylTCNBXTrapODMR}. In contrast, the present work demonstrates continuous-wave, zero-field ODMR from \emph{thin MAS-grown microcrystals} at room temperature under 405~nm excitation with fluorescence collection above 463~nm. This establishes that (i) triplet-state ODMR contrast survives the MAS growth process and (ii) the ODMR readout is compatible with the micron-scale thin-crystal geometries produced by microspacing-in-air sublimation.

\textbf{An:TCNB.} Prior work on bulk An:TCNB crystals includes room-temperature high-field RYDMR measurements (Frankevich \textit{et al.}) \cite{Frankevich1978ATCNBRYDMRTripletPairs} as well as low-temperature zero-field delayed-fluorescence ODMR (Sch\"utz \textit{et al.}, 1.2~K) \cite{VonSchutz1980ATCNBDelayedFluorescenceODMR}, with complementary optical spectroscopy emphasizing triplet formation on the anthracene sublattice and TTA-mediated delayed fluorescence (Table~\ref{tab:SI_ODMR_comparison_ATCNB}). Our work extends these measurements to MAS-grown An:TCNB \emph{microcrystals} by demonstrating room-temperature, near-zero-field cw ODMR under 405~nm excitation with total fluorescence detection above 463~nm. Importantly, the MAS crystals are optically clear and thin, which is the geometry of interest for wide-field quantum-sensing implementations.

Overall, the combination of (i) MAS fabrication in a micro-confined glass-sandwich geometry, (ii) micron-scale thin-crystal form factors on transparent substrates, and (iii) room-temperature, near-zero-field cw ODMR constitutes the key novelty relative to the bulk-crystal literature summarized in Tables~\ref{tab:SI_ODMR_comparison} and~\ref{tab:SI_ODMR_comparison_ATCNB}. To our knowledge, this is the first demonstration of ODMR in MAS-fabricated crystals \emph{at room temperature}, and it provides a practical route to ODMR-active organic triplet layers without vacuum deposition, inert-gas crystal growth, or bulk single-crystal handling.
\section{Constrained DFT identifies optically accessible triplet states in An:TCNB and Bi:TCNB}\label{sec:SI_TDDFT}

\begin{figure}[!t]
    \centering
    \includegraphics[width=0.49\textwidth]{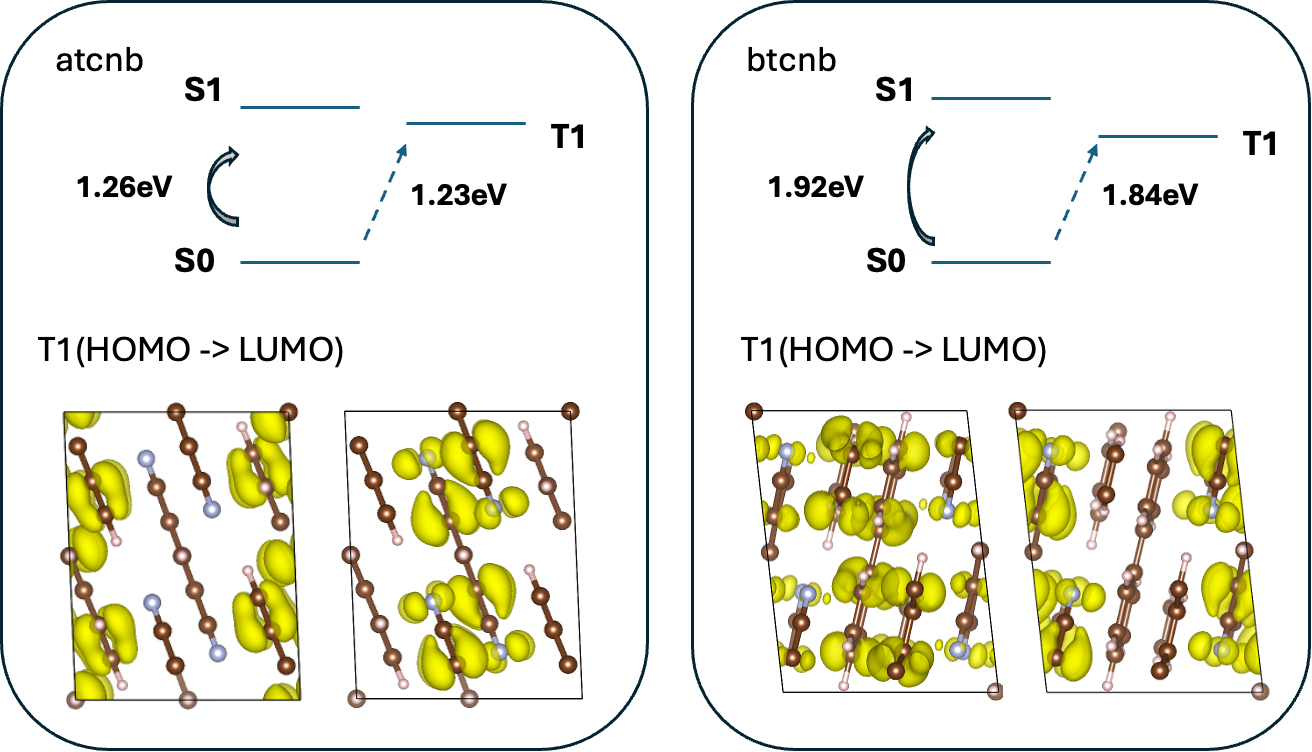}
    \caption{\justifying
    \textbf{Constrained DFT simulations of An:TCNB and Bi:TCNB electronic structure.}
    The $S_0$, $S_1$, and $T_1$ states were modeled using constrained DFT by imposing the corresponding electronic occupation configurations. \textbf{(a) An:TCNB.} The calculated $S_1$ and $T_1$ energies are approximately $1.26$ and $1.23$~eV, respectively. \textbf{(b) Bi:TCNB.} The corresponding energies are approximately $1.92$ and $1.84$~eV. In both systems, the computed $T_1$ state is described by a HOMO-to-LUMO excitation.}
    \label{fig:dft_atcnb_btcnb}
\end{figure}

The DFT computational setup was identical to that described in the main text. The $S_0$, $S_1$, and $T_1$ states were modeled using constrained DFT by imposing the following electronic occupation configurations. 
S0: HOMO = ↑↓, LUMO = empty. 
S1: HOMO = ↑,  LUMO = ↓    opposite spin (singlet) . 
T1: HOMO = ↑,  LUMO = ↑    parallel spin (triplet).
The calculations were performed to examine the energetic accessibility of triplet-state formation and exciton mobility in An:TCNB and Bi:TCNB MAS-triplet crystals.

For An:TCNB (Fig.~\ref{fig:dft_atcnb_btcnb}a), the calculated $S_1$ and $T_1$ excitation energies are approximately $1.26$ and $1.23$~eV, respectively. For Bi:TCNB (Fig.~\ref{fig:dft_atcnb_btcnb}b), the corresponding $S_1$ and $T_1$ excitation energies are approximately $1.92$ and $1.84$~eV, respectively. In both cases, the triplet state is energetically close to the singlet excited state, supporting the formation of optically accessible triplet states in these co-crystals. The computed $T_1$ state is described by a HOMO-to-LUMO excitation.

These results are in agreement with Attwood \textit{et al.}, who reported energetically accessible triplet states near the singlet charge-transfer state in anthracene:TCNB \cite{Attwood2025SpinRelaxationTripletMedia}. Quantitative differences are expected because the present work uses periodic PBE calculations, whereas Attwood \textit{et al.} used CAM-B3LYP/6-311+G(d) TD-DFT on finite dimers \cite{Attwood2025SpinRelaxationTripletMedia}. PBE is expected to underestimate charge-transfer excitation energies.

The higher apparent spin mobility in Bi:TCNB may arise from the more delocalized orbital character compared to An:TCNB (Fig.~\ref{fig:main_fig_4}), facilitating inter-molecular exciton hopping.

\section{Room-temperature ODMR is reproducible across MAS microcrystals and bulk controls}
\label{sec:SI_additional_ODMR}

\begin{figure*}[!t]
    \centering
    \includegraphics[width=0.95\textwidth]{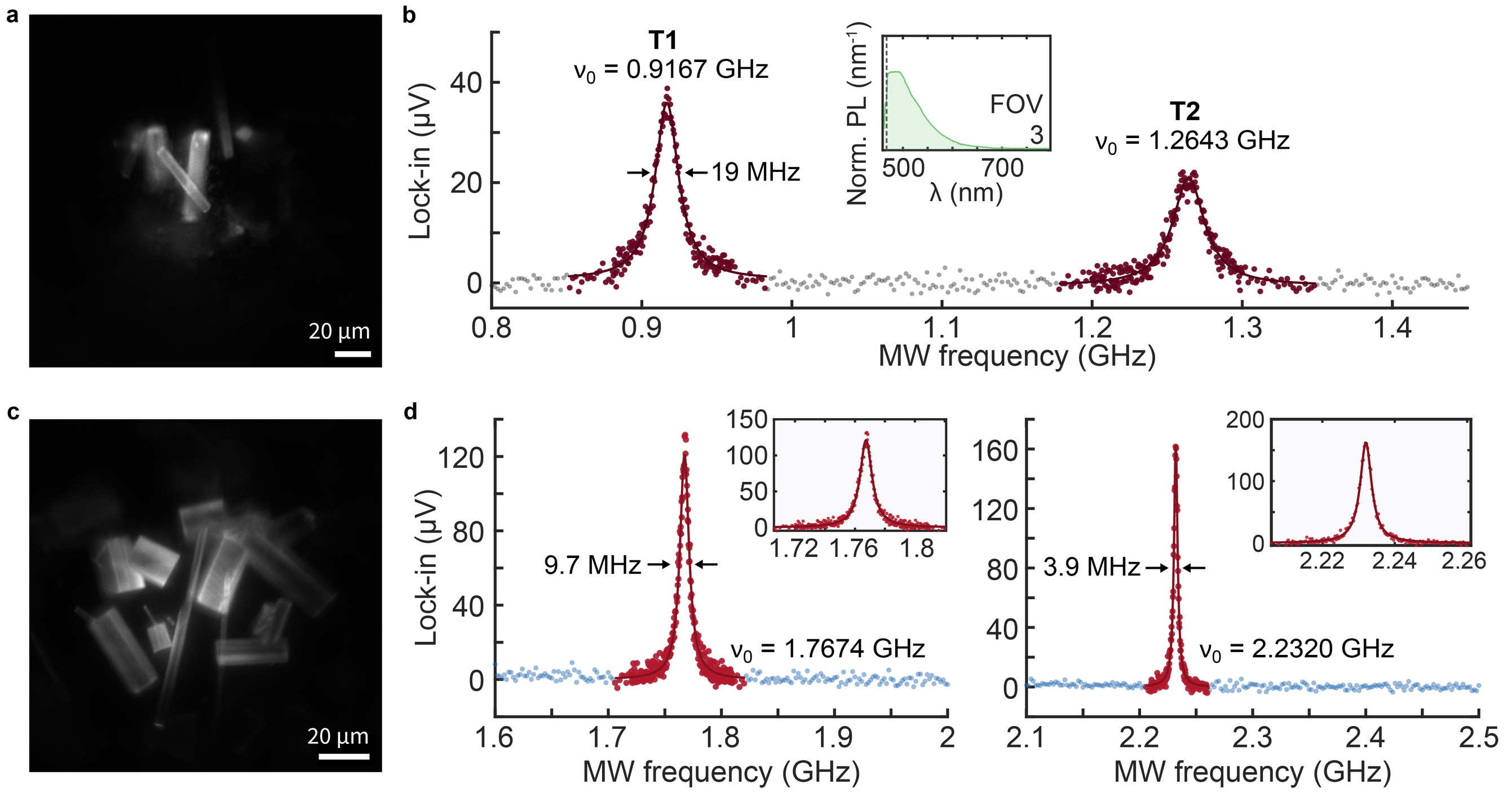}
    \caption{\justifying{\T{Additional MAS-grown Bi:TCNB and An:TCNB fields of view showing room-temperature ODMR.}
    \textbf{(a,b) MAS-grown Bi:TCNB microcrystals.}
    (a) Fluorescence image of MAS-grown Bi:TCNB microcrystals under 405~nm excitation. Scale bar, 20~\textmu m.
    (b) Room-temperature near-zero-field ODMR spectrum from the field of view in (a), showing two visible triplet-manifold transitions near $\nu_0 \sim 0.91$~GHz and $\nu_0 \sim 1.264$~GHz. The inset shows the corresponding normalized PL emission spectrum.
    \textbf{(c,d) MAS-grown An:TCNB microcrystals.}
    (c) Fluorescence image of MAS-grown An:TCNB microcrystals with clustered needle- and plate-like morphologies under 405~nm excitation. Scale bar, 20~\textmu m.
    (d) Room-temperature near-zero-field ODMR spectra from the field of view in (c), showing transitions at $\nu_0=1.7674$~GHz and $\nu_0=2.2320$~GHz. Arrow markers indicate example linewidths of 9.7~MHz and 3.9~MHz, respectively; insets show zoomed-in views of the corresponding ODMR resonances.}}
    \label{fig:supp_additional_btcnb_atcnb_fov}
\end{figure*}

\begin{figure}[!t]
    \centering
    \includegraphics[width=0.45\textwidth]{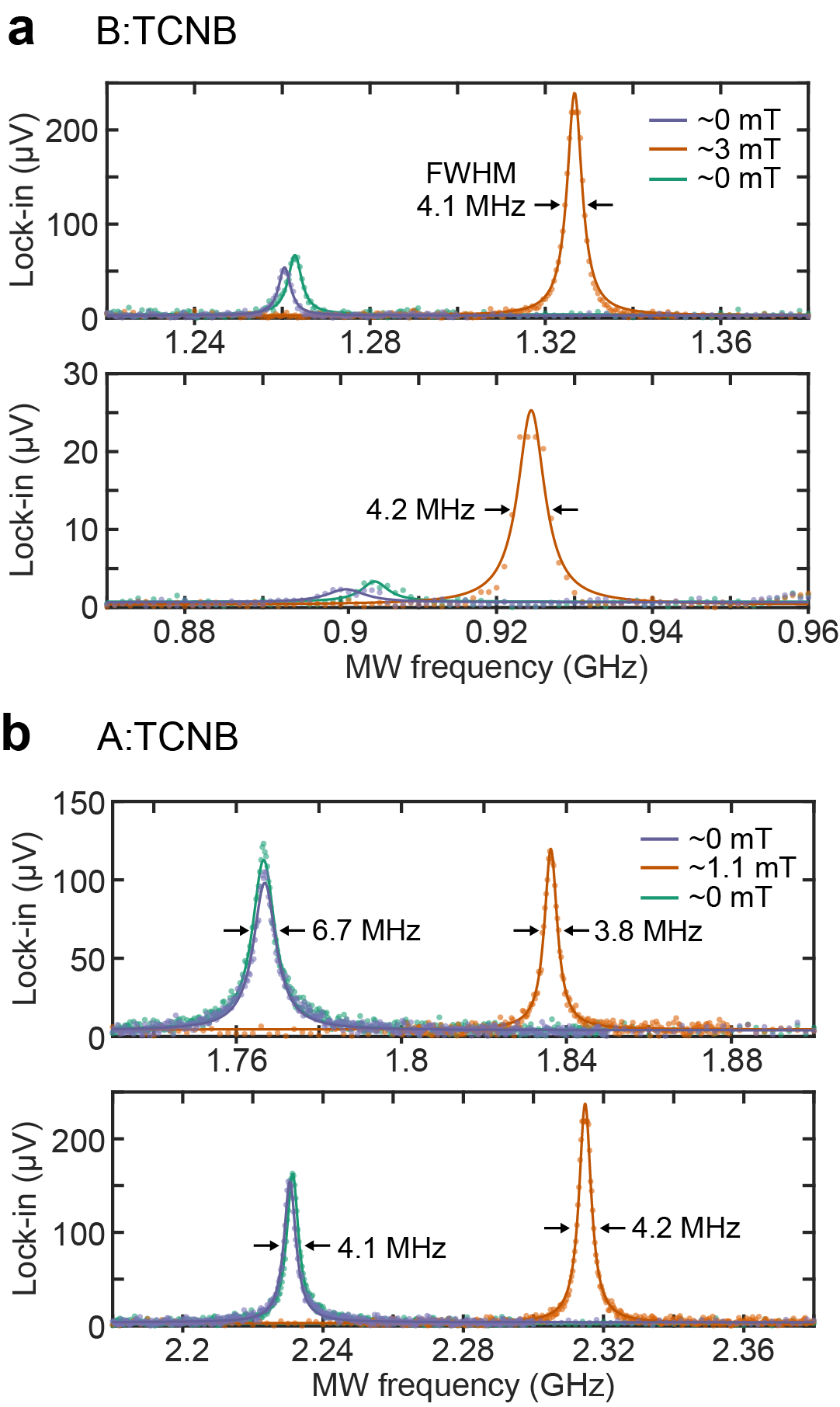}
    \caption{\justifying{\T{Room-temperature ODMR from solvent-grown bulk Bi:TCNB and An:TCNB crystals.}
    \T{\I{Bulk Bi:TCNB ODMR.}} (a) ODMR spectra from acetone-grown millimetre-scale Bi:TCNB bulk crystals under 405~nm excitation. The upper and lower panels show two observed triplet-manifold transitions in separate microwave-frequency windows.
    \T{\I{Bulk An:TCNB ODMR.}} (b) ODMR spectra from acetone-grown millimetre-scale An:TCNB bulk crystals, showing two observed triplet-manifold transitions near zero field.
    For both systems, a weak magnetic field was applied to perturb the ODMR resonances and subsequently removed, after which the resonances returned to their near-zero-field room-temperature positions. Solid lines are Lorentzian fits, and arrow markers indicate representative linewidths.}}
    \label{fig:supp_bulk_atcnb_btcnb_odmr}
\end{figure}

\begin{figure*}[!t]
    \centering
    \includegraphics[width=0.85\textwidth]{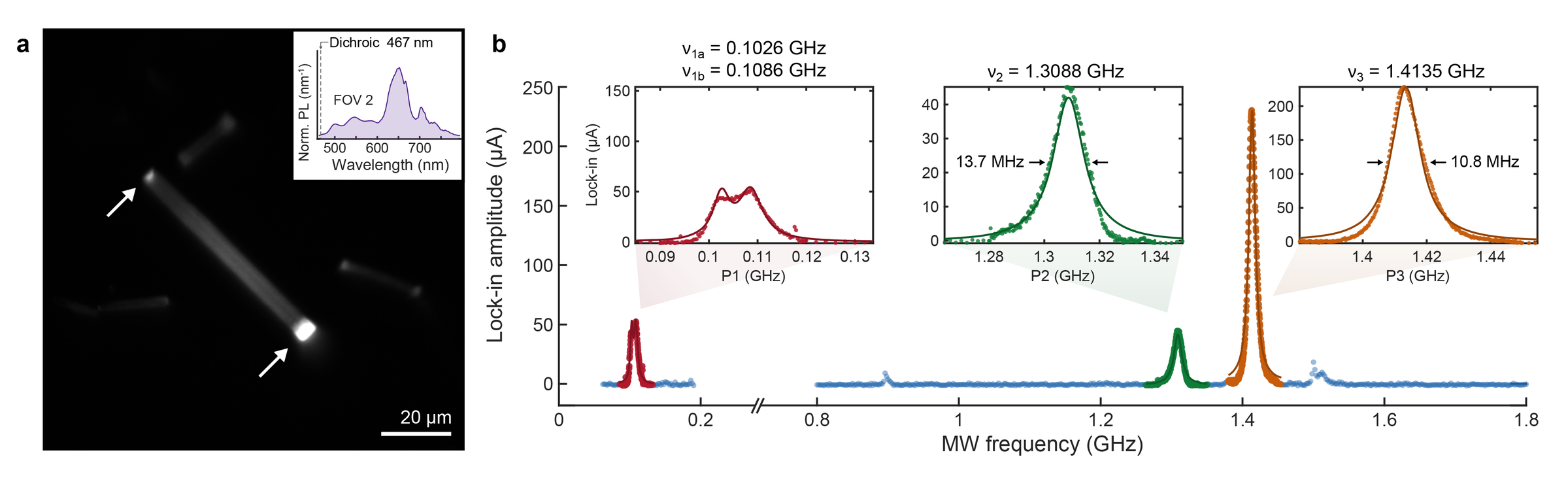}
    \caption{\justifying{\T{Additional example of MAS-grown P:PQ microneedle field of view showing room-temperature ODMR.}
    This supplementary field of view complements the P:PQ microneedle ODMR data shown in Fig.~\ref{fig:main_fig_5}.
    (a) Fluorescence image of MAS-grown pentacene:pentacenequinone (P:PQ) microneedles under 405~nm excitation. White arrows indicate waveguided emission visible through the ends of the MAS-grown microneedles. The inset shows the corresponding PL emission spectrum collected above the 463~nm dichroic cutoff. Scale bar, 20~\textmu m.
    (b) Room-temperature near-zero-field ODMR spectrum from the field of view in (a), showing a bimodal low-frequency transition at $\nu_{1a}=0.1026$~GHz and $\nu_{1b}=0.1086$~GHz, together with higher-frequency resonances at $\nu_2=1.3088$~GHz and $\nu_3=1.4135$~GHz. Insets show zoomed-in views of the ODMR features; arrow markers indicate example linewidths of 13.7~MHz and 10.8~MHz for the higher-frequency transitions.}}
    \label{fig:supp_ppq_additional_fov}
\end{figure*}

Reference Bi:TCNB and An:TCNB bulk crystals were grown by slow solvent evaporation and measured under room-temperature 405~nm excitation to benchmark the MAS-grown microcrystal response. The microwave drive was amplitude modulated, and the lock-in signal was detected from the total collected photoluminescence. Bi:TCNB bulk crystals showed near-zero-field ODMR resonances near $\sim$0.905 and $\sim$1.260~GHz (Fig.~\ref{fig:supp_bulk_atcnb_btcnb_odmr}a), whereas An:TCNB bulk crystals showed resonances near $\sim$1.77 and $\sim$2.23~GHz (Fig.~\ref{fig:supp_bulk_atcnb_btcnb_odmr}b). Weak applied-field perturbations were used only as additional verification of ODMR transitions and not as angular ODMR measurements.

Additional MAS-grown Bi:TCNB and An:TCNB microcrystal fields of view were measured on crystals grown on glass substrates. In addition to the FOVs shown in Fig.~\ref{fig:main_fig_4}, Fig.~\ref{fig:supp_additional_btcnb_atcnb_fov} shows regions containing multiple MAS-grown microcrystals within the same optical field of view. These measurements show that the characteristic PL emission and near-zero-field ODMR transitions are consistent across dense MAS-grown microcrystal regions on glass substrates as well as from the more isolated fields shown in the main text. For Bi:TCNB, the additional FOV shows two visible triplet-manifold transitions near $\sim$0.91~GHz and $\sim$1.264~GHz (Fig.~\ref{fig:supp_additional_btcnb_atcnb_fov}a,b). For An:TCNB, the additional FOV shows transitions near $\sim$1.767~GHz and $\sim$2.232~GHz, consistent with the main-text MAS-grown An:TCNB measurements (Fig.~\ref{fig:supp_additional_btcnb_atcnb_fov}c,d).

Together, the solvent-grown bulk-crystal ODMR spectra and the additional MAS-grown microcrystal FOVs connect the reference bulk response to the substrate-bound MAS geometry. The bulk spectra establish the room-temperature near-zero-field transition frequencies for Bi:TCNB and An:TCNB under the same optical readout configuration, while the additional MAS FOVs show that these characteristic ODMR transitions are retained in MAS-grown microcrystal regions on glass substrates, with small local differences in ODMR frequencies arising from temperature and microscopic strain.

Further, MAS-grown P:PQ microneedle field of view is shown in Fig.~\ref{fig:supp_ppq_additional_fov}a,b. This field of view complements the P:PQ data in Fig.~\ref{fig:main_fig_5} and shows waveguided emission from the microneedle ends, the corresponding PL emission spectrum, and reproducible room-temperature ODMR resonances. These supplementary measurements therefore connect the reference solvent-grown Bi:TCNB/An:TCNB bulk crystals, additional MAS-grown Bi:TCNB/An:TCNB microcrystal FOVs, and an additional MAS-grown P:PQ microneedle FOV under similar room-temperature ODMR experiment conditions.

\section{Delayed emission supports mobile triplet excitons at room temperature}\label{sec:SI_DelayEmmission}
An:TCNB and Bi:TCNB donor--acceptor stacks are known to support mobile triplet excitons at room temperature. Under sufficiently dense pulsed optical excitation, mobile triplet excitons can encounter one another and undergo triplet--triplet annihilation, producing delayed fluorescence after the prompt optical excitation pulse. To test whether our bulk An:TCNB and Bi:TCNB crystals show this room-temperature delayed-emission signature, we performed time-resolved photon counting following pulsed 405~nm excitation.

\begin{figure*}[!t]
    \centering
    \includegraphics[width=0.7\textwidth]{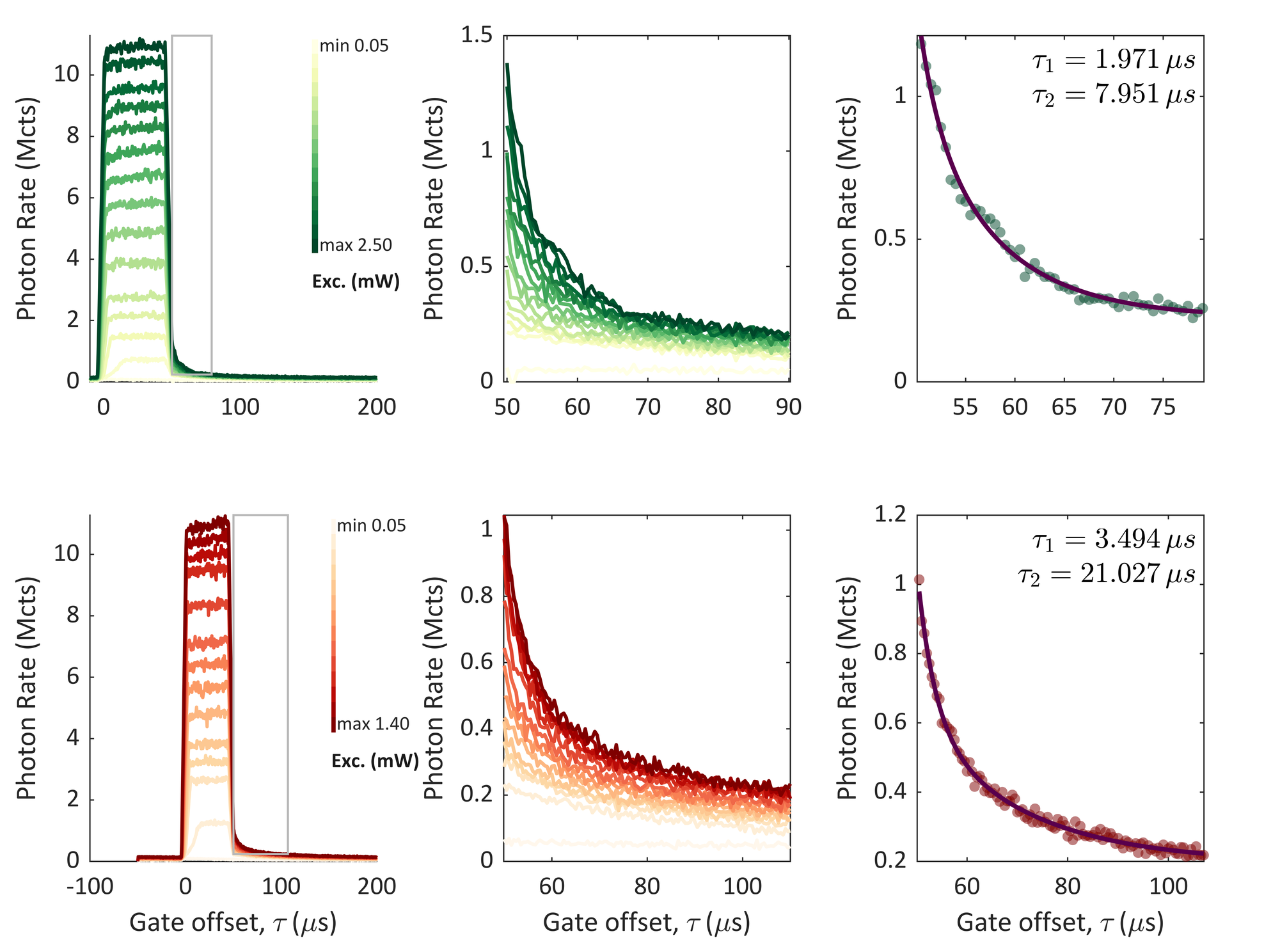}
    \caption{\justifying
    \textbf{Room-temperature delayed emission from bulk Bi:TCNB and An:TCNB organic crystals.}
    \textbf{(a--c) Bi:TCNB}, plotted in green to match its typical blue-green emission. \textbf{(d--f) An:TCNB}, plotted in orange. Bulk crystals were excited with pulsed 405~nm light over the excitation-power ranges indicated by the colour bars. \textbf{(a,d)} Full photon-counting traces showing the large prompt response and the weaker delayed-emission decay in the shaded gate region. \textbf{(b,e)} Expanded hardware-gated traces resolving the power-dependent delayed emission. \textbf{(c,f)} Representative biexponential fits, yielding $\tau_1=1.971~\mu$s and $\tau_2=7.951~\mu$s for Bi:TCNB, and $\tau_1=3.494~\mu$s and $\tau_2=21.027~\mu$s for An:TCNB, at the indicated excitation conditions.}
    \label{fig:supp_delayed_emission}
\end{figure*}

Bulk Bi:TCNB crystals were excited over the optical-power range shown in Fig.~\ref{fig:supp_delayed_emission}a, while bulk An:TCNB crystals were excited over the range shown in Fig.~\ref{fig:supp_delayed_emission}d. Emission from the bulk crystals was coupled into a multimode fiber ($200~\mu$m core) and detected using a fast single-photon detector. An OD3 neutral-density filter was used to avoid saturation of the avalanche photodiode during the prompt emission. The detector was hardware gated to enable microsecond-scale photon counting after the excitation pulse.

Both Bi:TCNB and An:TCNB show delayed emission persisting for tens of microseconds at room temperature. The pulsed 405~nm laser produces a large step-like response at the $\sim 10$~Mcts level, followed by a much weaker decay in the shaded gate region (Fig.~\ref{fig:supp_delayed_emission}a,d). This signal occurs after the excitation pulse rather than during the rising prompt response, consistent with delayed fluorescence from triplet-mediated intermolecular processes. The expanded traces resolve power-dependent microsecond tails for both crystals (Fig.~\ref{fig:supp_delayed_emission}b,e), while biexponential fits in Fig.~\ref{fig:supp_delayed_emission}c,f give $\tau_1=1.971~\mu$s and $\tau_2=7.951~\mu$s for Bi:TCNB, and $\tau_1=3.494~\mu$s and $\tau_2=21.027~\mu$s for An:TCNB. The longer decay constants in An:TCNB further distinguish its room-temperature triplet dynamics.

\section{Unencapsulated An:TCNB retains ODMR for over one hour}\label{sec:SI_odmr_stable}

To evaluate ODMR stability under repeated optical and microwave interrogation, we recorded sequential ODMR spectra from an An:TCNB MAS-grown microcrystal field of view on glass substrate. The measured field of view contained multiple directly deposited An:TCNB microcrystals over an area of approximately $50~\mu$m. The crystals were not encapsulated and were measured without any encapsulation after direct transfer of MAS-grown glass slides from the fume-hood MAS-growth setup to the ODMR microscope.

The An:TCNB field of view was excited with $\sim 3~\mathrm{mW}$ of 405~nm light. Individual ODMR single-shot spectra were acquired, with an approximately $50~\mathrm{s}$ waiting period between spectra to allow heat dissipation. The elapsed time for each spectrum was recorded using the PC clock, allowing the ODMR response to be tracked over more than one hour. For analysis, each spectrum was baseline-corrected to zero, and the peak response of the initially selected ODMR spectrum was used to normalize the full time series.

Figure~\ref{fig:odmr_stable}a shows the ODMR amplitude as a function of microwave frequency and measurement time, with selected time points marked by white circles. The resonance remains visible throughout the measurement, but a gradual decrease in peak ODMR amplitude is observed. Selected profiles and Lorentzian fits from the marked time points (Fig.~\ref{fig:odmr_stable}b) illustrate this progressive reduction under repeated laser and microwave exposure. These data indicate that MAS-grown An:TCNB microcrystals remain ODMR-active over extended measurements, although the ODMR contrast decreases under the stated non-encapsulated measurement conditions. Growth of MAS crystals on more thermally conductive substrates, such as sapphire, could enable prolonged high-SNR ODMR measurements. Notably, P:PQ microneedles grown on glass under similar laser and microwave conditions showed no visible ODMR-amplitude decay over several days, indicating that photostability varies across co-crystal systems.

\section{Coherent spin control is reproducible across independent P:PQ microneedles}\label{sec:SI_PPQ_additional_coherence}

\begin{figure*}[t]
    \begin{minipage}[t]{0.45\textwidth}
        \strut\vspace{-\baselineskip}
        \centering
        \includegraphics[width=\linewidth]{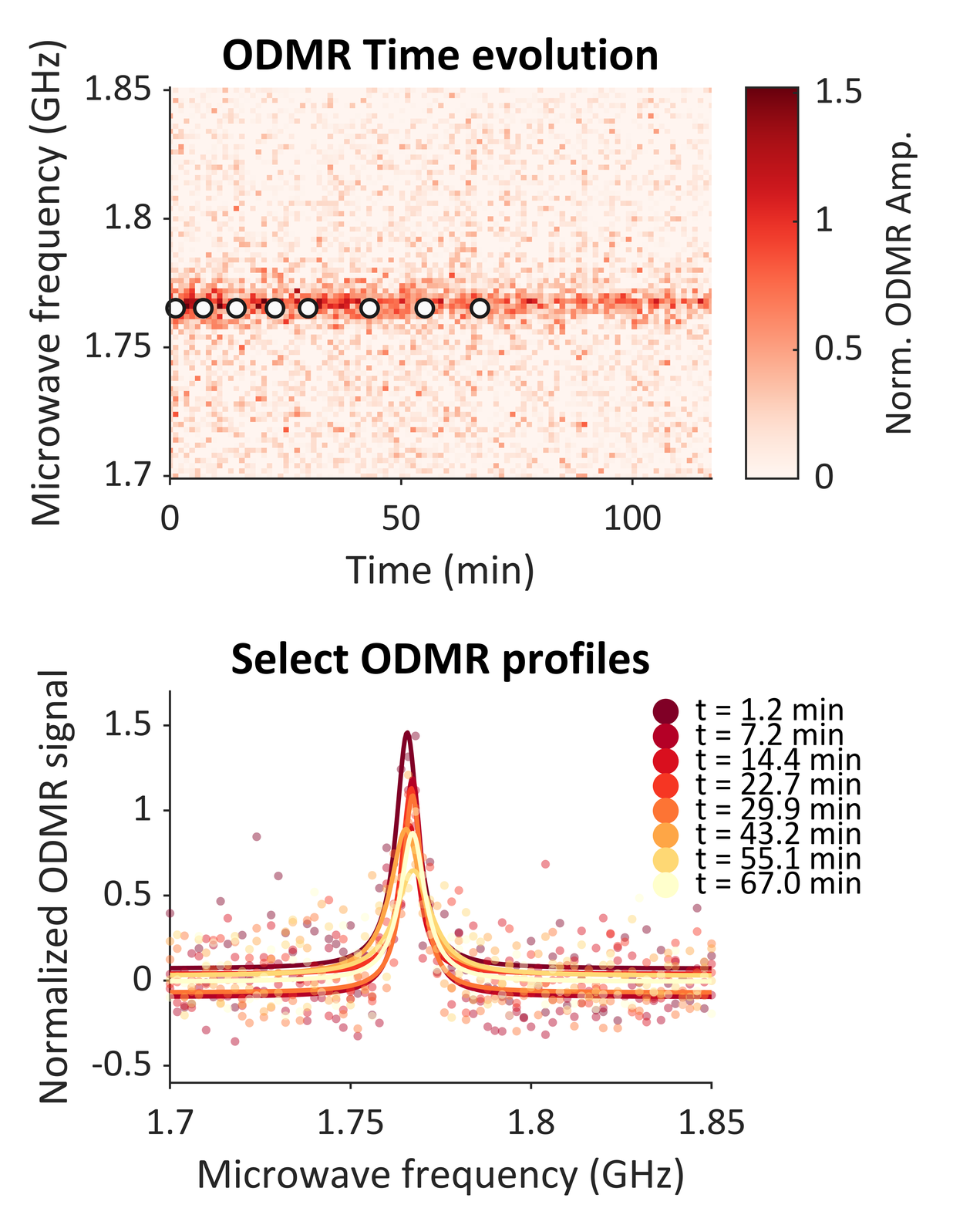}
        \captionof{figure}{\justifying
        \textbf{Long-term ODMR stability of MAS-grown An:TCNB microcrystals.}
        \textbf{(a)} Normalized ODMR amplitude as a function of microwave frequency and measurement time, with selected time points marked by white circles, showing that MAS-grown An:TCNB remains ODMR-active for at least $\sim 60$~min on a thermally low-conductivity glass substrate. \textbf{(b)} Selected ODMR profiles and Lorentzian fits showing the gradual decrease in ODMR peak amplitude under repeated 405~nm excitation and microwave interrogation.}
        \label{fig:odmr_stable}
    \end{minipage}
    \hfill
    \begin{minipage}[t]{0.48\textwidth}
        \centering
        \captionof{table}{Comparison of zero-field splitting parameters: MAS-grown P:PQ (Fig.~\ref{fig:main_fig_5}) vs.\ bulk PDP reference \cite{Singh2025OrganicCrystalQuantumSensing}}
        \label{tab:SI_PPQ_ZFS_comparison_table}
        \small
        \renewcommand{\arraystretch}{1.5}
        \begingroup
        \arrayrulecolor{SITableRule}
        \setlength{\arrayrulewidth}{0.45pt}
        \begin{tabular}{|l|>{\columncolor{SITableHighlight}}l|l|}
        \hline
        \rowcolor{SITableHeader}
        \multicolumn{1}{|c|}{\textbf{Property}} &
        \multicolumn{1}{c|}{\textbf{P:PQ (Fig.~\ref{fig:main_fig_5})}} &
        \multicolumn{1}{c|}{\textbf{PDP} \cite{Singh2025OrganicCrystalQuantumSensing}} \\
        \hline
        $2|E|$ transition & 0.103/0.109~GHz & 0.108~GHz \\
        \hline
        $D-|E|$ transition & 1.308~GHz & 1.340~GHz \\
        \hline
        $D+|E|$ transition & 1.413~GHz & 1.448~GHz \\
        \hline
        $D$ (derived) & $\approx$1361~MHz & $\approx$1392~MHz \\
        \hline
        $|E|$ (derived) & $\approx$53~MHz & $\approx$53~MHz \\
        \hline
        \end{tabular}
        \endgroup

        \vspace{1.5em}

        \captionof{table}{Comparison of coherence lifetimes: MAS-grown P:PQ (Fig.~\ref{fig:main_fig_6}) vs.\ bulk PDP reference (Fig.~\ref{fig:supp_pdp_reference})}
        \label{tab:SI_PQ_PDP_comparison_table}
        \small
        \renewcommand{\arraystretch}{1.5}
        \begingroup
        \arrayrulecolor{SITableRule}
        \setlength{\arrayrulewidth}{0.45pt}
        \begin{tabular}{|l|>{\columncolor{SITableHighlight}}l|l|}
        \hline
        \rowcolor{SITableHeader}
        \multicolumn{1}{|c|}{\textbf{Property}} &
        \multicolumn{1}{c|}{\textbf{P:PQ (Fig.~\ref{fig:main_fig_6})}} &
        \multicolumn{1}{c|}{\shortstack{\textbf{PDP}\\(Fig.~\ref{fig:supp_pdp_reference})}} \\
        \hline
        ODMR transition driven & 1.306~GHz & 1.34~GHz \\
        \hline
        ODMR linewidth (1.3~GHz) & 17.0~MHz & 8.8~MHz \\
        \hline
        Ramsey $T_2^{*}$ & 123~ns & 246~ns \\
        \hline
        Ramsey detuning & 17~MHz & 15.6~MHz \\
        \hline
        Hahn-echo $T_2$ & 0.56~\textmu s & 1.4~\textmu s \\
        \hline
        CPMG $T_2$ (4-$\pi$-pulse) & 1.35~\textmu s & n/a \\
        \hline
        Ground-state recovery & 82.8~\textmu s & 58.6~\textmu s \\
        \hline
        Stretched-exponential $\beta$ & 1.01 & 1.15 \\
        \hline
        \end{tabular}
        \endgroup
    \end{minipage}
\end{figure*}

\begin{figure*}[!t]
    \centering
    \includegraphics[width=0.85\textwidth]{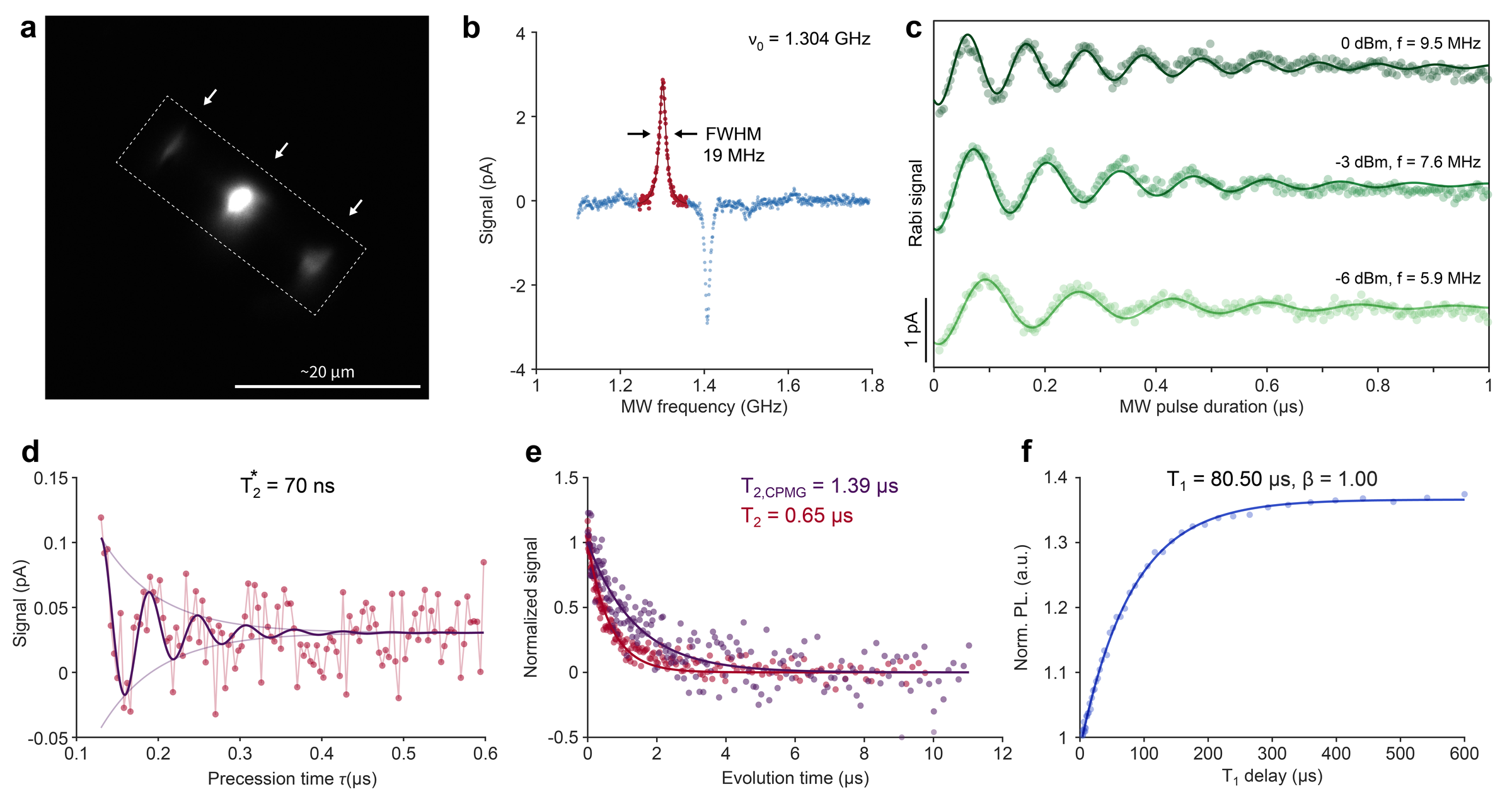}
    \caption{\justifying{\T{Additional example field of view of room-temperature optically detected coherent control of photoexcited triplet spins in an MAS-grown P:PQ microneedle.}
    (a) \T{\I{Fluorescence field of view.}} Fluorescence image of an additional P:PQ microneedle field of view under 405~nm excitation; arrows mark needles identified for measurement. Scale bar, 20~\textmu m.
    (b) \T{\I{Selected transition.}} ODMR spectrum showing the transition selected for coherent driving, $\nu_0=1.3039$~GHz (FWHM = 19.0~MHz), fit with a Lorentzian line shape (red).
    (c) \T{\I{Rabi oscillations.}} Coherent Rabi driving of the emissive spin ensemble at three microwave powers (before a 30~W amplifier): $+0$~dBm (9.48~MHz), $-3$~dBm (7.57~MHz) and $-6$~dBm (5.93~MHz).
    (d) \T{\I{Ramsey fringes.}} Ramsey fringes give $T_2^{*}=70$~ns for this microneedle.
    (e) \T{\I{Hahn-echo and CPMG.}} Hahn-echo gives $T_2=0.65~\mu$s; CPMG noise rejection extends this to $T_{2,\mathrm{CPMG}}=1.389~\mu$s.
    (f) \T{\I{Ground-state recovery lifetime.}} Ground-state recovery lifetime is $80.50~\mu$s (stretched-exponential $\beta=1.00$).}}
    \label{fig:supp_ppq_fov2_full_coherence}
    \vspace{0.8em}
    \includegraphics[width=0.85\textwidth]{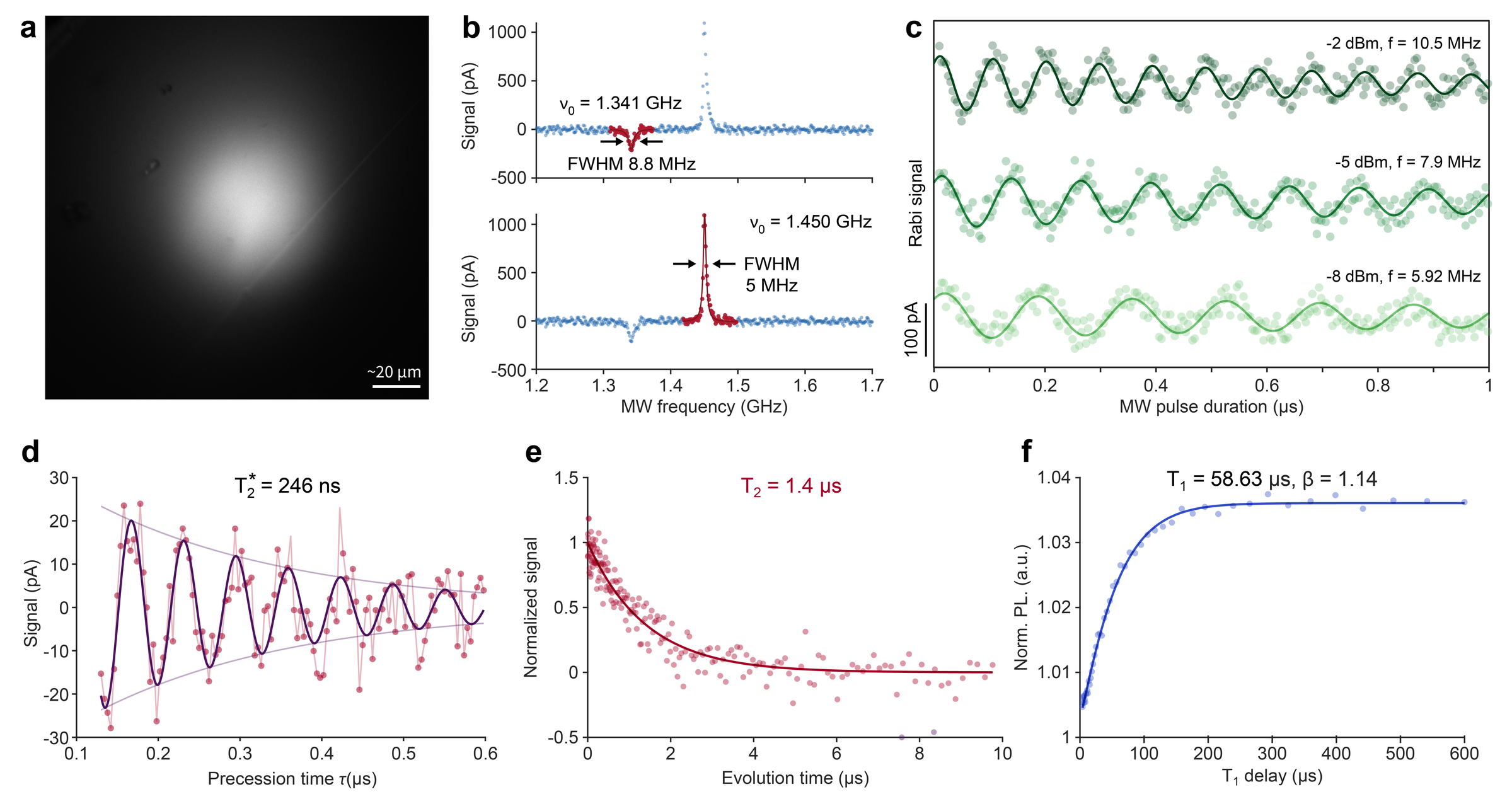}
    \caption{\justifying{\T{Room-temperature optically detected coherent control of photoexcited triplet spins in bulk pentacene-doped para-terphenyl (PTP), a reference sample for the MAS-grown P:PQ measurements in Fig.~\ref{fig:main_fig_6}.}
    (a) \T{\I{Fluorescence field of view.}} Sample image of the PTP reference crystal under 532~nm excitation. Scale bar, 20~\textmu m.
    (b) \T{\I{ODMR spectra.}} Continuous-wave ODMR spectra showing two resonances: the 1.34~GHz transition (FWHM = 8.83~MHz), which is the transition driven in the coherent-control measurements shown in (c)--(f), and a second transition at 1.45~GHz (FWHM = 5.02~MHz), each fit with a Lorentzian line shape (red).
    (c) \T{\I{Rabi oscillations.}} Coherent Rabi driving of the photoexcited spin ensemble at three microwave powers: $-2$~dBm ($f=10.5$~MHz), $-5$~dBm ($f=8.0$~MHz) and $-8$~dBm ($f=5.9$~MHz).
    (d) \T{\I{Ramsey fringes.}} Ramsey fringes give $T_2^{*}=246$~ns for this PTP reference, measured at a detuning of 15.6~MHz from resonance.
    (e) \T{\I{Hahn-echo decay.}} Hahn-echo sequence gives $T_2=1.400~\mu$s.
    (f) \T{\I{Ground-state recovery lifetime.}} Ground-state recovery lifetime $T_1=58.629~\mu$s (stretched-exponential $\beta=1.143$), as contrast recovers with increasing delay between initialization and readout pulses.}}
    \label{fig:supp_pdp_reference}
\end{figure*}

\begin{figure*}[!t]
    \begin{minipage}[t]{0.48\textwidth}
        \centering
       \includegraphics[width=\textwidth]{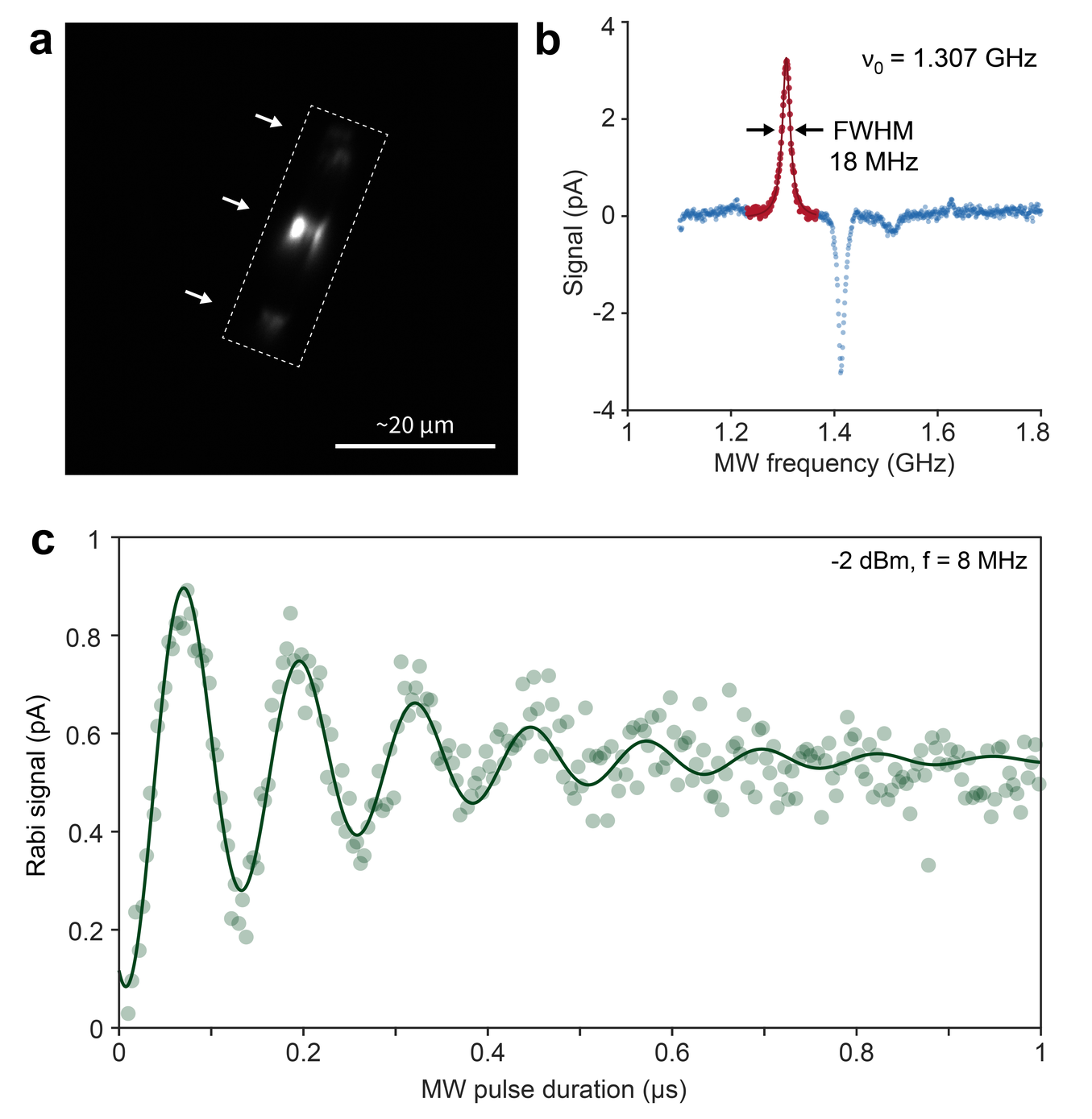}
        \caption{\justifying{\T{Additional example of coherent Rabi driving in an MAS-grown P:PQ microneedle, from two microneedles grown in close proximity.}
        (a) \T{\I{Sample image.}} Fluorescence field of view of two closely grown P:PQ microneedles under 405~nm excitation; arrows mark the excitation spot and waveguided emission, collected jointly from the combined spin ensemble. Scale bar, 20~\textmu m.
        (b) \T{\I{Selected transition.}} ODMR spectrum showing the transition selected for Rabi driving, $\nu_0=1.3070$~GHz (FWHM = 18.0~MHz), fit with a Lorentzian line shape (red).
        (c) \T{\I{Rabi oscillations.}} Rabi oscillations of the emissive spin ensemble at $-2$~dBm ($f=7.97$~MHz).}}
        \label{fig:supp_ppq_fov_rabi_only}
    \end{minipage}
    \hfill
    \begin{minipage}[t]{0.48\textwidth}
        \centering
       \includegraphics[width=0.6\textwidth]{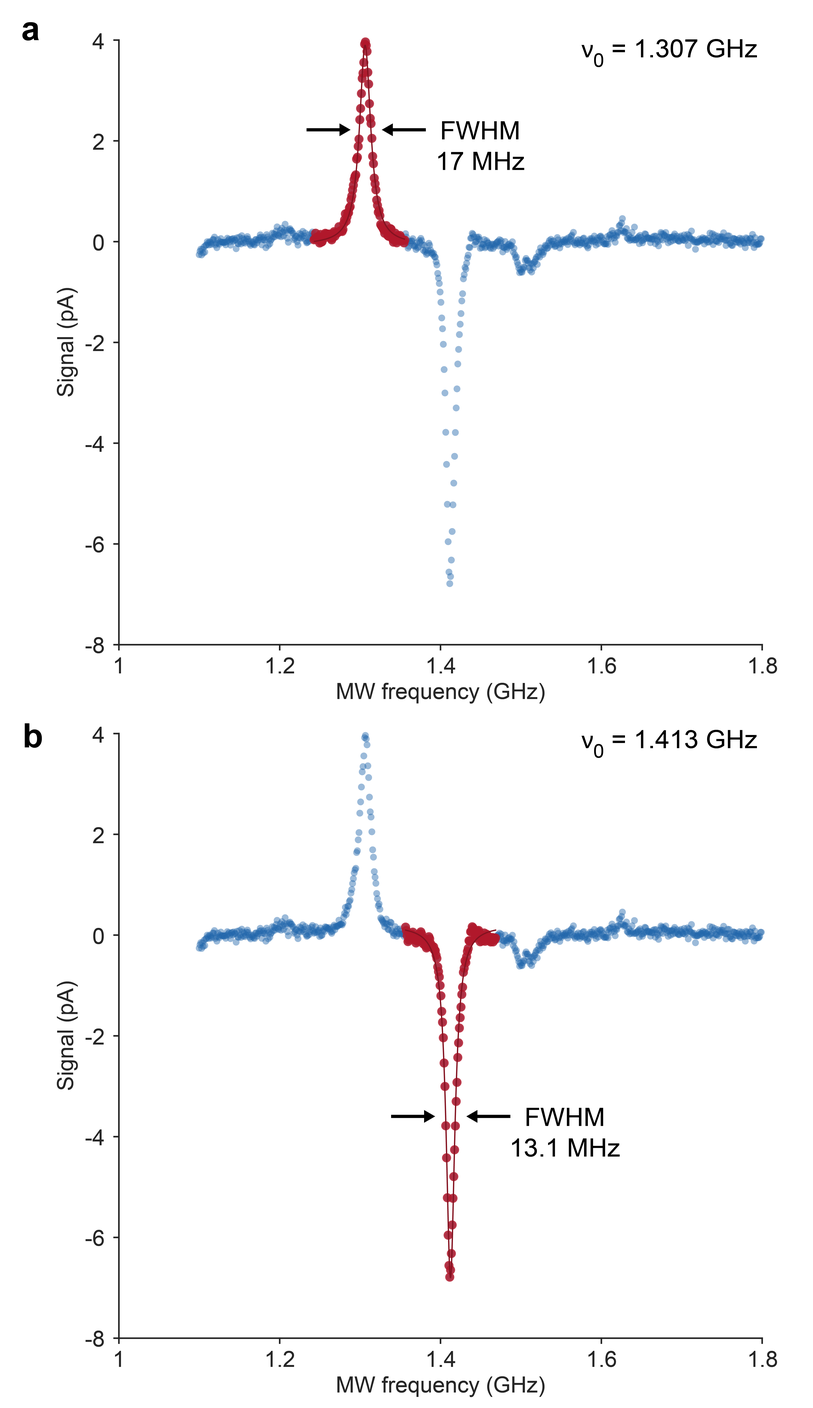}
        \caption{\justifying{\T{ODMR spectra of the same MAS-grown P:PQ microneedle characterized in Figure 6; the resonance in (a) corresponds to the transition driven in the Figure 6 coherence measurements.}
        (a) \T{\I{Lower-frequency resonance.}} ODMR spectrum with the lower-frequency transition highlighted and fit with a Lorentzian line shape (red), $\nu_0=1.307$~GHz (FWHM = 17.0~MHz).
        (b) \T{\I{Higher-frequency resonance.}} Same ODMR spectrum, with the higher-frequency transition highlighted and fit with a Lorentzian line shape (red), $\nu_0=1.413$~GHz (FWHM = 13.1~MHz).}}
        \label{fig:supple_ppq_odmr_figure6}
    \end{minipage}
\end{figure*}

To further support the reproducibility of the P:PQ coherence measurements shown in Figure 6, we characterized an additional MAS-grown P:PQ microneedle under the same conditions. The fluorescence field of view and selected 1.3039~GHz transition are shown in Fig.~\ref{fig:supp_ppq_fov2_full_coherence}a,b. Power-dependent Rabi oscillations establish coherent driving (Fig.~\ref{fig:supp_ppq_fov2_full_coherence}c), while Ramsey fringes give $T_2^{*}=70$~ns (Fig.~\ref{fig:supp_ppq_fov2_full_coherence}d), modestly reduced relative to the Figure 6 crystal ($T_2^{*}=123$~ns). This variation is consistent with Ramsey measurements being particularly sensitive to static disorder and strain between individual MAS-grown crystals. Hahn echo and CPMG give $T_2=0.65~\mu$s and $T_{2,\mathrm{CPMG}}=1.389~\mu$s (Fig.~\ref{fig:supp_ppq_fov2_full_coherence}e), similar to $0.562~\mu$s and $1.346~\mu$s for Figure 6. The ground-state recovery is likewise comparable, at $80.50~\mu$s (Fig.~\ref{fig:supp_ppq_fov2_full_coherence}f) versus $82.8~\mu$s. For reference, Lorentzian fits to the lower- and higher-frequency ODMR resonances of the Figure 6 microneedle are shown in Fig.~\ref{fig:supple_ppq_odmr_figure6}a,b.

In the same ODMR and spin-coherence setup, we additionally measured a bulk pentacene-doped para-terphenyl (PDP) reference crystal under 532~nm excitation, using the same microstrip resonator and microwave switching and phase-control circuitry, following prior established measurements in this system \cite{Singh2025OrganicCrystalQuantumSensing,Mena2024MolecularSpinCoherentControl}. The reference field of view and selected 1.34~GHz transition are shown in Fig.~\ref{fig:supp_pdp_reference}a,b. Power-dependent Rabi oscillations again establish coherent driving (Fig.~\ref{fig:supp_pdp_reference}c). Ramsey fringes give $T_2^{*}=246$~ns (Fig.~\ref{fig:supp_pdp_reference}d), and Hahn echo gives $T_2=1.400~\mu$s (Fig.~\ref{fig:supp_pdp_reference}e). The ground-state recovery is $58.629~\mu$s (Fig.~\ref{fig:supp_pdp_reference}f). The measured $T_2^{*}$ is of the same order as, though somewhat reduced relative to, prior reference measurements, which we attribute to strain in the single-crystal fragment broken from a larger PDP crystal. A full comparison with the Figure 6 P:PQ crystal is provided in Table~\ref{tab:SI_PQ_PDP_comparison_table}.

To further test the reproducibility of MAS-grown P:PQ microneedles, we performed Rabi driving on two microneedles deposited in close proximity. Figure~\ref{fig:supp_ppq_fov_rabi_only}a shows the shared excitation spot and waveguided emission, while Fig.~\ref{fig:supp_ppq_fov_rabi_only}b identifies the 1.3070~GHz transition selected for driving. Rabi nutations at $-2$~dBm ($f=7.97$~MHz) were observed from the joint emission of the two spin ensembles (Fig.~\ref{fig:supp_ppq_fov_rabi_only}c).

Together, these additional measurements corroborate the reproducibility of P:PQ spin coherence at room temperature and provide a referential comparison to pentacene doped in para-terphenyl.

\FloatBarrier
\section{Structural resolution of novel yellow microneedle phase in P:PQ MAS}\label{sec:SI_PPQ_structure}
\begin{figure}[!t]
    \centering
    \includegraphics[width=0.45\textwidth]{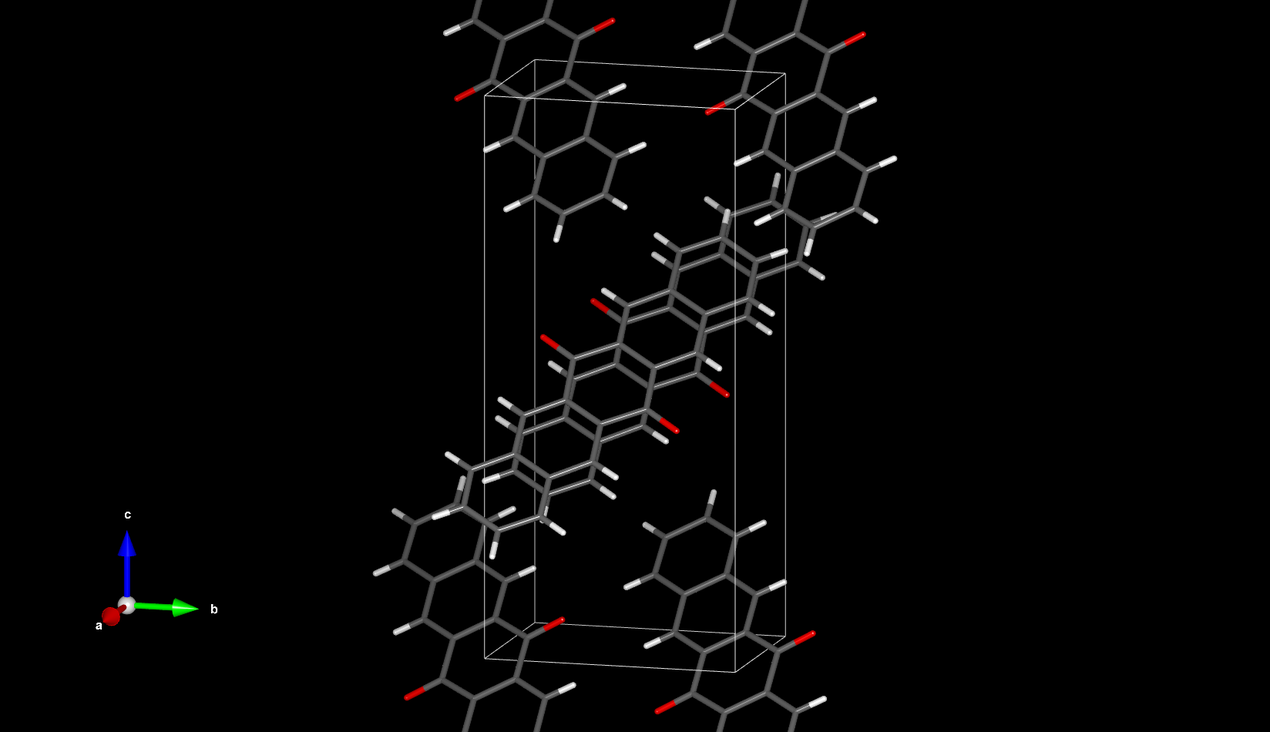}
    \caption{\justifying{\T{Structural resolution of the P:PQ microneedle phase.} Single-crystal X-ray diffraction resolves the MAS-grown yellow microneedle phase as monoclinic $P2_1/c$ pentacenequinone (PQ), $a=4.9024(4)$~\AA, $b=8.1293(7)$~\AA, $c=17.5622(15)$~\AA, $\beta=94.115(7)^\circ$, $Z=2$. Crystal structure rendered in VESTA as a stick visualization with the unit cell outlined and the $a$, $b$, $c$ axes indicated in the image corner. O atoms are red, C atoms grey, H atoms white.}}
    \label{fig:SI_PQ_structure}
\end{figure}

MAS-grown pentacene:pentacenequinone (P:PQ) crystals for single-crystal X-ray diffraction study were prepared under growth conditions similar to the ODMR-active microneedles shown in Figs.~\ref{fig:main_fig_5} and~\ref{fig:main_fig_6}. Isolated yellow microneedles, selected to avoid the blue pure-pentacene crystals also present in the heterogeneous MAS growth, were visually identified on the MAS-grown slide under the microscope and transferred to the X-ray crystallography facility for structure determination.
A single yellow needle (0.06 $\times$ 0.03 $\times$ 0.02~mm) was mounted in a cryoloop with Paratone oil, and data were collected at 100~K. The data were integrated using the CrysAlisPro 1.172.43.143a software program and scaled using the SCALE3 ABSPACK scaling algorithm. The structure was solved by intrinsic phasing (SHELXT) and refined anisotropically by full-matrix least-squares (SHELXL).
The refined structure is monoclinic $P2_1/c$ (No.~14), with $a=4.9024(4)$~\AA, $b=8.1293(7)$~\AA, $c=17.5622(15)$~\AA, $\beta=94.115(7)^\circ$ and $Z=2$, consistent with the unit cell rendered in Fig.~\ref{fig:SI_PQ_structure}. The asymmetric unit comprises half of the pentacenequinone molecule (C$_{22}$H$_{12}$O$_2$); the remaining half, generated by the crystallographic inversion centre coincident with the molecular centre, is completed in the VESTA rendering shown in Fig.~\ref{fig:SI_PQ_structure}. Single-crystal XRD resolves this example yellow, novel microneedle phase formed during the heterogeneous MAS growth of pentacene and pentacenequinone as a single-crystal lattice of pentacenequinone, consistent with the pentacene-doped-lattice picture discussed in the main text. Prior to this MAS-grown single crystal, pentacenequinone had been reported via other growth methods, including phase-separated domains within vacuum-codeposited pentacene/pentacenequinone thin films \cite{Salzmann2007PhaseSeparationPentacenePQ} and a surface-induced polymorph grown as a vacuum-deposited thin film and resolved by combined X-ray diffraction reciprocal-space mapping and theoretical structure modeling \cite{Salzmann2011StructureSolutionPentacenequinone}.

\FloatBarrier
\section{Sub-microtesla field sensitivity in MAS triplet-spin layers}\label{sec:SI_field_sensitivity}
We estimated the preliminary static (DC) magnetic-field sensitivity of representative MAS-triplet crystal fields of view from the single-shot ODMR spectra shown in Figs.~\ref{fig:main_fig_4} and~\ref{fig:main_fig_5}. The analysis was performed for one representative field of view from Bi:TCNB, An:TCNB, and P:PQ. These estimates assume a locally linear magnetic-field response of the ODMR resonance near the steepest point of the fitted lineshape. A more specific analysis would first identify the regime of linear resonance-frequency dependence on the applied magnetic field---itself set by the $D/E$ values (crystal-specific) and each crystal's field-of-view-specific orientation (roughly, whether the chromophore aromatic rings lie parallel or perpendicular to the growth substrate)---and then solve for the spin-Hamiltonian eigenfrequencies as a function of that field.

\begin{figure}[!t]
    \centering
    \includegraphics[width=0.40\textwidth]{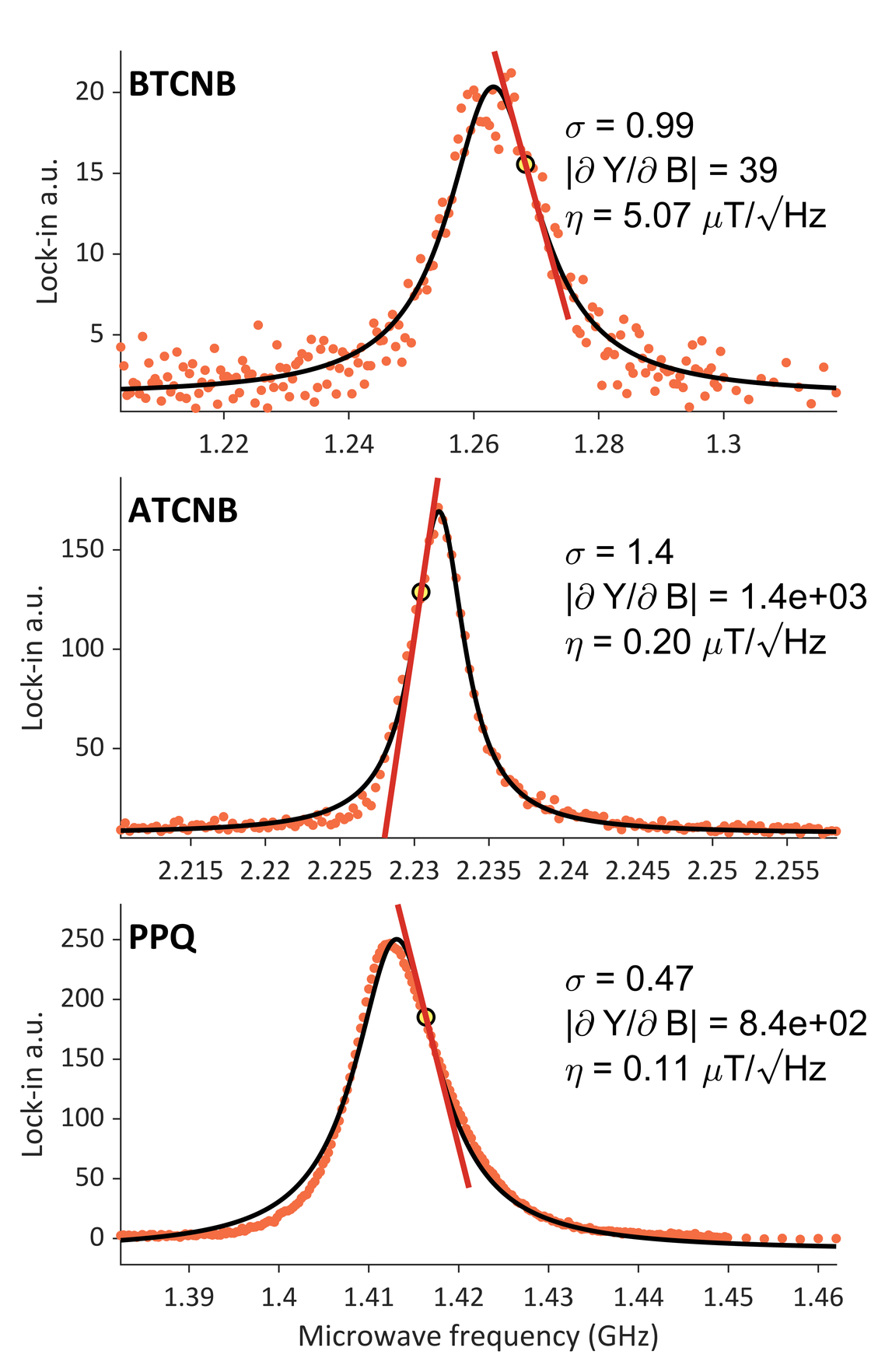}
    \caption{\justifying
    \textbf{Preliminary DC magnetic-field sensitivity estimates from representative MAS-triplet ODMR spectra.}
    Single-shot ODMR spectra from representative \textbf{(a)} Bi:TCNB, \textbf{(b)} An:TCNB, and \textbf{(c)} P:PQ fields of view were fit with Lorentzian profiles, and the steepest fitted slope was used with the off-resonance lock-in noise to estimate the magnetic-field sensitivity $\eta$. The extracted preliminary sensitivities are $5.07~\mu\mathrm{T}/\sqrt{\mathrm{Hz}}$ for Bi:TCNB, $0.20~\mu\mathrm{T}/\sqrt{\mathrm{Hz}}$ for An:TCNB, and $0.11~\mu\mathrm{T}/\sqrt{\mathrm{Hz}}$ for P:PQ.}
    \label{fig:fig_field_sensitivity}
\end{figure}

For each spectrum, the ODMR resonance was fit with a Lorentzian profile. A dense fitted curve was then generated, and the maximum absolute slope of the ODMR response with respect to microwave frequency, $\left|dY/df\right|_{\max}$, was extracted. The frequency-to-field conversion used the electron gyromagnetic ratio, $df/dB=28~\mathrm{MHz/mT}=0.028~\mathrm{GHz/mT}$, so the maximum magnetic-field slope was calculated as $\left|dY/dB\right|_{\max}=\left|dY/df\right|_{\max}(df/dB)$.
The noise level, $\sigma$, was calculated from an off-resonance baseline region of the ODMR spectrum, with optical excitation on but away from the resonance. Because the lock-in amplifier internally averages according to the low-pass filter response, the effective integration time was taken as four times the low-pass filter time constant. For these measurements, the low-pass filter was $10~\mathrm{ms}$. The preliminary magnetic-field sensitivity was then estimated from the noise, effective integration time, and maximum magnetic-field slope as $\eta=\sigma\sqrt{\tau}/\left|dY/dB\right|_{\max}$.

Figure~\ref{fig:fig_field_sensitivity}a shows the Bi:TCNB trace and fit, giving the least favorable preliminary sensitivity, $5.07~\mu\mathrm{T}/\sqrt{\mathrm{Hz}}$, consistent with its broader linewidth and smaller slope. An:TCNB gives $0.20~\mu\mathrm{T}/\sqrt{\mathrm{Hz}}$ (Fig.~\ref{fig:fig_field_sensitivity}b), reflecting its narrower linewidth and larger extracted slope. P:PQ gives the best preliminary value, $0.11~\mu\mathrm{T}/\sqrt{\mathrm{Hz}}$ (Fig.~\ref{fig:fig_field_sensitivity}c), despite linewidths in the tens-of-megahertz range, because the ODMR signal response is large. Further, because the crossover field---above which the resonance frequency depends linearly on $B$ rather than quadratically near $B=0$---scales with the zero-field splitting parameter $E$ ($\lvert E\rvert\approx457$, $232$, and $53$~MHz for Bi:TCNB, An:TCNB, and P:PQ, respectively; Figs.~\ref{fig:main_fig_4} and~\ref{fig:main_fig_5}), the large-$E$ Bi:TCNB and An:TCNB estimates likely overestimate the true DC sensitivity, whereas the small-$E$ P:PQ value should lie closer to the valid regime. This behavior is highly cocrystal-specific, but MAS's ability to readily grow a broad class of such cocrystals offers access to a chemically tunable, variable-sensing-regime class of materials. These are preliminary single-shot field-of-view estimates for these thin triplet-spin layers, at low optical excitation of $500~\mu\mathrm{W}$ for Bi:TCNB and P:PQ, and $3~\mathrm{mW}$ for An:TCNB. They account for the total volume of photoexcited triplet spins in the field of view, and are intended to give first-order preliminary estimates of MAS crystals as magnetic-field-sensitive spin layers.

\end{document}

%% file: Commands3.tex
\newcommand{\beq}{\begin{equation}}
\newcommand{\eeq}{\end{equation}}
                  
\newcommand{\benum}{\begin{enumerate}}
\newcommand{\eenum}{\end{enumerate}}
                    
\newcommand{\bit}{\begin{itemize}}
\newcommand{\eit}{\end{itemize}}

\newcommand{\bea}{\begin{eqnarray}}
\newcommand{\eea}{\end{eqnarray}}

\newcommand{\bev}{\begin{verbatim}}
\newcommand{\eev}{\end{verbatim}}

\newcommand{\noi}{\noindent}

\newcommand{\T}[1]{\textbf{#1}}
\newcommand{\I}[1]{\textit{#1}}

\newcommand{\ba}{\left\{ \begin{array}{lr}}
\newcommand{\ea}{\end{array}\right.}

\newcommand{\blist}[1]{
 \begin{list}{#1}%
 \begin{align}
	 arrow
 \end{align}
 $\checkmark\star
  { \setlength{\itemsep}{3pt}
     \setlength{\parsep}{2pt}
     \setlength{\topsep}{3pt}
     \setlength{\partopsep}{0pt}
     \setlength{\leftmargin}{1em}
     \setlength{\labelwidth}{1em}
     \setlength{\labelsep}{0.5em} } }
\newcommand{\elist}{
  \end{list}  }

\DeclareMathSymbol{\vartheta}{\mathalpha}{letters}{"12}
\DeclareMathSymbol{\theta}{\mathalpha}{letters}{"23}
\DeclareMathSymbol{\phi}{\mathalpha}{letters}{"27}
\DeclareMathSymbol{\varphi}{\mathalpha}{letters}{"1E}

\newcommand{\bef}
{
\begin{figure}[htbp]
\centering
}

\newcommand{\eef}{\end{figure}}